\documentclass[aip,jcp,twocolumn,showpacs,superscriptaddress,reprint]{revtex4-1}
\usepackage{graphicx}
\usepackage{siunitx}
\usepackage{bm}
\usepackage{hyperref}
\usepackage[caption=false]{subfig}
\usepackage{mathtools}
\usepackage{multirow}
\usepackage{amssymb}
\usepackage{braket}
\usepackage{amsmath}
\usepackage{placeins}
\usepackage{cancel}
\usepackage{xcolor}
\usepackage{spreadtab}

\begin{document}

\title{Temperature dependence of the electronic structure of semiconductors and insulators}

\author{S. Ponc\'e}
\email{samuel.pon@gmail.com}
\affiliation{%
European Theoretical Spectroscopy Facility and Institute of Condensed Matter and Nanosciences, Universit\'e catholique de Louvain, Chemin des \'etoiles 8, bte L07.03.01, B-1348 Louvain-la-neuve, Belgium.
}%
\author{Y. Gillet}
\affiliation{%
European Theoretical Spectroscopy Facility and Institute of Condensed Matter and Nanosciences, Universit\'e catholique de Louvain, Chemin des \'etoiles 8, bte L07.03.01, B-1348 Louvain-la-neuve, Belgium.
}%
\author{J. Laflamme Janssen}
\affiliation{%
European Theoretical Spectroscopy Facility and Institute of Condensed Matter and Nanosciences, Universit\'e catholique de Louvain, Chemin des \'etoiles 8, bte L07.03.01, B-1348 Louvain-la-neuve, Belgium.
}%
\author{A. Marini}
\affiliation{%
Consiglio Nazionale delle Ricerche (CNR),Via Salaria Km 29.3, CP 10, 00016, Monterotondo Stazione, Italy
}%
\author{M. Verstraete}
\affiliation{%
European Theoretical Spectroscopy Facility and Physique des mat\'eriaux et nanostructures, Universit\'e de Li\`ege, All\'ee du 6 Ao\^ut 17, B-4000 Li\`ege, Belgium. 
}%
\author{X. Gonze}
\affiliation{%
European Theoretical Spectroscopy Facility and Institute of Condensed Matter and Nanosciences, Universit\'e catholique de Louvain, Chemin des \'etoiles 8, bte L07.03.01, B-1348 Louvain-la-neuve, Belgium.
}%

\date{\today}

\begin{abstract}

The renormalization of electronic eigenenergies due to electron-phonon coupling (temperature dependence and zero-point motion effect) is sizeable in many materials with light atoms.
This effect, often neglected in \textit{ab-initio} calculations, can be computed using the perturbation-based Allen-Heine-Cardona theory in the adiabatic or non-adiabatic harmonic approximation. After a short description of the numerous recent progresses in this field, and a brief overview of the theory, we focus on the issue of phonon wavevector sampling convergence, until now poorly understood.
Indeed, the renormalization is obtained numerically through a (usually slowly converging) $\mathbf{q}$-point sampling inside the Brillouin Zone. For $\mathbf{q}$-points close to $\Gamma$, we show that a divergence due to non-zero Born effective charge appears in the electron-phonon matrix elements, leading to a divergence of the integral over the (phonon) Brillouin zone for band extrema.
Although it should vanish for non-polar materials, unphysical residual Born effective charges are usually present in \textit{ab-initio} calculations.
Here, we propose a solution that improves the coupled (electronic) $\mathbf{k}$-point convergence dramatically. 
For polar materials, the problem is more severe: the divergence of the integral does not disappear in the adiabatic harmonic approximation, but only in the non-adiabatic harmonic approximation. 
In all cases, we study in detail the convergence behavior of the renormalization as the $\mathbf{q}$-point sampling goes to infinity and the imaginary broadening broadening parameter goes to zero. This allows extrapolation, thus enabling a systematic way to converge the renormalization for both polar and non-polar materials. 
Finally, the adiabatic and non-adiabatic theory, with corrections for the divergence problem, are applied to the study of five semiconductors and insulators:  $\alpha$-AlN, $\beta$-AlN, BN, diamond and silicon. 
For these five materials, we present the zero-point renormalization, temperature dependence, phonon-induced lifetime broadening and the renormalized electronic bandstructure.

\end{abstract}


\maketitle

\section{Introduction}
\label{Introduction}

The theoretical understanding of the effects of the electron-phonon coupling on the electronic structure and the  capability to compute them have a long and chaotic history that started in the early fifties. 
Over the years, these effects have been computed using three types of methods, with different advantages and drawbacks: 
(1) as a time average of the bandgap using first-principles molecular dynamics (MD) simulations; 
(2) through the frozen-phonon (FP) method, which weights the eigenenergy change along the phonon modes with a Bose-Einstein distribution; 
(3) thanks to the diagrammatic method of many-body perturbation theory. For a historical review, the reader can consult Ref.~\onlinecite{Ponce2014a},
in which these three types of methods are compared to each other, at the harmonic level.

In the present contribution, we rely on the Allen-Heine-Cardona (AHC) theory~\cite{Allen1976,Allen1981,Allen1983} to compute the zero-point motion renormalization as well as the temperature dependence of electronic eigenenergies. The AHC theory originates from the diagrammatic method of many-body perturbation theory. It has been applied in several recent milestone contributions in the field, including the computation of temperature-dependence of the optical properties~\cite{Marini2008}, 
the computation of the surprisingly large zero-point renormalization (ZPR) of the diamond bandgap~\cite{Giustino2010,Ponce2014}, 
the demonstration of large non-rigid ion corrections for molecules~\cite{Gonze2011}, the inclusion of dynamical effects beyond the adiabatic approximation~\cite{Cannuccia2011, Cannuccia2012, Marini2015}, the study of the anharmonic electron-phonon contribution to the indirect bandgap of diamond~\cite{Monserrat2013}, and the inclusion of electronic many-body effects (in the GW approximation) in diamond~\cite{Antonius2014}, noticing a large increase of the renormalization with respect to density-functional theory (DFT). Also, we think that the confusion in the theoretical understanding of the relationship between MD, FP, and AHC as well as the inaccuracies in first-principles 
software implementations of AHC have been largely eliminated in two recent publications~\cite{Ponce2014,Ponce2014a}.

One of the major issues when performing AHC calculations is the slow convergence with respect to phonon wavevector sampling of the Brillouin Zone (BZ)~\cite{Ponce2014}, refereed to as $\mathbf{q}$-point sampling from now on.
To accelerate this convergence, a small imaginary component $i\delta$ 
(which can be inferred as a finite lifetime for the unoccupied electronic states due to thermal effects) 
is often used. However, this imaginary parameter is \textit{ad hoc} rather than \textit{ab-initio}. 
Also, the convergence problem is even more severe with the MD and FP methods, as supercells have to be used to sample the phonons wavevectors, thus dramatically 
increasing the computational time and memory required. Actually, numerical convergence for the MD and FP methods cannot really be reached in three-dimensional solids, in contrast with finite systems~\cite{Ponce2014,Antonius2014,Ramirez2008}.

In this paper, we highlight that when trying to converge the ZPR with respect to $\mathbf{q}$-point sampling for vanishing $i\delta$ in the AHC simulations, the ZPR diverges. 
For non-polar materials, such unphysical divergence is attributed to a residual Born effective charge, which stems from the finite $\mathbf{k}$-point sampling. 
We propose a solution to this problem and devise a systematic way to converge the ZPR for vanishing $i\delta$.  For polar materials,
the problem is more profound. Indeed, the divergence in the adiabatic AHC approach is not simply numerical, but indicates a breakdown of the AHC approach.
A similar problem should also be present in the MD and FT methods.
On the other hand, the non-adiabatic AHC theory naturally leads to non-diverging quantities.
 
This paper is organized as follow. First, a short reminder of the AHC theory is presented in section~\ref{static}. In Sec.~\ref{convergence}, the bottleneck of the $\mathbf{q}$-point convergence is discussed, the divergence problem of the ZPR at large $\mathbf{q}$-point density is explored and a solution is proposed. 
We also device in sections~\ref{qconv} and \ref{idconv} a systematic and parameter free way to extrapolate the ZPR (without $i\delta$).
Finally in section~\ref{results}, we present the temperature dependences, the zero-point motion renormalizations, as well as the phonon-induced lifetimes for five semiconductors: $\alpha$-AlN, $\beta$-AlN, BN, diamond and silicon.

\section{Review of the Allen-Heine-Cardona formalism}
\label{static}
\subsection{The AHC theory within the adiabatic harmonic approximation}

The temperature-dependent renormalization of the electronic eigenenergy $\varepsilon_{n\mathbf{k}}$ for band $n$ and wavevector $\mathbf{k}$ can be written in the adiabatic harmonic approximation as a sum over the BZ of the phonon contributions for each wavevector $\mathbf{q}$~\cite{Ponce2014a}
\begin{equation}\label{Denk(T)}
 \Delta\varepsilon_{n\mathbf{k}}(T) = \frac{1}{N_q}\sum_{\mathbf{q}} \sum_{m}^{3N} \frac{\partial \varepsilon_{n\mathbf{k}}}{\partial n_{m\mathbf{q}}}\Big(n_{m\mathbf{q}}(T)+\frac{1}{2}\Big),
\end{equation}
with
\begin{multline}\label{denk_dnmq}
 \frac{\partial \varepsilon_{n\mathbf{k}}}{\partial n_{m\mathbf{q}}} = 
    \frac{1}{2\omega_{m\mathbf{q}}}  \sum_{\substack{\kappa\alpha\\ \kappa'\gamma}}
 \sum_{ll'}\frac{\partial^2 \varepsilon_{n\mathbf{k}}}{\partial R_{l\kappa\alpha}\partial R_{l'\kappa'\gamma}}\\
 e^{-i\mathbf{q}\cdot (\mathbf{R}_{l}-\mathbf{R}_{l'})}
U_{m,\kappa'\gamma}^*(\mathbf{q})U_{m,\kappa\alpha}(\mathbf{q}),
\end{multline} 
where $m$ is the phonon branch, $T$ is the temperature, $N_q$ is the number of wavevectors used to sample the BZ, $\omega_{m\mathbf{q}}$ is the phonon frequency, $n_{m\mathbf{q}}(T)= \frac{1}{e^{\frac{\omega_{m\mathbf{q}}}{k_B T}}-1}$ is the Bose-Einstein distribution, $U_{m,\kappa\alpha}(\mathbf{q})$ is the eigendisplacement vector of atom $\kappa$ in direction $\alpha$ associated to the phonon mode, and $\partial /\partial R_{l\kappa\alpha}$ is the derivative of a quantity with respect to the displacement of atom $\kappa$ of the unit cell $l$ in the direction $\alpha$.

The quantity $\Delta\varepsilon_{n\mathbf{k}}(T)\triangleq \varepsilon_{n\mathbf{k}}(T)-\varepsilon_{n\mathbf{k}}[0]$ is the difference between the temperature-dependent eigenenergy and the eigenenergy at ground-state atomic positions. We distinguish the temperature-dependent $\varepsilon_{n\mathbf{k}}$ from the atomic-position-dependent $\varepsilon_{n\mathbf{k}}$ by using $()$ in the former and $[]$ in the latter. The difference between $\varepsilon_{n\mathbf{k}}(T=0)$ and $\varepsilon_{n\mathbf{k}}[0]$ is called the zero-point motion renormalization (ZPR). 

In a mean field approximation like the Density Functional Theory (DFT), the eigenenergies are the expectation values of the Hamiltonian $\hat{H}_{\mathbf{k,k}}$ of the system
\begin{equation}\label{eigenenergy}
\varepsilon_{n\mathbf{k}} = \Bra{u_{n\mathbf{k}}^{(0)}}\hat{H}_{\mathbf{k,k}}\Ket{u_{n\mathbf{k}}^{(0)}},
\end{equation}
with $u_{n\mathbf{k}}^{(0)}$ the periodic part of the electronic wavefunctions. Using perturbation theory to obtain the second-order derivative with respect to atomic displacements of such eigenenergies, Eq.~\eqref{denk_dnmq} can be rewritten as
\begin{multline}\label{zprequation}
 \frac{\partial \varepsilon_{n\mathbf{k}}}{\partial n_{m\mathbf{q}}} = 
 \frac{1}{2\omega_{m\mathbf{q}}}
 \sum_{\substack{\kappa\alpha\\ \kappa'\gamma}} U_{m,\kappa'\gamma}^*(\mathbf{q})U_{m,\kappa\alpha}(\mathbf{q}) \\
\bigg\{ \Bra{ u_{n\mathbf{k}}^{(0)}} \frac{\partial^2 \hat{H}_{\mathbf{k,k}}}{\partial R_{\kappa\alpha}(-\mathbf{q}) \partial R_{\kappa'\gamma}(\mathbf{q})} \Ket{ u_{n\mathbf{k}}^{(0)}} \\
  + \frac{1}{2} \bigg( \Big( \Big\langle \frac{\partial u_{n\mathbf{k}}}{\partial R_{\kappa\alpha}(\mathbf{q})} \Big|\frac{\partial \hat{H}_{\mathbf{k,k}}}{\partial R_{\kappa'\gamma}(\mathbf{q})} \Big|u_{n\mathbf{k}}^{(0)}\Big\rangle \\
   + (\kappa\alpha)\leftrightarrow  (\kappa'\gamma)\Big) + (c.c.)\bigg)\bigg\},
\end{multline}
where $(c.c.)$ stands for the complex conjugate of the previous terms within parenthesis (), and where we use the following notation for the derivative with respect to atomic positions of an arbitrary quantity $X$
\begin{equation}\label{defbasederiv}
\frac{\partial X}{\partial R_{\kappa\alpha}(\mathbf{q})} = \frac{1}{N_{BvK}}\sum_l e^{i\mathbf{q}\cdot \mathbf{R}_l}\frac{\partial X}{\partial R_{l\kappa\alpha}},
\end{equation}
$N_{BvK}$ being the number of primitive cells of the periodic system defined by the Born-von Karman boundary conditions~\cite{Born1954}.

The first term within $\{ \}$ in Eq.~\eqref{zprequation} is called the Debye-Waller (DW) term 
\begin{equation}\label{DW}
\mathcal{D}_{\substack{\kappa\alpha\\ \kappa'\gamma}}(\mathbf{q}) \triangleq \Big\langle u_{n\mathbf{k}}^{(0)} \Big| \frac{\partial^2 \hat{H}_{\mathbf{k,k}}}{\partial R_{\kappa\alpha}(-\mathbf{q})\partial R_{\kappa'\gamma}(\mathbf{q})}\Big| u_{n\mathbf{k}}^{(0)}\Big\rangle,
\end{equation}
while the remainder constitutes the Fan term 
\begin{equation}\label{Fan}
\begin{split}
\mathcal{F}_{\substack{\kappa\alpha\\ \kappa'\gamma}}(\mathbf{q}) \triangleq 
\frac{1}{2} \bigg[\bigg( \Big( 
\Big\langle \frac{\partial u_{n\mathbf{k}}}{\partial R_{\kappa\alpha}(\mathbf{q})} \Big| \frac{\partial \hat{H}_{\mathbf{k,k}}}{\partial R_{\kappa'\gamma}(\mathbf{q}) }\Big| u_{n\mathbf{k}}^{(0)}\Big\rangle \\
 + (\kappa\alpha) \leftrightarrow (\kappa'\gamma) \Big) + (c.c.)\bigg) \bigg].
\end{split}
\end{equation}

The change of eigenenergy due to a specific phonon mode (e.g. Eq. \eqref{zprequation}) thus becomes
\begin{equation}\label{denk_dnmq_abbrev}
 \frac{\partial \varepsilon_{n\mathbf{k}}}{\partial n_{m\mathbf{q}}} \triangleq \frac{\partial \varepsilon_{n\mathbf{k}}^{\text{FAN}}}{\partial n_{m\mathbf{q}}} +\frac{\partial \varepsilon_{n\mathbf{k}}^{\text{DW}}}{\partial n_{m\mathbf{q}}}, 
\end{equation} 
with the Fan contribution given by
\begin{multline}\label{denk_dnmq_FAN}
\frac{\partial \varepsilon_{n\mathbf{k}}^{\text{FAN}}}{\partial n_{m\mathbf{q}}} 
\triangleq
\frac{1}{2\omega_{m\mathbf{q}}} \sum_{\substack{\kappa\alpha\\ \kappa'\gamma}} \mathcal{F}_{\substack{\kappa\alpha\\ \kappa'\gamma}}(\mathbf{q})U_{m,\kappa'\gamma}^*(\mathbf{q})U_{m,\kappa\alpha}(\mathbf{q}),
\end{multline}
and the Debye-Waller contribution given by
\begin{multline}\label{denk_dnmq_DW}
\frac{\partial \varepsilon_{n\mathbf{k}}^{\text{DW}}}{\partial n_{m\mathbf{q}}} 
\triangleq
\frac{1}{2\omega_{m\mathbf{q}}} \sum_{\substack{\kappa\alpha\\ \kappa'\gamma}} \mathcal{D}_{\substack{\kappa\alpha\\ \kappa'\gamma}}(\mathbf{q})U_{m,\kappa'\gamma}^*(\mathbf{q})U_{m,\kappa\alpha}(\mathbf{q}).
\end{multline}
At this point, no approximations beyond  the adiabatic and harmonic ones were made. However, the calculation of the Debye-Waller term (Eq.~\eqref{DW}) requires the second-order derivative of the Hamiltonian, which is a computational bottleneck within the density functional perturbation theory (DFPT) framework. To overcome this issue, we make the rigid-ion approximation (RIA) as is usual within the AHC theory
~\cite{Allen1976,Allen1981,Allen1983}. 
We begin by splitting the DW term into two parts. The first part contains all contributions that can be computed using first-order derivatives of the Hamiltonians, while the second part contains the remaining contributions
\begin{equation}
\frac{\partial \varepsilon_{n\mathbf{k}}^{\text{DW}}}{\partial n_{m\mathbf{q}}}
= \frac{\partial \varepsilon_{n\mathbf{k}}^{\text{DW}_{\text{RIA}}}}{\partial n_{m\mathbf{q}}}
+ \frac{\partial \varepsilon_{n\mathbf{k}}^{\text{DW}_{\text{NRIA}}}}{\partial n_{m\mathbf{q}}},
\end{equation}
with 
\begin{multline}\label{denk_dnmq_DDW_RIA}
\frac{\partial \varepsilon_{n\mathbf{k}}^{\text{DW}_{\text{RIA}}}}{\partial n_{m\mathbf{q}}} 
= \frac{-1}{4\omega_{m\mathbf{q}}}\sum_{\substack{\kappa\alpha\\ \kappa'\gamma}}   \mathcal{F}_{\substack{\kappa\alpha\\ \kappa'\gamma}}(\boldsymbol{\Gamma}) \\
  \Big( U_{m,\kappa\gamma}^*(\mathbf{q})U_{m,\kappa\alpha}(\mathbf{q}) + U_{m,\kappa'\gamma}^*(\mathbf{q})U_{m,\kappa'\alpha}(\mathbf{q})\Big),
\end{multline} 
and
\begin{multline}\label{NDDW}
\frac{\partial \varepsilon_{n\mathbf{k}}^{\text{DW}_{\text{NRIA}}}}{\partial n_{m\mathbf{q}}}
 = \\
  \frac{1}{2\omega_{m\mathbf{q}}}\sum_{\substack{\kappa\alpha\\ \kappa'\gamma}}
 \bigg[ \mathcal{D}_{\substack{\kappa\alpha\\ \kappa'\gamma}}(\mathbf{q})
    U_{m,\kappa'\gamma}^*(\mathbf{q})U_{m,\kappa\alpha}(\mathbf{q}) 
     -  \frac{1}{2}\mathcal{D}_{\substack{\kappa\alpha\\ \kappa'\gamma}}(\boldsymbol{\Gamma}) \\
   \Big( U_{m,\kappa\gamma}^*(\mathbf{q})U_{m,\kappa\alpha}(\mathbf{q})
                   + U_{m,\kappa'\gamma}^*(\mathbf{q})U_{m,\kappa'\alpha}(\mathbf{q})\Big) \bigg].
\end{multline}

Moreover, in our calculations, all Fan-like contributions are obtained within DFPT and can thus be written as follow 
\begin{multline}\label{matrix_element}
\Big\langle \frac{\partial u_{n\mathbf{k}}}{\partial R_{\kappa\alpha}(\mathbf{q})} \Big| \frac{\partial \hat{H}_{\mathbf{k,k}}}{\partial R_{\kappa'\gamma}(\mathbf{q}) }\Big| u_{n\mathbf{k}}^{(0)}\Big\rangle
=\\
\sideset{}{'}\sum_{n'=1}^{M} 
\frac{
\Big\langle u_{n\mathbf{k}}^{(0)}\Big| \frac{\partial \hat{H}_{\mathbf{k,k}}}{\partial R_{\kappa\alpha}(-\mathbf{q})}  \Big| u_{n'\mathbf{k+q}}^{(0)} \Big\rangle
\Big\langle u_{n'\mathbf{k+q}}^{(0)}\Big| \frac{\partial \hat{H}_{\mathbf{k,k}}}{\partial R_{\kappa'\gamma}(\mathbf{q})}  \Big| u_{n\mathbf{k}}^{(0)} \Big\rangle}
{\varepsilon_{n\mathbf{k}}^{(0)}-\varepsilon_{n'\mathbf{k+q}}^{(0)}}
\\
+
\Big\langle {\hat P}_{c\mathbf{k+q}}{\frac{\partial u_{n\mathbf{k}}}{\partial R_{\kappa\alpha}(\mathbf{q})}}\Big| \frac{\partial \hat{H}_{\mathbf{k,k}}}{\partial R_{\kappa'\gamma}(\mathbf{q})}  \Big| u_{n\mathbf{k}}^{(0)} \Big\rangle,
\end{multline}
where the summation over energetic bands (above $M$) has been replaced by the solution $\big| {\hat P}_{c\mathbf{k+q}}{\frac{\partial u_{n\mathbf{k}}}{\partial R_{\kappa\alpha}(\mathbf{q})}}\big\rangle$ of a linear equation as proposed by Sternheimer~\cite{Sternheimer1954} and applied to this problem in Ref.~\onlinecite{Gonze2011}. The definition for the projector ${\hat P}_{c\mathbf{k+q}}$  and active space $M$ as well as the description of the linear equation to be solved can be found in the appendix of Ref.~\onlinecite{Ponce2014a}. 

We finally obtain the adiabatic temperature-dependent renormalization in the RIA by neglecting the non-RIA contribution as defined by Eq.~\eqref{NDDW}, which yields
\begin{widetext}
\begin{multline}\label{equationsumoverstate_id}
 \Delta\varepsilon_{n\mathbf{k}}^{(\text{adiabatic,RIA})}(T) = \frac{1}{N_q}\sum_{\mathbf{q}} \sum_{m}^{3N} \Big(n_{m\mathbf{q}}(T)+\frac{1}{2}\Big)  \frac{1}{4\omega_{m\mathbf{q}}} \sum_{\substack{\kappa\alpha\\ \kappa'\gamma}} 
\Bigg\{\Bigg(\bigg[\bigg[ \sideset{}{'}\sum_{n'=1}^{M} 
\frac{
\Big\langle u_{n\mathbf{k}}^{(0)}\Big| \frac{\partial \hat{H}_{\mathbf{k,k}}}{\partial R_{\kappa\alpha}(-\mathbf{q})}  \Big| u_{n'\mathbf{k+q}}^{(0)} \Big\rangle
\Big\langle u_{n'\mathbf{k+q}}^{(0)}\Big| \frac{\partial \hat{H}_{\mathbf{k,k}}}{\partial R_{\kappa'\gamma}(\mathbf{q})}  \Big| u_{n\mathbf{k}}^{(0)} \Big\rangle}
{\varepsilon_{n\mathbf{k}}^{(0)}-\varepsilon_{n'\mathbf{k+q}}^{(0)}+i\delta}
\\
+
\Big\langle \hat{P}_{c\mathbf{k+q}}{\frac{\partial u_{n\mathbf{k}}}{\partial R_{\kappa\alpha}(\mathbf{q})}}\Big| \frac{\partial \hat{H}_{\mathbf{k,k}}}{\partial R_{\kappa'\gamma}(\mathbf{q})}  \Big| u_{n\mathbf{k}}^{(0)} \Big\rangle \bigg] + (\kappa\alpha) \leftrightarrow (\kappa'\gamma) \bigg] + (c.c.)  \Bigg) U_{m,\kappa'\gamma}^*(\mathbf{q})U_{m,\kappa\alpha}(\mathbf{q}) \\
-\frac{1}{2}\Bigg(\bigg[\bigg[\sideset{}{'}\sum_{n'=1}^{M} 
\frac{
\Big\langle u_{n\mathbf{k}}^{(0)}\Big| \frac{\partial \hat{H}_{\mathbf{k,k}}}{\partial R_{\kappa\alpha}(\boldsymbol{\Gamma})}  \Big| u_{n'\mathbf{k}}^{(0)} \Big\rangle
\Big\langle u_{n'\mathbf{k}}^{(0)}\Big| \frac{\partial \hat{H}_{\mathbf{k,k}}}{\partial R_{\kappa'\gamma}(\boldsymbol{\Gamma})}  \Big| u_{n\mathbf{k}}^{(0)} \Big\rangle}
{\varepsilon_{n\mathbf{k}}^{(0)}-\varepsilon_{n'\mathbf{k}}^{(0)}+i\delta}
+
\Big\langle \hat{P}_{c\mathbf{k}}{\frac{\partial u_{n\mathbf{k}}}{\partial R_{\kappa\alpha}(\boldsymbol{\Gamma})}}\Big| \frac{\partial \hat{H}_{\mathbf{k,k}}}{\partial R_{\kappa'\gamma}(\boldsymbol{\Gamma})}  \Big| u_{n\mathbf{k}}^{(0)} \Big\rangle  \bigg]  \\
+(\kappa\alpha) \leftrightarrow (\kappa'\gamma)\bigg] + (c.c.)  \Bigg)  \Big( U_{m,\kappa\gamma}^*(\mathbf{q})U_{m,\kappa\alpha}(\mathbf{q}) 
+ U_{m,\kappa'\gamma}^*(\mathbf{q})U_{m,\kappa'\alpha}(\mathbf{q})\Big)\Bigg\}, 
\end{multline}
\end{widetext}
where a small imaginary component $i\delta$ is usually introduced in the AHC equation to smooth the energy denominators. For example, in the case of diamond, several authors have used an $i\delta$ of 100~meV to account for the finite lifetimes of the electronic states~\cite{Zollner1992,Giustino2010,Antonius2014,Ponce2014}. However, the theory must also be valid (apart from controlled numerical instabilities) for vanishing $i\delta$. This point will be further discussed in section~\ref{convergence}.  

\subsection{Beyond the Rayleigh-Schr\"odinger perturbation theory}

Phonons alter the one-electron energy bands $\varepsilon_{n\mathbf{k}}$ in two ways: there is a shift $\Delta \varepsilon_{n\mathbf{k}}$ and a lifetime broadening $1/\tau_{n\mathbf{k}}$. As seen in the previous sub-section, the adiabatic approximation leads to a real renormalization of the eigenstates. The study of the lifetime broadening requires an extension of the adiabatic theory. 

In 1978, Allen generalized his earlier work~\cite{Allen1976} derived within the standard Rayleigh-Schr\"odinger perturbation theory to include finite phonon frequencies using many-body perturbation techniques~\cite{Allen1978}.
These techniques describe excitations in terms of spectral functions~\cite{Cannuccia2012}, where quasiparticules cannot always be unambiguously identified,  with the associated well defined eigenenergies. In this work,  following Allen~\cite{Allen1978}, rather than obtaining the full spectral function to describe the electronic excitation, we suppose that their description in terms of quasiparticles is still valid and evaluate the associated eigenenergies by correcting the DFT eigenvalues to first-order in perturbation theory, taking the self-energy evaluated at $\omega = \varepsilon_{n\mathbf{k}}^{(0)}$ as the perturbation. 
Complex eigenenergies are obtained within this generalization, that we refer to as the ``non-adiabatic" extension of the AHC theory $\varepsilon_{n\mathbf{k}}(T,\omega)$~\cite{Allen1978,Grimvall1981}.

%
%
 
We therefore obtain the following equation, based on electron-phonon matrix elements already calculated for the adiabatic renormalization
\begin{widetext}
\begin{multline}\label{equation_dynamical}
 \Delta\varepsilon_{n\mathbf{k}}^{(\text{non-adiabatic,RIA})}(T) = \Re  \frac{1}{N_q}\sum_{\mathbf{q}} \sum_{m}^{3N}   \frac{1}{4\omega_{m\mathbf{q}}} \sum_{\substack{\kappa\alpha\\ \kappa'\gamma}} 
\Bigg\{\Bigg(\bigg[\bigg[ \sum_{n'=1}^{M} 
\Big\langle u_{n\mathbf{k}}^{(0)}\Big| \frac{\partial \hat{H}_{\mathbf{k,k}}}{\partial R_{\kappa\alpha}(-\mathbf{q})}  \Big| u_{n'\mathbf{k+q}}^{(0)} \Big\rangle
\Big\langle u_{n'\mathbf{k+q}}^{(0)}\Big| \frac{\partial \hat{H}_{\mathbf{k,k}}}{\partial R_{\kappa'\gamma}(\mathbf{q})}  \Big| u_{n\mathbf{k}}^{(0)} \Big\rangle \\
\frac{1}{2}\bigg( \frac{n_{m\mathbf{q}}(T)+f_{n'\mathbf{k+q}}}{\varepsilon_{n\mathbf{k}}^{(0)}-\varepsilon_{n'\mathbf{k+q}}^{(0)}+\omega_{m\mathbf{q}}+i\delta\text{sgn}(\varepsilon_{n\mathbf{k}}-\mu)} 
+ \frac{n_{m\mathbf{q}}(T)+1-f_{n'\mathbf{k+q}}}{\varepsilon_{n\mathbf{k}}^{(0)}-\varepsilon_{n'\mathbf{k+q}}^{(0)}-\omega_{m\mathbf{q}}+i\delta\text{sgn}(\varepsilon_{n\mathbf{k}}-\mu)} \bigg)
\\
+\Big\langle P_{c\mathbf{k+q}}{\frac{\partial u_{n\mathbf{k}}}{\partial R_{\kappa\alpha}(\mathbf{q})}}\Big| \frac{\partial \hat{H}_{\mathbf{k,k}}}{\partial R_{\kappa'\gamma}(\mathbf{q})}  \Big| u_{n\mathbf{k}}^{(0)} \Big\rangle \Big(n_{m\mathbf{q}}(T)+\frac{1}{2}\Big) \bigg] +(\kappa\alpha) \leftrightarrow (\kappa'\gamma)\bigg] + (c.c.)  \Bigg)  
U_{m,\kappa'\gamma}^*(\mathbf{q})U_{m,\kappa\alpha}(\mathbf{q}) \\
-\frac{1}{2} \Bigg( \bigg[\bigg[ \sum_{n'=1}^{M} 
\frac{
\Big\langle u_{n\mathbf{k}}^{(0)}\Big| \frac{\partial \hat{H}_{\mathbf{k,k}}}{\partial R_{\kappa\alpha}(\boldsymbol{\Gamma})}  \Big| u_{n'\mathbf{k}}^{(0)} \Big\rangle
\Big\langle u_{n'\mathbf{k}}^{(0)}\Big| \frac{\partial \hat{H}_{\mathbf{k,k}}}{\partial R_{\kappa'\gamma}(\boldsymbol{\Gamma})}  \Big| u_{n\mathbf{k}}^{(0)} \Big\rangle}
{\varepsilon_{n\mathbf{k}}^{(0)}-\varepsilon_{n'\mathbf{k}}^{(0)}+i\delta}
+ \Big\langle P_{c\mathbf{k}}{\frac{\partial u_{n\mathbf{k}}}{\partial R_{\kappa\alpha}(\boldsymbol{\Gamma})}}\Big| \frac{\partial \hat{H}_{\mathbf{k,k}}}{\partial R_{\kappa'\gamma}(\boldsymbol{\Gamma})}  \Big| u_{n\mathbf{k}}^{(0)} \Big\rangle \bigg] \\
+(\kappa\alpha) \leftrightarrow (\kappa'\gamma)\bigg] + (c.c.)  \Bigg) \Big( U_{m,\kappa\gamma}^*(\mathbf{q})  U_{m,\kappa\alpha}(\mathbf{q}) + U_{m,\kappa'\gamma}^*(\mathbf{q})U_{m,\kappa'\alpha}(\mathbf{q})\Big)\Bigg\},
\end{multline}
\end{widetext}
where $\mu$ is the chemical potential, $f_{n\mathbf{k}}$ is the electronic occupation of the wavevector $\mathbf{k}$ at band $n$ and  
where a convergence study on $M$ is required for the Fan term due to the fact that the Sternheimer solution neglects the phonon frequency $\omega_{m\mathbf{q}}$ while the sum over the active space does not. 

The phonon-induced lifetime broadening $1/\tau_{n\mathbf{k}}$ is the imaginary part of the complex Fan self-energy 
\begin{multline}\label{phonon_induced_lifetime_eq_non_adia}
\frac{1}{2\tau_{n\mathbf{k}}^{(\text{non-adiabatic,RIA})}} \\
= \frac{\pi}{N_q}\sum_{\mathbf{q}}\sum_{m}^{3N}\frac{1}{8\omega_{m\mathbf{q}}}   \sum_{\substack{\kappa\alpha\\ \kappa'\gamma}} U_{m,\kappa'\gamma}^*(\mathbf{q})U_{m,\kappa\alpha}(\mathbf{q}) \sideset{}{'}\sum_{n'=1}^{M} \bigg( \Big[    \\
\Big[ \Big\langle u_{n\mathbf{k}}^{(0)}\Big| \frac{\partial \hat{H}_{\mathbf{k,k}}}{\partial R_{\kappa\alpha}(-\mathbf{q})}  \Big| u_{n'\mathbf{k+q}}^{(0)} \Big\rangle\Big\langle u_{n'\mathbf{k+q}}^{(0)}\Big| \frac{\partial \hat{H}_{\mathbf{k,k}}}{\partial R_{\kappa'\gamma}(\mathbf{q})}  \Big| u_{n\mathbf{k}}^{(0)} \Big\rangle \Big]\\
+(\kappa\alpha) \leftrightarrow (\kappa'\gamma)\Big] + (c.c.)  \bigg) \\
\Big( \big( n_{m\mathbf{q}}(T) + f_{n'\mathbf{k+q}} \big) \delta(\varepsilon_{n\mathbf{k}}^{(0)}-\varepsilon_{n'\mathbf{k+q}}^{(0)}+\omega_{m\mathbf{q}})+\\
\big( n_{m\mathbf{q}}(T) +1- f_{n'\mathbf{k+q}} \big)\delta(\varepsilon_{n\mathbf{k}}^{(0)}-\varepsilon_{n'\mathbf{k+q}}^{(0)}-\omega_{m\mathbf{q}})\Big),
\end{multline}
where $\delta$ is the Dirac delta (broadened for numerical reasons). 

The phonon-induced lifetime broadening in the adiabatic limit ($\omega_{m\mathbf{q}} << \varepsilon_{n\mathbf{k}}^{(0)} - \varepsilon_{n'\mathbf{k+q}}^{(0)}$) is
\begin{multline}\label{phonon_induced_lifetime_eq}
\frac{1}{2\tau_{n\mathbf{k}}^{(\text{adiabatic,RIA})}} = \frac{\pi}{N_q}\sum_{\mathbf{q}}\sum_{m}^{3N}\frac{1}{4\omega_{m\mathbf{q}}} \Big(n_{m\mathbf{q}}(T)+\frac{1}{2}\Big) \\
 \sum_{\substack{\kappa\alpha\\ \kappa'\gamma}} \sideset{}{'}\sum_{n'=1}^{M} \bigg( \Big[  
\Big[\Big\langle u_{n\mathbf{k}}^{(0)}\Big| \frac{\partial \hat{H}_{\mathbf{k,k}}}{\partial R_{\kappa\alpha}(-\mathbf{q})}  \Big| u_{n'\mathbf{k+q}}^{(0)} \Big\rangle \\
\Big\langle u_{n'\mathbf{k+q}}^{(0)}\Big| \frac{\partial \hat{H}_{\mathbf{k,k}}}{\partial R_{\kappa'\gamma}(\mathbf{q})}  \Big| u_{n\mathbf{k}}^{(0)} \Big\rangle \Big] +(\kappa\alpha) \leftrightarrow (\kappa'\gamma)\Big] + (c.c.)  \bigg) \\
  \delta (\varepsilon_{n\mathbf{k}}^{(0)}-\varepsilon_{n'\mathbf{k+q}}^{(0)})U_{m,\kappa'\gamma}^*(\mathbf{q})U_{m,\kappa\alpha}(\mathbf{q}) .
\end{multline}

The adiabatic and non-adiabatic renormalizations, Eqs.~\eqref{equationsumoverstate_id} and~\eqref{equation_dynamical}, as well as the 
adiabatic lifetime Eq.~\eqref{phonon_induced_lifetime_eq} have been coded in the ABINIT software (v7.11), and will be used in the following sections. 

\section{Phonon wavevector sampling and the divergence problem}
\label{convergence}

\subsection{Potential breakdown of perturbation theory}

Quantum mechanical perturbation theory can breakdown when vanishing denominators appear in the perturbation series. This can happen in the present case,
as the short-hand form of Eq.~\eqref{equationsumoverstate_id} is
\begin{equation}\label{propintegral}
\Delta\varepsilon_{n\mathbf{k}}^{(\text{adiabatic,RIA})}(T) \propto \sum_\mathbf{q} \frac{|GKK(\mathbf{q})|^2}{\varepsilon_{\mathbf{k}}^{(0)}-\varepsilon_{\mathbf{k+q}}^{(0)}}.
\end{equation}

Actually, there are two types of potential divergences in Eq.~\eqref{propintegral}: (i) when $\varepsilon_{\mathbf{k}}^{(0)}=\varepsilon_{\mathbf{k+q}}^{(0)}$ and (ii) when the electron-phonon matrix elements $GKK(\mathbf{q})$ diverge, which happens when the sum of Born effective charges does not vanish,
as we shall see ($GKK(\mathbf{q})$ is then proportional to $\frac{1}{\mathbf{q}}$).

In practical calculations, the $\mathbf{q=0}$ contribution from the same band (the denominator being thus zero) is not included in the summation. Also, in case of degeneracies,
the terms with zero denominators are ignored.
However, the integral of these divergences still needs to be obtained through the $\mathbf{q}$-point summation. 
For this reason, the numerical convergence of the adiabatic ZPR of diamond with respect to $\mathbf{q}$-point density is slow and requires large $\mathbf{q}$-point grids~\cite{Ponce2014}. This problem is often assessed in practice by adding an ad-hoc $i\delta$ to the denominator of Eq.~\eqref{propintegral}. 

The dipoles present in polar materials induce a Born effective charge, which describes the coupling between the electric field generated by the dipoles and the ionic motion. Such Born effective charges lead to a $\frac{1}{\mathbf{q}}$ behavior of the electron-phonon matrix elements (GKK). Divergences of type (ii) are therefore present in these materials. 
However,  in non-polar materials, there are theoretically no such effective charges and there should therefore be no divergence of type (ii). Since divergences of type (i) have a finite integral when no divergence of type (ii) are present, the $\mathbf{q}$-point sum should convergence to a finite value for non-polar materials.  

\begin{figure}[ht]
\begin{center}
  \includegraphics[width=0.99\linewidth]{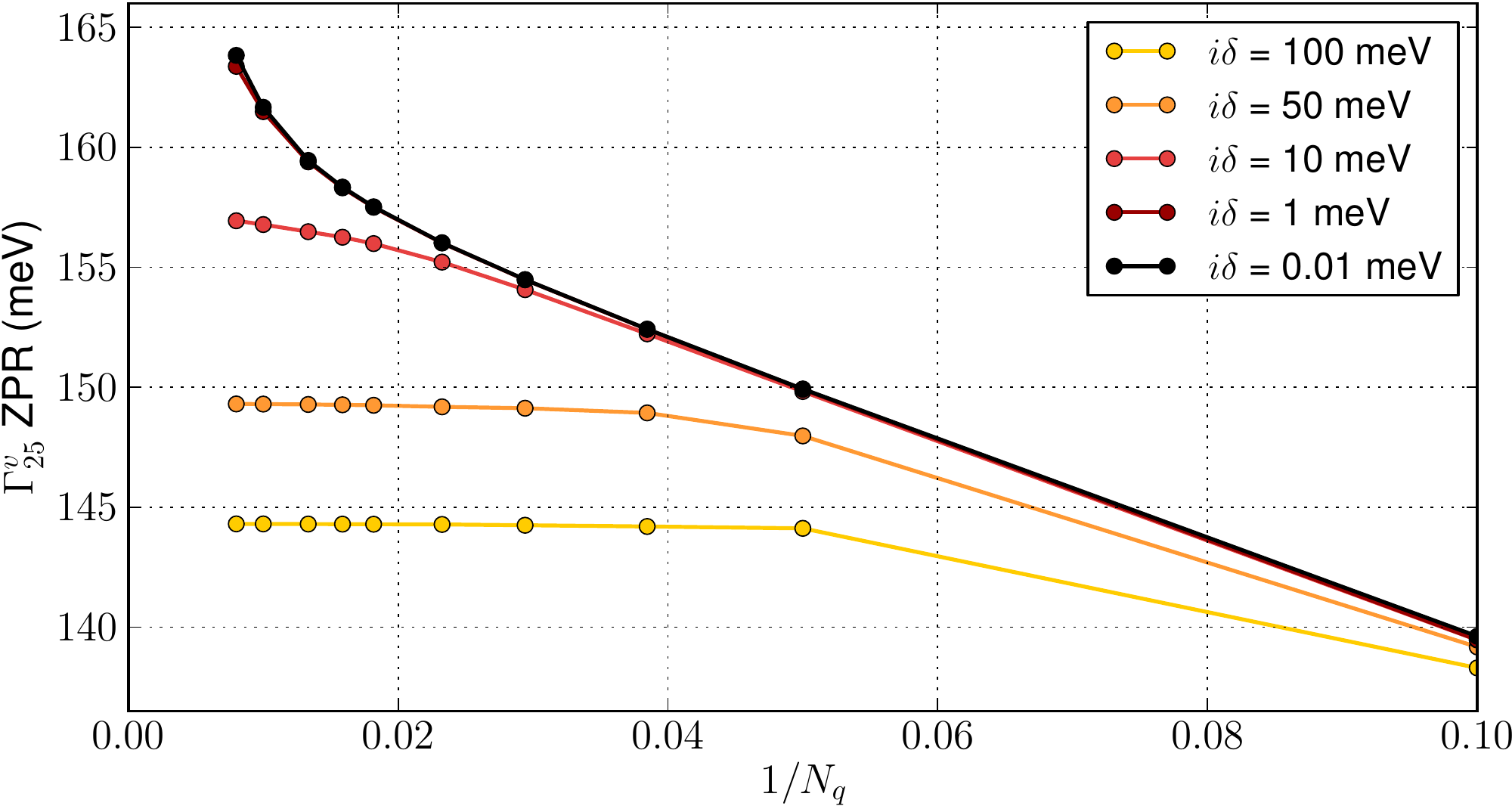}
   \caption{\label{fig:C-div} (color online) Adiabatic ZPR of the $\Gamma_{25}^v$ state of diamond with respect to the $\mathbf{q}$-point grid density for decreasing values of $i\delta$ before restoration of the charge neutrality. The size of the $\mathbf{q}$-point grid is $N_q \times N_q\times N_q$. }
\end{center}
\end{figure}

In practice, however, we observed a non-physical divergence of the ZPR for large $\mathbf{q}$-points densities. 
This effect can be clearly seen in Figure~\ref{fig:C-div}, where the $\mathbf{q}$-point density dependence of the adiabatic $\Gamma_{25}^v$ ZPR of diamond (calculated using Eq.~\eqref{equationsumoverstate_id}) for vanishing $i\delta$ exhibits a divergent behavior when $\delta$ is 1 meV or 0.01 meV, hardly seen for $\delta$ equal to 50 meV or 100 meV.

Actually, this divergence is attributed to a residual electric field connected to the breaking of the Born effective charge neutrality sum rule in non-polar periodic solids. This leads to the presence of divergences of type (ii), which combined to these of type (i), give an infinite integral around $\mathbf{q=0}$, thus making the $\mathbf{q}$-point sum diverge.

 This residual electric field is due to the finite $\mathbf{k}$-point grid used within DFT. Indeed, the first-order density is obtained thanks to a discretized integral on the BZ (see Eq.~(B9) of Ref.~\onlinecite{Gonze1997}). This first-order density in turn determines the electric field and the Born effective charges (see Eq.~(42) of Ref.~\onlinecite{Gonze1997a}). Such residual electric field that breaks the charge neutrality is found to converge to zero exponentially but is nonetheless substantial at $\mathbf{k}$-point grid usually sufficient to converge other relevant quantities. An example of this slow convergence is given in Table~\ref{table:convergence_Born} for the case of diamond (a non-polar material).

\begin{table}[ht]
\begin{footnotesize}
\begin{ruledtabular}
\begin{tabular}{c  c }
Number of $\mathbf{k}$-points & Born effective charge  \\
\hline
8    & -2.5406 \\
64   & -0.3514 \\
216  & -0.0534 \\
512  & -0.0080 \\
1000 & -0.0011\\
\end{tabular}          
\caption{Born effective charge of one carbon atom in diamond, for different electronic wavevector samplings.}
\label{table:convergence_Born}
\end{ruledtabular}
\end{footnotesize}
\end{table}

\subsection{Restoration of the charge neutrality}
In this section, we present a scheme to numerically remove this spurious electric field, and thus considerably speed up the ZPR convergence with respect to the electronic wavevector sampling. To this end, let us study the impact of a small Born effective charge on the matrix elements of $\frac{\partial \hat{H}_{\mathbf{k,k}}}{\partial R_{\kappa\alpha}(\mathbf{q})}$ present in Eq.~\eqref{equationsumoverstate_id}. 
We introduce the following short-hand notation for derivatives, which matches the one used in Ref.~\onlinecite{Gonze1997}
\begin{equation}
\frac{\partial \hat{H}_{\mathbf{k,k}}}{\partial R_{\kappa\alpha}(\mathbf{q})} \triangleq H_{\mathbf{k+q,k}}^{(1)}.
\end{equation}

As mentioned in Ref.~\onlinecite{Gonze1997}, within the pseudopotential framework, $H_{\mathbf{q}}^{(1)}$ can be decomposed into a first-order change of the non-local, local, Hartree and exchange-correlation potentials:
\begin{equation}
H_{\mathbf{k+q,k}}^{(1)}= v_{sep,\mathbf{k+q,k}}^{(1)}+\bar{v}_{loc,\mathbf{q}}^{(1)}+\bar{v}_{H,\mathbf{q}}^{(1)}+\bar{v}_{xc,\mathbf{q}}^{(1)}, 
\end{equation}
where the bar symbol above a quantity $\bar{X}$ means that it is the periodic part of $X$
\begin{equation}
\bar{X}_{\mathbf{q}}^{(1)}(\mathbf{r})=e^{-i\mathbf{q}\cdot \mathbf{r}} X_{\mathbf{q}}^{(1)}(\mathbf{r}),
\end{equation}
and where both  $\bar{v}_{loc,\mathbf{q}}^{(1)}$ and $\bar{v}_{H,\mathbf{q}}^{(1)}$ diverge as $\frac{1}{\mathbf{|q|}}$ with opposite signs. 

To make this more explicit, we express $\bar{v}_{loc,\mathbf{q}}^{(1)}$ as  
\begin{equation}\label{localpotential1}
\bar{v}_{loc,\mathbf{q}}^{(1)}(\mathbf{G})=\frac{-i}{\Omega_0}(\mathbf{G+q})_{\alpha}e^{-i(\mathbf{G+q})\cdot \boldsymbol{\tau}_\kappa}v_{\kappa}^{loc}(\mathbf{G+q}),
\end{equation}
where $\Omega_0$ is the volume of the unperturbed unit cell, $\boldsymbol{\tau}_\kappa$ the vector position of the atom $\kappa$ in the unit cell and with
\begin{equation}\label{vloc}
v_{\kappa}^{loc}(\mathbf{q}\rightarrow 0)=-\frac{4\pi Z_\kappa}{q^2}+C_\kappa + \mathcal{O}(q^2).
\end{equation}
To also explicit the same behavior in $\bar{v}_{H,\mathbf{q}}^{(1)}$, we express it as
\begin{equation}\label{hatree1}
\bar{v}_{H,\mathbf{q}}^{(1)}(\mathbf{G})= 4\pi \frac{\bar{n}_{\mathbf{q}}^{(1)}}{|\mathbf{G+q}|^2},
\end{equation}
where $\bar{n}_{\mathbf{q}}^{(1)}\propto \mathbf{|q|}$ when $\mathbf{q}\rightarrow \mathbf{0}$.

The $\bar{v}_{loc,\mathbf{q}}^{(1)}(\mathbf{G})$ of Eq.~\eqref{localpotential1} has an explicit algebraic form for $\mathbf{q}\rightarrow 0$, where $Z_\kappa$ is the number of valence electrons of the atom $\kappa$ described in the pseudopotential. Therefore, a residual electric charge can only affect the first-order density $\bar{n}_{\mathbf{q}}^{(1)}$ in Eq.~\eqref{hatree1}. The derivation of the impact of a residual Born effective charge on the first-order density is presented in Appendix~\ref{Technicalities} of this paper and can be seen in Eq.~\eqref{renormdensity} in the $\mathbf{G}=\mathbf{0}$ limit.

Using this knowledge, we can renormalize the Hartree term as follow (see Eq.~\eqref{finalrenorm})
\begin{multline}\label{RenormEq}
\bar{v}_{H,\mathbf{q}}^{ren(1)}(\mathbf{0}) = \\
\bar{v}_{H,\mathbf{q}}^{(1)}(\mathbf{0})
\frac{\sum_\gamma q_\gamma \Big (Z_{\kappa}\delta_{\alpha\gamma} -\frac{(Z_{\kappa\alpha,\gamma}^{*} -\bar{Z}_{\alpha\gamma}) }{\frac{1}{q^2} \sum_{\delta,\xi}q_{\delta}\epsilon_{\delta\xi}q_{\xi}}\Big)}
{\sum_\gamma q_\gamma \Big(Z_{\kappa}\delta_{\alpha\gamma}-\frac{Z_{\kappa\alpha,\gamma}^{*}}{ \frac{1}{q^2} \sum_{\delta,\xi}q_{\delta}\epsilon_{\delta\xi}q_{\xi}}\Big)},
\end{multline}
%
where $Z_{\kappa,\alpha\beta}^{*}$ is the Born effective charge, $\epsilon_{\gamma\xi}$ is the macroscopic static dielectric tensor for the electronic system (where the ions are considered fixed) and  $\bar{Z}_{\alpha\beta}$ is the averaged Born effective charge
\begin{equation}
\bar{Z}_{\alpha\beta} = \frac{1}{N_{at}}\sum_\kappa Z_{\kappa,\alpha\beta}^{*},
\end{equation} 
where $N_{at}$ is the number of atoms in the primitive cell. If the charge neutrality sum rule was fulfilled, the averaged Born effective charge should be exactly zero.

\begin{figure}[ht]
\begin{center}
  \includegraphics[width=0.98\linewidth]{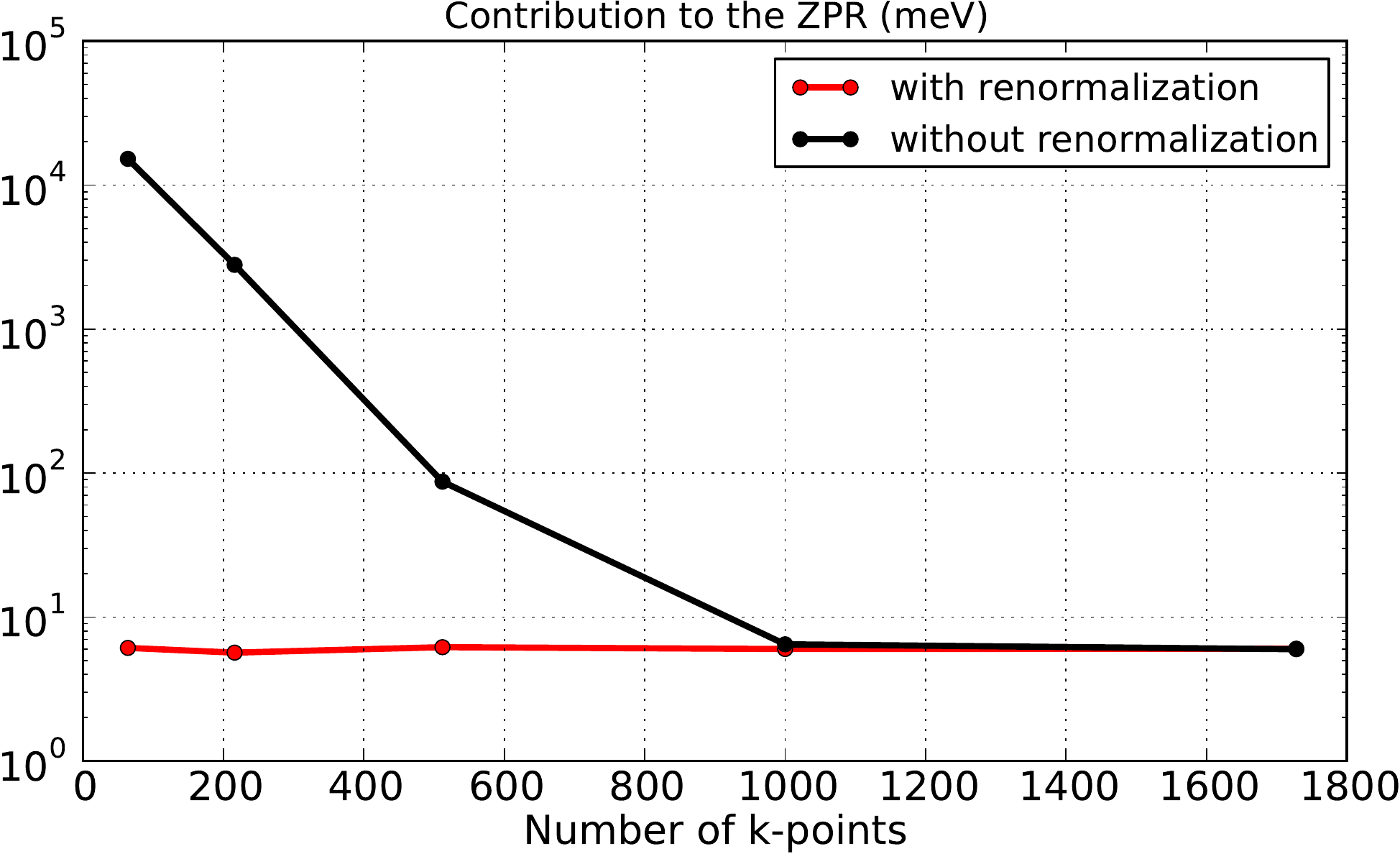}
   \caption{\label{comparison-200} (color online) Example of the contribution to the ZPR of 6 $\mathbf{q}$-points with respect to the densification of the $\mathbf{k}$-point grid of the first band of diamond at $\mathbf{k}=\mathbf{L}$. Only 6 symmetry equivalent $\frac{1}{100}\mathbf{X}$ $\mathbf{q}$-points were used to calculate the ZPR. The calculations were done with and without the renormalization of Eq.~\eqref{RenormEq}. }
\end{center}
\end{figure}

With this renormalization, the $\bar{v}_{H,\mathbf{q}}^{ren(1)}$ term correctly cancels the $\bar{v}_{loc,\mathbf{q}}^{(1)}$ when $\mathbf{q} \rightarrow 0$. Figure~\ref{comparison-200} clearly shows the faster convergence rate of the ZPR with respect to the density of the $\mathbf{k}$-point grid obtained with  this renormalization for the specific case of the first band of diamond at $\mathbf{k}=\mathbf{L}$. To highlight the divergent behavior of the ZPR with respect to $\mathbf{q}$-point grid density without the associated high computational cost for $\mathbf{q}$-point integration, only 6 symmetry equivalent $\mathbf{q}$-points are used in the sum. The $\mathbf{q}$-points are chosen close enough to zero to show the divergence: $\mathbf{q}=\frac{1}{100}\mathbf{X}$.

\section{Behavior of the $\mathbf{q}$-point convergence}
\label{qconv}

After enforcing the charge neutrality by application of Eq.~\eqref{RenormEq}, the theoretical rate of convergence of the ZPR can be analyzed when the number of $\mathbf{q}$-points along a side of the Brillouin Zone $N_q$ increases (the total number of q-points in the Brillouin Zone is $N_q^3$). After isolating the divergent behavior, we will analyze analytically, and numerically on simple models, the rate of convergence of the ZPR. 

We will observe that the $\mathbf{q}$-point convergence can be either constant, linear ($1/N_q$) or divergent ($N_q$) depending on the state that is renormalized, the use of the adiabatic (Eq.~\eqref{equationsumoverstate_id}) or non-adiabatic (Eq.~\eqref{equation_dynamical}) equation as well as the polar or non-polar nature of  the material.  The table~\eqref{table:convergence_behavior1} gives a summary of those behavior.


\begin{table}[ht]
\begin{footnotesize}
\begin{ruledtabular}
\begin{tabular}{c | c c c  }
\multicolumn{2}{c}{}   & \multicolumn{2}{c}{q-convergence}  \\
 \multicolumn{1}{c}{} &Cases   &  Adiabatic        &  Non-adiabatic   \\
\cline{2-4}
\multirow{2}{*}{Non-polar}     &  VBM/CBM      & $1/N_q$ (\ref{linear-q1}) & flat (\ref{constantq1})  \\
                       &   other & flat (\ref{constantq2}) & flat (\ref{constantq2})     \\
\cline{3-4}
\multirow{2}{*}{Polar}   &  VBM/CBM & $N_q$ (\ref{sec:diverge-q})  & $1/N_q$ (\ref{linear-q2}) \\   
                         &     other   &  $1/N_q$ (\ref{linear-q2})  &  $1/N_q$ (\ref{linear-q2})  \\                              
\end{tabular}          
\caption{Convergence behavior with the densification of the $\mathbf{q}$-grid and behavior for vanishing $i\delta$ at converged or extrapolated $\mathbf{q}$-grid. The only case that diverges is a polar material at the VMB/CBM using the adiabatic equation. The referenced sub-sections are given in parenthesis.}
\label{table:convergence_behavior1}
\end{ruledtabular}
\end{footnotesize}
\end{table}

\subsection{Rapid convergence with $N_q$}
\label{constantq}
The corrected (using Eq.~\eqref{RenormEq}) electron-phonon matrix elements of non-polar materials have no strong $\mathbf{q}$-point dependence. 
Also, if the state of interest ($n\mathbf{k}$) is a valence band maximum (VBM) or a conduction band minimum (CBM), the band dispersion is quadratic in reciprocal space around $\mathbf{k}$, and therefore $\varepsilon_{n\mathbf{k}}-\varepsilon_{n'\mathbf{k+q}}$ behaves as $q^2$.

\subsubsection{Non-polar materials in the non-adiabatic approximation at VBM/CBM \label{constantq1}} For a non-polar material within the non-adiabatic approximation (Eq.~\eqref{equation_dynamical}), we can model the $\mathbf{q}$-point behavior of the ZPR of the VBM with
\begin{equation}\label{firstintegralq}
 \lim_{\delta \rightarrow 0} \iiint_0^{q_c} d^3q \frac{1}{\varepsilon_{n\mathbf{k}}-\varepsilon_{n'\mathbf{k+q}}+\omega_{\mathbf{q}}+i\delta},
\end{equation}
where $q$ is integrated in a sphere of radius $q_c$. The same derivation applies for the CBM with $-\omega_{\mathbf{q}}$, the energy difference $\varepsilon_{n\mathbf{k}}-\varepsilon_{n\mathbf{k+q}}$ being negative. The phonon frequency shifts the poles of the function, so that the integrand is analytic over the domain of integration. The parabolic behavior of the extrema leads to
\begin{align}\label{nonadiaEq1}
 &=  \lim_{\delta \rightarrow 0} \iiint_0^{q_c} d^3q \frac{1}{q^2+\omega+i\delta} \\
 &=   \lim_{\delta \rightarrow 0} \int_{0}^{q_c} dq \int_{-\pi}^{\pi} d\phi \int_{0}^{\pi} d\theta \frac{q^2 \sin \theta   }{q^2+\omega+i\delta} \\
 &= \int_{0}^{q_c} dq 4\pi \frac{q^2}{q^2+\omega} \label{neweqintegral} \\
 &= 4\pi \sqrt{\omega} \Big(\frac{q_c}{\sqrt{\omega}} -\tan^{-1}\big(\frac{q_c}{\sqrt{\omega}}\big) \Big).
\end{align}

A Taylor expansion around $q_c = 0$ reveals that 
\begin{multline}
4\pi \sqrt{\omega} \Big(\frac{q_c}{\sqrt{\omega}} -\tan^{-1}\big(\frac{q_c}{\sqrt{\omega}}\big) \Big) \\
= \frac{4\pi \sqrt{\omega}}{3}\Big( \frac{q_c}{\sqrt{\omega}} \Big)^3 + \mathcal{O}(q_c^5),
\end{multline}
which means that the contribution from the integration around $q=0$ is simply proportional to the volume of integration, as expected given the non-divergent nature of the integrand. 

Thus, neglecting the $q=0$ contribution in the $\mathbf{q}$-point sum of the non-adiabatic ZPR for a band extrema of a non-polar material causes an error proportional to $\frac{1}{N_q^3}$ whereas discretization of the $\mathbf{q}$-point integration over the Brillouin Zone with the rectangle method causes an error proportional to $\frac{1}{N_q^2}$. Therefore, the error caused by the neglected $q=0$ contribution is not visible in the global convergence behavior for this ZPR. 
This behavior will be referred to as ``flat" convergence with respect to $\mathbf{q}$-point grid density from now on.  

\begin{figure}[ht]
\begin{center}
  \includegraphics[width=0.99\linewidth]{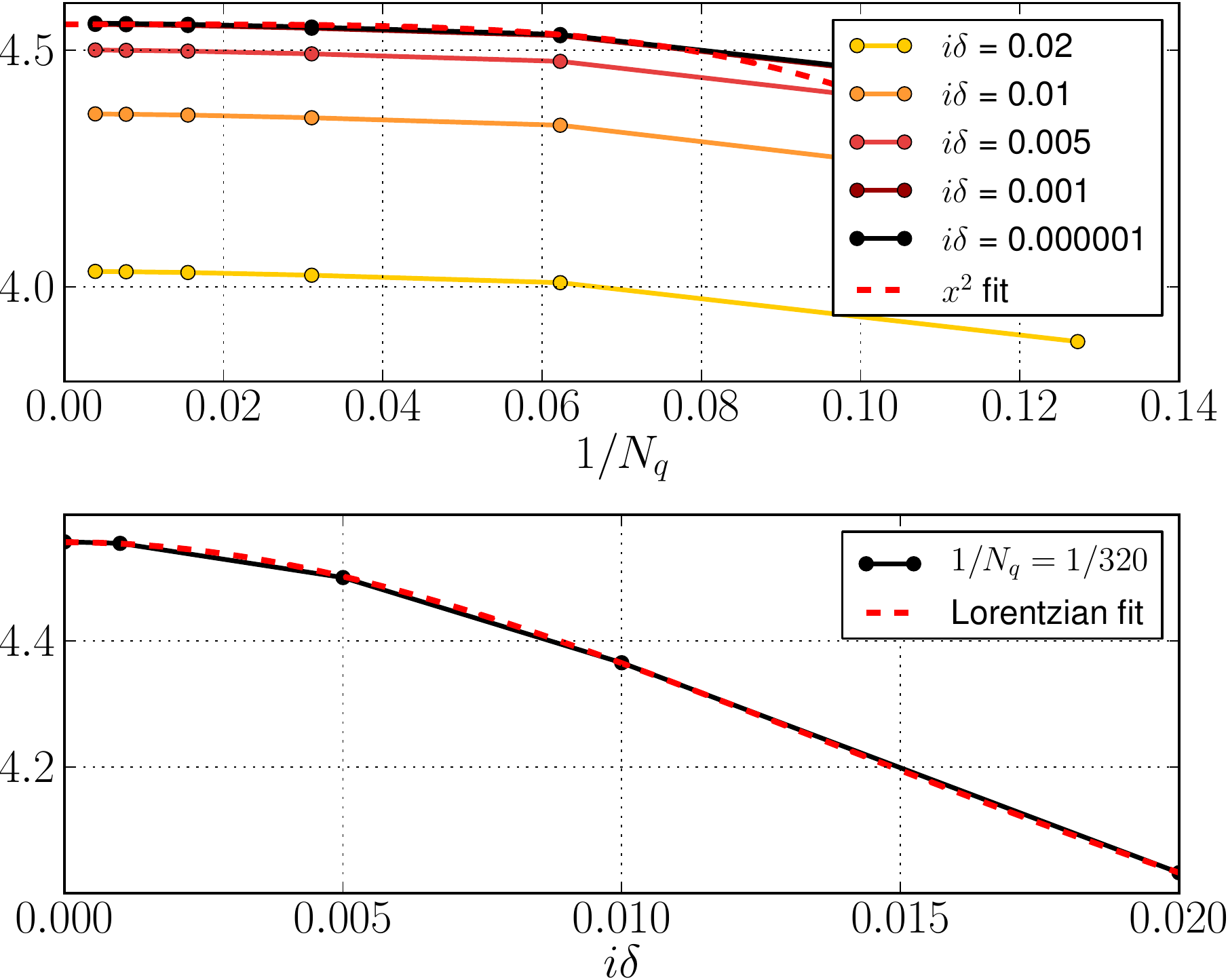}
   \caption{\label{conv-numerical-b} (color online) Behavior of the numerical integral of Eq.~\eqref{numericalintegralfirst} with respect to the $\mathbf{q}$-point density and $\delta$ for $q_c=0.5$ and $\omega=0.01$.  }
\end{center}
\end{figure}

Additionally, we can numerically integrate Eq.~\eqref{firstintegralq} on a three dimensional grid of $\mathbf{q}$-points using
\begin{equation}\label{numericalintegralfirst}
\frac{V}{N_q^3}\sum_{q} \frac{1}{q^2+\omega+i\delta},
\end{equation}
where $V$ is the volume of integration and where the element of volume is inversely proportional to  the number of $\mathbf{q}$-points $N_q^3$ needed to discretized the grid. 
An example of the convergence of Eq.~\eqref{numericalintegralfirst} with $\frac{1}{N_q}$ is shown on the top of Figure~\ref{conv-numerical-b} for $q_c=0.5$ and $\omega=0.01$.

This function converges very quickly with increasing $\mathbf{q}$-sampling as can be seen e.g. on the center of Figure~\ref{C1bis-conv1} for the VBM of diamond using the non-adiabatic equation.

\subsubsection{Non-polar materials in the adiabatic or non-adiabatic approximation for a non-extremal band \label{constantq2}}
If the state that we would like to renormalize is not a VBM nor a CBM, the denominator of the adiabatic (Eq.~\eqref{equationsumoverstate_id}) or non-adiabatic (Eq.~\eqref{equation_dynamical}) equations will be small when the state that we consider ($\varepsilon_{n\mathbf{k}}$) has almost the same energy as another state  ($\varepsilon_{n'\mathbf{k}+\mathbf{q}}$), minus a phonon frequency ($\omega_{m\mathbf{q}}$) in the non-adiabatic framework. 
As a result, the integrand in Eq.~\eqref{firstintegralq} is not analytic anymore in these cases and a non-zero imaginary $i\delta$ is required to avoid numerical instabilities. 

To get a deeper understanding of the behavior of non-extremal points, we will model the energy difference ($\varepsilon_{n\mathbf{k}}-\varepsilon_{n'\mathbf{k+q}}$) by a shifted parabola with its minimum at $\mathbf{q}_0$ 
\begin{equation}\label{eq:parabola_shifted}
\varepsilon_{\mathbf{k}} - \varepsilon_{\mathbf{k}+\mathbf{q}}  = (\mathbf{q}-\mathbf{q}_0)^2 - q_0^2.
\end{equation}

The ZPR has poles on a sphere of radius $q_0$ centered on $\mathbf{q} = \mathbf{q_0}$ (chosen to be on the z-axis) and passing through the origin.
The integral on the spherical shell between radii of values $q_0 - \Delta$ and $q_0 + \Delta$ gives a contribution that is linear with $\Delta$ (see Appendix \ref{annex_spher_poles}), as would any regular function when integrated over a spherical shell. This indicates that the integration of these poles will not
contribute an error of higher order than the reminder of the numerical integration. 
Neglecting the $\mathbf{q} = 0$ contribution in the numerical integration leads to an error proportional to $\frac{1}{N_q^3}$ in the non-polar case (see Appendix \ref{annex_int_0}). This leads to a $\mathbf{q}$-point grid convergence that is flat for the non-polar case in the adiabatic framework.

In the non-adiabatic case, the $\omega$ to be added to the right-hand side of Eq.~\eqref{eq:parabola_shifted} only slightly reduces the radius of the sphere at $\mathbf{q}_0$ (which thus does not touch the origin anymore) and the conclusion for the adiabatic case remains valid for the numerical integration over this sphere. 
The integrand at $\mathbf{q}=\mathbf{0}$ becomes analytical in this case, so that the convergence behavior remains effectively flat. 
Therefore, in practice, the $\mathbf{q}$-point integration required to evaluate the ZPR can be considered converged when the ZPR does not change significantly with denser grids. An example of this type of convergence is given at the top of Figure~\ref{C1-conv3} for diamond in the adiabatic framework at a non-extremal energy.

In other cases, the discretized integral does not converge as quickly as the rectangle method for an analytical integrand. It sometimes converges linearly ($\propto \frac{1}{N_q}$) (see subsection \ref{sec:linear-q}) or even diverges ($\propto N_q$)  (see subsection \ref{sec:diverge-q}).

\subsection{Convergence proportional to the inverse of $N_q$}
\label{sec:linear-q}
\subsubsection{Non-polar materials in the adiabatic approximation at the VBM/CBM }
\label{linear-q1}
Non-polar materials with a parabolic energy dispersion (VBM or CBM) have a $\mathbf{q}$-dependence for the adiabatic ZPR that behaves as

\begin{equation}\label{integralqconv}
\iiint_0^{q_c} d^3q \frac{1}{q^2+i\delta} =4\pi q_c \propto \frac{1}{N_q},
\end{equation}
when $i\delta=0$. Therefore, neglecting the $q=0$ contribution in the numerical integration yields an error proportional to $\frac{1}{N_q}$ that dominates the $\frac{1}{N_q^2}$ error of the rectangle method. We will call this type of convergence ``linear" here.

\begin{figure}[ht]
\begin{center}
  \includegraphics[width=0.99\linewidth]{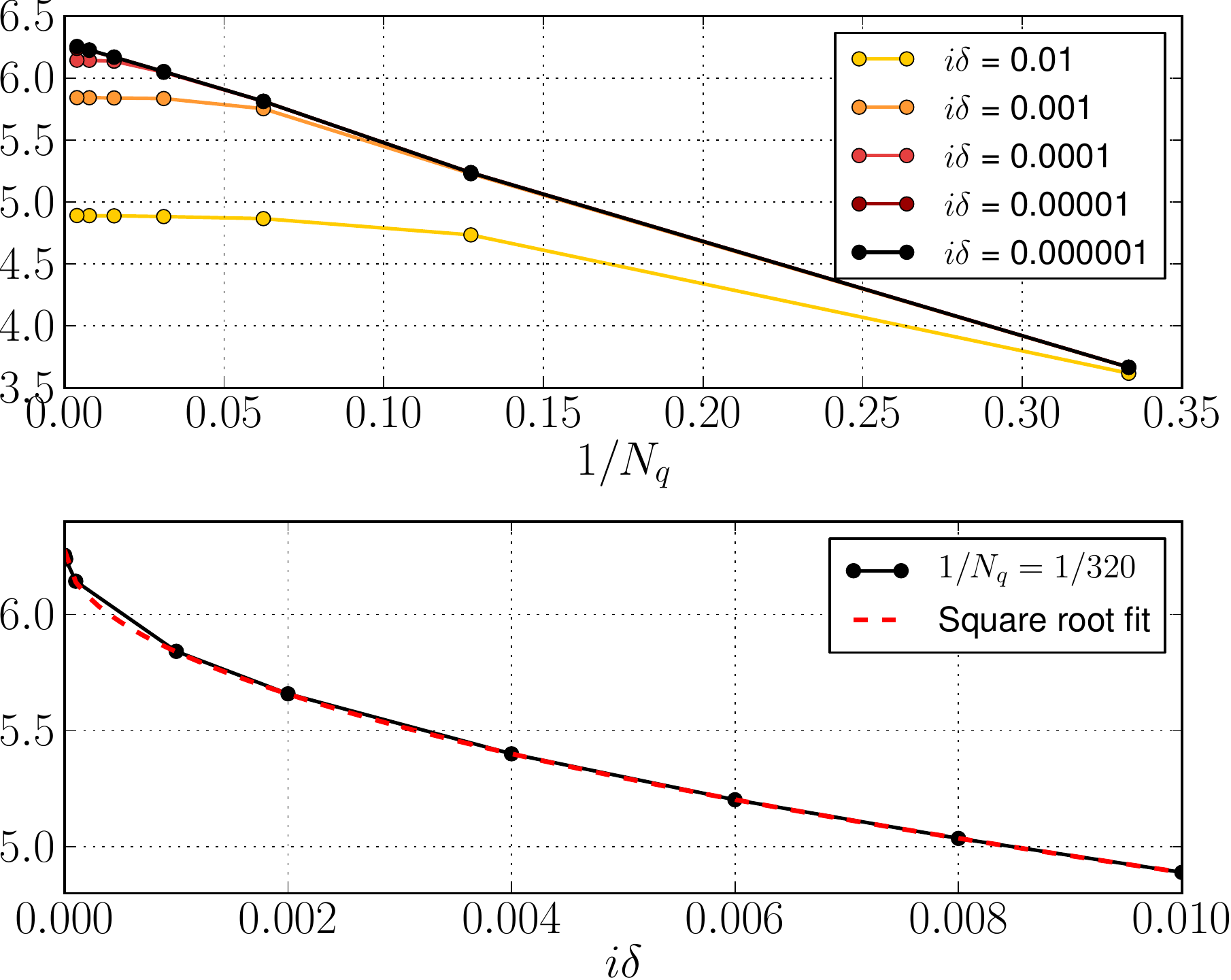}
   \caption{\label{conv-numerical-a} (color online) The numerical integral of Eq.~\eqref{numerical_inte} converges toward $2\pi$ as expected. We can see that the convergence in $\mathbf{q}$ points and $i\delta$ is rather slow. }
\end{center}
\end{figure}

The rate of convergence with $\mathbf{q}$-densification can be numerically tested by summing this function on a three dimensional grid of $\mathbf{q}$-points
\begin{equation}\label{numerical_inte}
\frac{1}{N_q^3}\sum_{q\neq 0} \frac{1}{q^2+i\delta},
\end{equation}
where the $q=0$ term has been omitted in the sum for numerical reasons (as the expression must stand for vanishing $\delta$). The numerical integral of Eq.~\eqref{numerical_inte} is shown on Figure~\ref{conv-numerical-a} and converges towards $2\pi$ for $q_0=0.5$, as expected.

\subsubsection{Polar materials in the adiabatic (non-extremal point) and non-adiabatic approximation} 
\label{linear-q2}
In the case of polar materials, the GKK behaves as $1/q^2$ for small $q$, but this time, the divergence is physical and comes from the physical finite Born effective charges. In this case, at the VBM or CBM, the non-adiabatic Eq.~\eqref{equation_dynamical} shifts the pole inside the bandgap. Therefore, the parabolic eigenenergy dispersion is not problematic  anymore. This can be represented by
\begin{equation}
\Re \iiint_0^{q_c}d^3q \frac{1}{q^2(\omega+i\delta)}.
\end{equation}
This case behaves as Eq.~\eqref{numerical_inte} and we have the same linear behavior because the diverging $\mathbf{q=0}$ term is removed from the numerical sum.

In the case of points that are not at a VBM or a CBM, the integration of the ZPR in the adiabatic and non-adiabatic approximations exhibits a behavior that is linear with $1/N_q$. This is explained by the removed $\mathbf{q=0}$ contribution and the integral is linear with the radius of the sphere (as shown in Appendix \ref{annex_int_0}), which is proportional to $1/N_q$.

\subsection{Increasing $N_q$ does not lead to convergence}
\label{sec:diverge-q}
In the case of polar materials, when we consider the renormalization of a state at the VBM or the CBM, the adiabatic equation diverges. Indeed,
\begin{equation}\label{firsttherela}
\frac{V}{N_q^3}\sum_{q} \frac{1}{q^2(q^2+i\delta)}
\end{equation}
diverges as $N_q$. An example of this case is given in the middle left Figure~\ref{BN1-conv1} for the VBM of boron nitride using the adiabatic equation.

More specifically, neglecting the $q=0$ contribution in the numerical integration of Eq.~\eqref{firsttherela} gives a convergence behavior that can be modeled by the 
 spherical integration of the summand of Eq.~\eqref{firsttherela} in a shell from $q_c/N_q$ to $q_c$. This gives, for $\delta = 0$
\begin{align}
\int_{-\pi}^{\pi} d\phi \int_{0}^{\pi}d\theta \int_{q_c/N_q}^{q_c}  dq \frac{1}{q^4} &= 4\pi \int_{q_c/N_q}^{q_c} dq \frac{1}{q^2} \nonumber \\
&= 4\pi \left( \frac{1}{q_c} - \frac{1}{q_c/N_q} \right).
\end{align}
As $1/N_q$ goes to 0, the value of this integral indeed diverges linearly with the number of division $N_q$. This shows that, \textit{for polar materials, only the non-adiabatic equation can be safely used}. 


\section{Behavior of the $i\delta$ convergences}
\label{idconv}

After enforcing the charge neutrality by application of Eq.~\eqref{RenormEq}, the theoretical rate of convergence of the ZPR can be analyzed when the small imaginary parameter $i\delta$ tends to 0. After isolating the divergent behavior, we will analyze analytically, and numerically on simple models, the rate of convergence of the ZPR. 

We will observe that the $\delta$ convergence can be square-root, linear or Lorentzian-like, depending on the state that is renormalized, the use of the adiabatic or non-adiabatic equation as well as the fact that the material is polar or not.  The table~\eqref{table:convergence_behavior2} gives a summary of those behavior.


\begin{table}[ht]
\begin{footnotesize}
\begin{ruledtabular}
\begin{tabular}{c | c c c  }
\multicolumn{2}{c}{}   & \multicolumn{2}{c}{q-convergence}  \\
 \multicolumn{1}{c}{} &Cases   &  Adiabatic        &  Non-adiabatic   \\
\cline{2-4}
\multirow{2}{*}{Non-polar}   &  VBM/CBM      & $\sqrt{\delta}$ (\ref{sqrtconv}) & Lorentzian (\ref{lorentzianconv1})  \\
                             &   other & $\delta$ (\ref{linearidelta1}) & $\delta$ (\ref{linearidelta1})     \\
\cline{3-4}
\multirow{2}{*}{Polar}   &  VBM/CBM & $1/\sqrt{\delta}$ (\ref{divergenceconv})  & Lorentzian (\ref{lorentzianconv2}) \\   
                         &     other   &  $\delta$  (\ref{linearidelta2})  &  $\delta$  (\ref{linearidelta2})  \\                                
\end{tabular}          
\caption{Convergence behavior for vanishing $i\delta$ at converged or extrapolated $\mathbf{q}$-grid. The only case that diverges is a polar material at the VMB/CBM using the adiabatic equation. The referenced sub-sections are given in parenthesis.}
\label{table:convergence_behavior2}
\end{ruledtabular}
\end{footnotesize}
\end{table}

\subsection{Convergence proportional to the square-root of $\delta$}
\label{sqrtconv}
For a non-polar material at the VBM or CBM, in the adiabatic framework, we can determine the ZPR dependence on the (finite) $\delta$ value by analytically integrating Eq.~\eqref{numerical_inte} 
\begin{align}
\Re \iiint d^3q \frac{1}{q^2 +i\delta} &= 4\pi \int_{0}^{q_c} dq \frac{q^4}{q^4+\delta^2} \nonumber \\
 &= 4\pi \sqrt{\delta} \int_{0}^{q_c} \frac{dq}{\sqrt{\delta}} \frac{\frac{q^4}{\delta^2}}{\frac{q^4}{\delta^2} +1}.
\end{align}
Carrying the integration and making a Taylor expansion for small $\frac{\sqrt{\delta}}{q_c}$ yields
\begin{equation} 
 4\pi q_c + 4\pi C \sqrt{\delta} +\mathcal{O}(\delta),
\end{equation}
 with $C$ a constant. This result matches the $\sqrt{\delta}$ behavior observed in our numerical integration in the bottom of Figure~\ref{conv-numerical-a}.

\subsection{Convergence proportional to $\delta$}
\label{linearidelta}

\subsubsection{Non-polar materials in the adiabatic and non-adiabatic approximation at a non-extremal point} 
\label{linearidelta1}
For non-polar materials, when considering a $\mathbf{k}$-point other than the VBM or CBM, the adiabatic equation can be modeled as 
\begin{equation}\label{Eqlinearindelta}
\Re \iiint_0^{q_c} d^3q \frac{1}{\varepsilon(\mathbf{q})+i\delta},
\end{equation}
where we have defined $\varepsilon(\mathbf{q}) = \varepsilon_{\mathbf{k}+\mathbf{q}} - \varepsilon_{\mathbf{k}}$. 

Since the small $i\delta$ will only affect the integrand around the pole at $\varepsilon (\mathbf{q}) = 0$, we can determine the $\delta$ dependence of this model ZPR by considering only a small range of energy  $\eta$ around it. We can then re-write Eq.~\eqref{Eqlinearindelta} as 
\begin{equation}
\int_{-\eta}^{\eta} dE g(E)f(E),
\end{equation}
with $ g(E) \triangleq \sum_{\mathbf{q}} \delta(E-\varepsilon(\mathbf{q}))$ the density of states and 
\begin{equation}
f(E) \triangleq \frac{1}{E+i\delta}.
\end{equation}

We can make a Taylor expansion of the density of states around $E=0$ (since $\eta$ is small) 
\begin{equation}\label{Taylor}
g(E) = g(0)+g'(0) E+ \mathcal{O}(E^2).
\end{equation}
The even terms in Eq.~\eqref{Taylor} will not contribute because $\Re f(E)$ is an odd function, therefore making no contribution to the integral. 
Therefore, the leading term is the first-order one. The integral can thus be re-written as
\begin{equation}\label{functiontangeante}
\delta \int_{-\eta}^{\eta} \frac{dE}{\delta} \frac{g'(0)\frac{E^2}{\delta^2}}{\frac{E^2}{\delta^2}+1} =  2 \delta g'(0)\Big( \frac{\eta}{\delta} - \tan^{-1} (\frac{\eta}{\delta})\Big).
\end{equation}

Making a Taylor expansion of Eq.~\eqref{functiontangeante} for small $\frac{\delta}{\eta}$ leads to 
\begin{equation}\label{etalinear}
2 g'(0) \eta - \delta g'(0)\pi,
\end{equation}
where we can see that this function is linear in $\delta$.

In the framework of the non-adiabatic theory, the integral is changed to
\begin{equation}
\int  dE f(E+\omega) g(E).
\end{equation}

As this function has poles when $E = -\omega$, we will now integrate around $-\omega$ 
\begin{equation}
\int_{-\omega-\eta}^{-\omega+\eta}  dE f(E+\omega) g(E)
\end{equation}
After the change of variable $u = E+\omega$, we obtain
\begin{equation}
\int_{-\eta}^{\eta} du f(u)g(u-\omega)
\end{equation}
The density of states can again be Taylor expanded around $-\omega$, giving
\begin{equation}
g(u-\omega) = g(-\omega) + u g'(-\omega) + \mathcal{O}(u^2).
\end{equation}
The same steps that led us from Eq.~\eqref{Taylor} to Eq.~\eqref{etalinear} now give us
\begin{equation}
2 g'(-\omega) \eta - \delta g'(-\omega)\pi,
\end{equation}
which is again linear in $\delta$.

In conclusion, the behavior of the ZPR of non-polar materials for non-extremum $\mathbf{k}$-points in the BZ is linear when $\delta \rightarrow 0$.

\subsubsection{Polar materials in the adiabatic and non-adiabatic approximation at a non-extremal point} 
\label{linearidelta2}

For polar materials, we can use the shifted parabola model (defined in Eq.~\eqref{eq:parabola_shifted}) in Eq.~\eqref{Eqlinearindelta}, multiply the integrand by $\frac{1}{q^2}$ and for the non-adiabatic framework, add $\omega$ to $i\delta$. Within this model, it is possible to show (see Appendix \ref{annexshifted}) that, for small $\delta$, the ZPR converges linearly with $\delta$. In practice, the value that is reached by the linear regime is very high and tends to infinity at VBM or CBM.

\subsection{Convergence in $\delta$ proportional to a Lorentzian  }
\label{lorentzianconv}

\subsubsection{Non-polar materials in the non-adiabatic approximation at the VBM/CBM} 
\label{lorentzianconv1}

When considering the non-adiabatic equation at the VBM or CBM of a non-polar material, the ZPR behavior with respect to $\delta$ is
\begin{equation}\label{eq:nonadianonpolar}
\Re \iiint d^3q \frac{1}{q^2 + \omega + i\delta} = 4\pi \Re \int_0^{q_c} dq \frac{q^2}{q^2+\omega+i\delta}, 
\end{equation}
which gives
\begin{equation}\label{lorentizannonpolar}
4\pi \Re \Big(q_c - \sqrt{\omega+i\delta} \tan^{-1}\Big( \frac{q_c}{\sqrt{\omega+i\delta}} \Big) \Big)
\end{equation}

Plotting Eq.~\eqref{lorentizannonpolar} reveals a Lorentzian-like shape centered at $\delta=0$. 
This can also be seen in the $\delta$-dependence of the numerical integration of Eq.~\eqref{eq:nonadianonpolar}
\begin{equation}\label{eqLorentzian}
\frac{1}{N_q^3}\sum_{S_{q_c}} \frac{(q^2+\omega)}{(q^2+\omega)^2 + \delta^2},
\end{equation}
where $S_{q_c}$ is a sphere of radius $q_c$.  This dependence is shown at the bottom of Figure~\ref{conv-numerical-b} for a cutoff radius $q_c = 0.5$ and $\omega = 0.01$.

In practice, we also observe that non-adiabatic ZPR for VBM/CBM of non-polar materials can be accurately fitted by a Lorentzian function. We thus use this type of functional dependence of the ZPR in $\delta$ to extrapolate the results at $\delta=0$.

\subsubsection{Polar materials in the non-adiabatic approximation at the VBM/CBM} 
\label{lorentzianconv2}

For polar materials, we get a supplementary $\frac{1}{q^2}$ factor in the integrand of Eq.~\eqref{eq:nonadianonpolar}, which thus becomes
\begin{multline}
\Re \iiint d^3 q \frac{1}{q^2} \frac{1}{q^2+\omega+i\delta} \\
= 4\pi \Re \int_0^{\infty} dq \frac{1}{q^2+\omega+i\delta}, 
\end{multline}
which becomes
\begin{equation}
2\pi \Re \int_{-\infty}^{\infty} dq \frac{1}{q^2+\omega+i\delta}.
\end{equation}

This integral can be performed by closing the contour of integration using a half-circle of infinite radius in the upper complex plane (which does not contribute to the integral) and then using the residue theorem. We obtain
\begin{multline}
2\pi \Re \int_{-\infty}^{\infty} dq \frac{1}{q^2+\omega+i\delta} \\
= 2\pi^2 \frac{1}{\sqrt{\omega} \big(1+(\frac{\delta}{\omega})^2\big)^{\frac{1}{4}}}\Re e^{-\frac{i}{2}\tan^{-1}(\frac{\delta}{\omega})}, 
\end{multline}
which gives
\begin{equation}\label{EqfinalLorentzian}
2\pi^2 \frac{1}{\sqrt{\omega} \big(1+(\frac{\delta}{\omega})^2\big)^{\frac{1}{4}}} \cos \Big( \frac{1}{2}\tan^{-1}\big(\frac{\delta}{\omega}\big)  \Big).
\end{equation}

Plotting Eq.~\eqref{EqfinalLorentzian} again reveals a Lorentzian-like shape centered at $\delta=0$. Accordingly, fitting the results using a Lorentzian is also found to be a good approximation in practice for polar materials.

\subsection{Decreasing $\delta$ does not lead to convergence}
\label{divergenceconv}

The adiabatic ZPR for the VBM/CBM of a polar material has already been shown to diverge with increasing $\mathbf{q}$-point sampling (see subsection \ref{sec:diverge-q}). We now examine the $\delta$-dependence of the ZPR, which has the form
\begin{align}
\Re \iiint d^3q \frac{1}{q^2(q^2+i\delta)} &= 4\pi \Re \int_0^{q_c} dq \frac{1}{q^2+i\delta} \nonumber \\
&= \frac{4\pi}{\sqrt{\delta}}\int_0^{q_c} \frac{dq}{\sqrt{\delta}} \frac{\frac{q^2}{\delta}}{\frac{q^4}{\delta^2}+1}. \label{eqdivergence}
\end{align}

When $q_c/\sqrt{\delta}$ is rather large (i.e. $> 10$), which happens for small $\delta$, the integral of Eq.~\eqref{eqdivergence} converges logarithmically to its value at infinity and we obtain $\frac{\sqrt{2}\pi^2}{\sqrt{\delta}}$.

We therefore see that Eq.~\eqref{eqdivergence} will numerically diverge as $\frac{1}{\sqrt{\delta}}$ for small values of $\delta$. An example of this kind of divergence is given in the middle right of Figure 9 of supplemental materials~\cite{supplemental} for the VBM of $\beta$-AlN (a polar material) using the adiabatic equation.


%

\section{Results on different semiconductors}
\label{results}

We will now examine five different semiconductors, two of which are non polar materials (diamond and silicon), and three of which are polar materials ($\alpha$-AlN, $\beta$-AlN and BN). We will be able to provide fully converged results, independent of any arbitrary parameter,
like an ad hoc broadening parameter. Of course, as outlined previously, in the case of polar materials, only the non-adiabatic theory can provide such results, as the standard adiabatic AHC theory breaks down for these.

\subsection{Non polar materials}
\subsubsection{Diamond} 
Diamond is a metastable allotrope of carbon where the C atoms are arranged into two interpenetrating face-centered cubic lattices shifted along the body diagonal by $\frac{1}{4}^{\text{th}}$ of its length. The space group associated with this spatial arrangement is Fd$\bar{3}$m (cubic, 227).
Diamond has the highest hardness and thermal conductivity of any bulk material~\cite{Ramdas2000}. It is therefore used as cutting and polishing tool in the industry. Even though the stable phase of bulk carbon under normal condition is graphite, we will focus on the diamond phase.

\begin{table*}[ht]
\begin{ruledtabular}
\begin{tabular}{l c c c c c c c}
  &     &  &  & \multicolumn{4}{c}{lattice parameters [Bohr]}  \\
\cline{5-8}  
  &  Space group   &  Ecut [Ha] & $\mathbf{k}$-grid &  this work (LDA) & other DFT (LDA) & other DFT (GGA) & experiment (300K) \\  
\hline
$\alpha$-AlN & P6$_3$mc [186]     & 35 & 6x6x6 & 5.783/9.255 & 5.820/9.335 \cite{Wright1995} &   5.913/9.481 \cite{Jain2013}                    & 5.881/9.415~\cite{Yu2001}   \\
             &                    &    &       &             & &                       & 5.880/9.409~\cite{Petrov1992}\\
             &                    &    &       &             &                              &                       & 5.877/9.411 \cite{Schulz1977}\\                    
$\beta$-AlN  & F$\bar{4}3$m [216] & 35 & 6x6x6 & 8.130       & 8.205 \cite{Wright1995}      & 8.317 \cite{Jain2013} & 8.258 \cite{Lambrecht1994}\\
             &                    &    &       &             & 8.164 \cite{Kim1996}         & 8.303 \cite{Stampfl1999} & 8.277 \cite{Petrov1992}\\                                                    
$c$-BN       & F$\bar{4}3$m [216] & 35 & 8x8x8 & 6.746       & 6.754 \cite{Cappellini2001}  & 6.852 \cite{Jain2013}    & 6.833 \cite{Soma1974} \\
             &                    &    &       &             & 6.752 \cite{Furthmuller1994} & 6.831 \cite{Madelung1972} &  \\
             &                    &    &       &             & 6.814 \cite{Wentzcovitch1986}&                       & \\
             &                    &    &       &             & 6.833 \cite{Xu1991}          &                       & \\   
C-d          & Fd$\bar{3}$m [227] & 30 & 6x6x6 & 6.652       & 6.652 \cite{Giustino2010}    & 6.756 \cite{Liu2013}  & 6.740 \cite{Gildenblat1996} \\     
Si           & Fd$\bar{3}$m [227] & 20 & 6x6x6 & 10.170      & 10.223 \cite{Patrick2014}    & 10.335 \cite{Jain2013}& 10.26 \cite{Wyckoff1963} \\ 
\end{tabular}          
\caption[Convergence and lattice parameters for the different studied compounds.]{Convergence parameters for the different studied compounds. The space groups are given in Hermann-Mauguin notation with the number in bracket being the crystallographic index number in international Tables and the homogeneous $\mathbf{k}$-points sampling are $\boldsymbol{\Gamma}$-centered. All the pseudopotentials in this work use the LDA exchange-correlation functional. }
\label{table:converegence_compounds}
\end{ruledtabular}
\end{table*}

The pseudopotential was generated using the \texttt{fhi98PP} code~\cite{Fuchs1999} with a 1.5 atomic unit cut-off radius for pseudization. The valence electrons of carbon, treated explicitly in the \textit{ab-initio} calculations, are the 2s$^{2}$2p$^{2}$3d$^{0}$ orbitals.

Careful convergence studies (error below 0.5~mHa per atom on the total energy) led to the use of a 6x6x6 $\boldsymbol{\Gamma}$-centered Monkhorst-Pack $\mathbf{k}$-point sampling~\cite{Monkhorst1976} of the BZ  and an energy cut-off of 30~Hartree for the truncation of the plane wave basis set. The Perdew and Zunger parametrization of LDA~\cite{Perdew1981} was used. The relaxed lattice parameter is calculated to be 6.652~Bohr, 1.3\% below the experimental value of 6.740~Bohr, measured at room temperature~\cite{Gildenblat1996} (see Table~\ref{table:converegence_compounds} for more information on the structural properties).

The electronic bandstructure was computed at the DFT level and gave a direct bandgap at $\boldsymbol{\Gamma}$ of 5.67~eV and an indirect $\boldsymbol{\Gamma}-0.727\mathbf{X}$ bandgap of 4.25~eV, intrinsically below the experimental bandgap of 5.48~eV at 0~K~\cite{Cardona2005a}(see Table~\ref{table:band_structure_compounds}).  

\begin{table*}[ht]
\begin{ruledtabular}
\begin{tabular}{l c c c c c c c c } 
  &  \multicolumn{4}{c}{direct gap [eV]} & \multicolumn{4}{c}{indirect gap [eV]}  \\
\cline{2-5} \cline{6-9}
  &  this work  & \multicolumn{2}{c}{other DFT} &      &  this work & \multicolumn{2}{c}{other DFT}  &      \\  
\cline{3-4} \cline{7-8}  
  & LDA & LDA & GGA & exp. & LDA & LDA & GGA & exp.  \\  
\hline
$\alpha$-AlN    & 4.691 & 4.2 \cite{Vogel1997}    & 4.056 \cite{Jain2013}       & 6.28* \cite{Perry1978}       & - & -     & -  & -  \\
                &       & 4.3 \cite{Litimein2002} &                             & 6.28* \cite{Christensen1994} & - & -     & -  & - \\ 
                &       & 4.41 \cite{Wright1995}  &                             & 6.3 \cite{Lambrecht1994}     & - & -     & -  & - \\    
                &       & 4.52 \cite{Christensen1993}&  &   & -     & -  & - &  -\\   
                &       & 4.74 \cite{Stampfl1999} &     &   & -     & -  & - & -\\   
$\beta$-AlN     & 4.677 & 4.2 \cite{Rubio1993}  & 3.995 \cite{Jain2013}    &  -  & 3.308 & & 3.306 \cite{Jain2013} & - \\
                &       & 4.2 \cite{Litimein2002} &     & -   &       & 3.2 \cite{Litimein2002} & & - \\ 
                &       & 4.35 \cite{Wright1995}  &     & -   &       & 3.2 \cite{Vogel1997}  & & - \\
                &       & 4.75 \cite{Stampfl1999} &     & -   &       &                      &  & - \\ 
$c$-BN          & 8.890 & 8.6 \cite{Prasad1984}   &     & 14.5 \cite{Madelung1972} & 4.446 & 4.4 \cite{Furthmuller1994} & 4.450 \cite{Jain2013} & 6.4 \cite{Coleburn1968}  \\
                &       & 8.7 \cite{Xu1991}       &     &   &       & 5.18 \cite{Xu1991}     & & 6.4 \cite{Chrenko1974}   \\
                &       & 8.8 \cite{Furthmuller1994} &  &   &      &  &  \\   
C-d             & 5.670 &   &  5.571 \cite{Jain2013}   & 7.3* \cite{Cardona2005a} & 4.250 & & 4.113* \cite{Jain2013} & 5.48* \cite{Cardona2005a} \\
                &       &               &            &                          &       &  & 4.12* \cite{Liu2013}  &   \\
Si              & 2.567 & 2.52 \cite{Gillet2013}  & 2.557* \cite{Jain2013} &  3.378* \cite{Jellison1983}    & 0.463 & 0.45 \cite{Gillet2013} & 0.612* \cite{Jain2013}& 1.17* \cite{Jellison1983}  \\
\end{tabular}
\caption[Direct and indirect DFT electronic bandgaps.]{\label{table:band_structure_compounds} The direct and indirect (when relevant) DFT electronic bandgaps are compared with other references (theoretical or experimental). The star * sign denotes low temperature experiment (below 10K) and no star means room temperature.}
\end{ruledtabular}
\end{table*}

For the calculations of the ZPR, we used 10 bands to describe the active space in Eqs.~\eqref{equationsumoverstate_id} and \eqref{equation_dynamical}. 

\begin{figure*}[ht]
\subfloat[Adiabatic ZPR of direct-gap diamond.\label{C1-conv3}]
{\includegraphics[width=.41\linewidth]{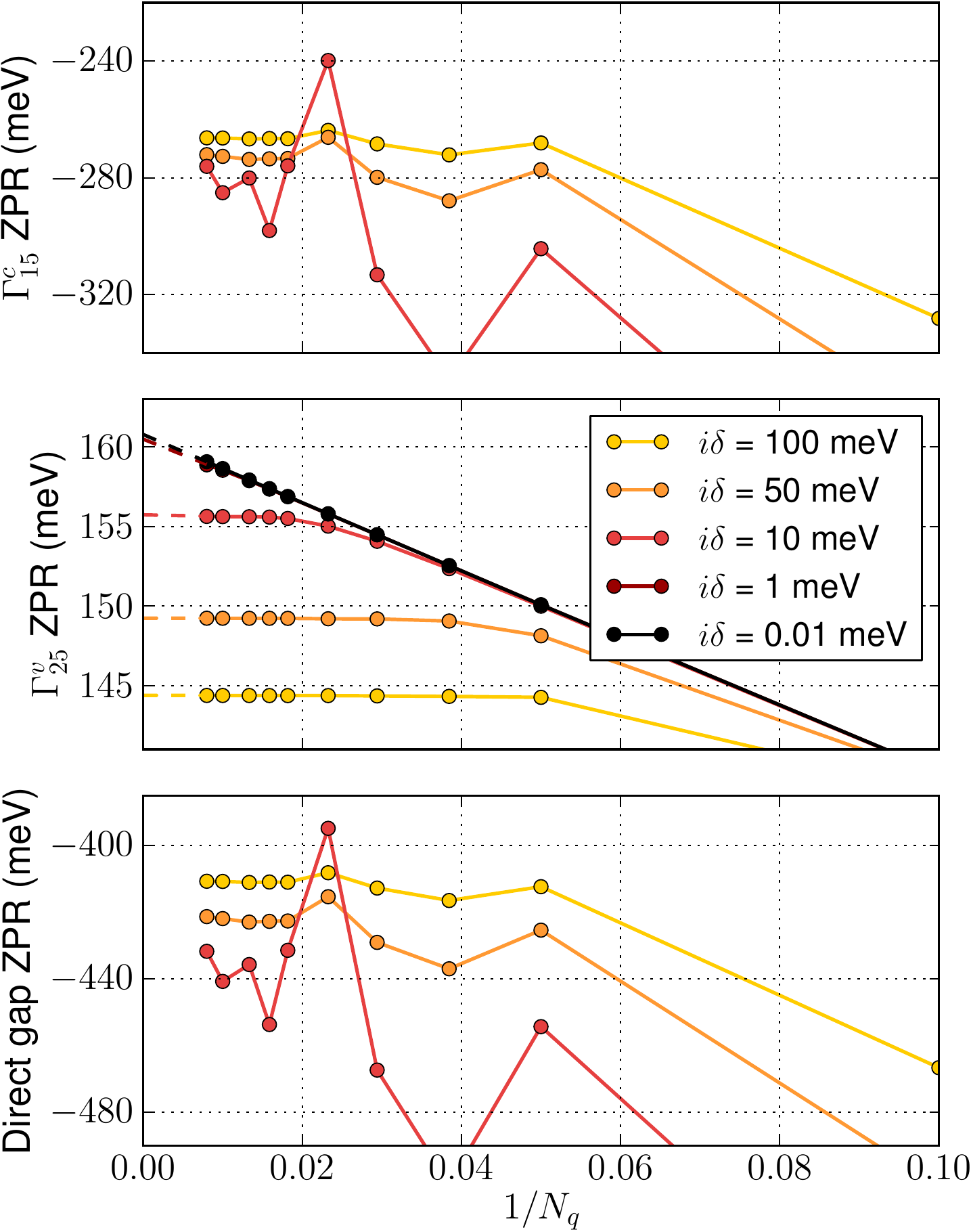}} 
\subfloat[$i\delta$ extrapolation of the ZPR.\label{C1-conv2}]
{\includegraphics[width=.41\linewidth]{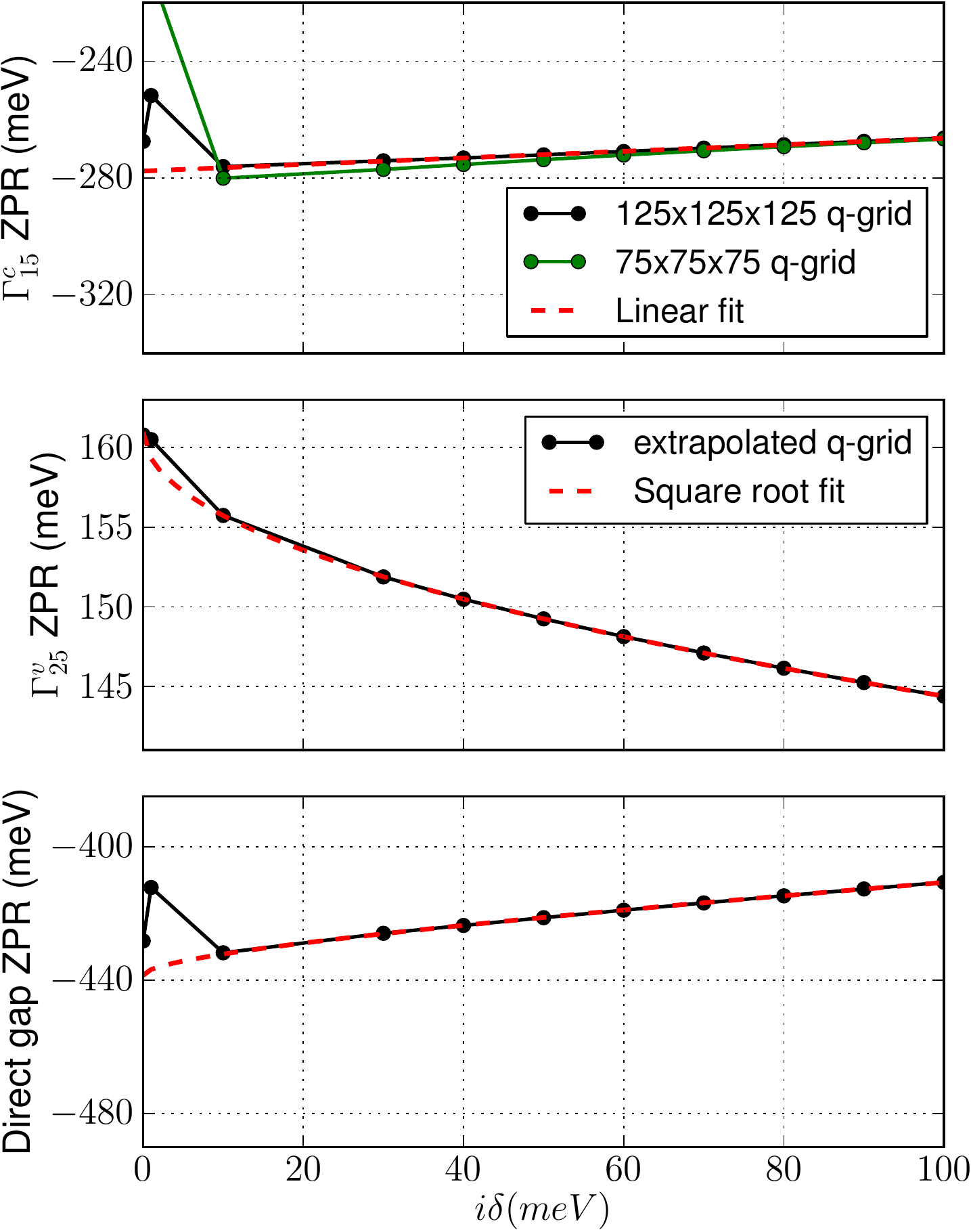}} 
\caption[Convergence study with respect to the $\mathbf{q}$-point grid density.]{\label{C1-conv} Convergence study for the adiabatic (a) $\mathbf{q}$-point grid density and (b) $i\delta$ parameter for the direct bandgap ZPR of diamond. The bottom Figures are the difference of the two Figures above them. The adiabatic ZPR of the direct bandgap of diamond is -438.6~meV.}
\end{figure*}

The convergences with respect to $\mathbf{q}$-point integration for the band edges and the direct band gap of diamond are shown on Figure~\ref{C1-conv3}, where the densest used grid is a 125x125x125 $\mathbf{q}$-grid (43680 $\mathbf{q}$-points in the irreducible Brillouin-Zone (IBZ)). The $\Gamma_{15}^c$ state is not the bottom of the conduction band, and therefore there are other states in the BZ with close energy. This leads to numerical instabilities as the denominator of the adiabatic Eq.~\eqref{equationsumoverstate_id} can diverge for small $i\delta$. For large enough imaginary component, the ZPR converges to approximatively -270~meV. The top of the valence band $\Gamma_{25}^v$ converges linearly and can be extrapolated to infinitely dense $\mathbf{q}$-grid. 
To obtain a definite value for the ZPR, we have to converge the ZPR for vanishing $\delta$. The convergence can be found on Figure~\eqref{C1-conv2} and shows that large $\mathbf{q}$-point grid are required to enter the expected linear regime (see section~\ref{linearidelta} for more information). The extrapolated ZPR is -277.61~meV. The VBM can be smoothly extrapolated using a square-root fit to zero value of $\delta$ to give 160.96~meV.   
The adiabatic direct bandgap ZPR of diamond is computed to be -438.6~meV.
\begin{figure*}[ht]
\subfloat[Non-adiabatic ZPR of direct-gap diamond.\label{C1bis-conv1}]
{\includegraphics[width=.41\linewidth]{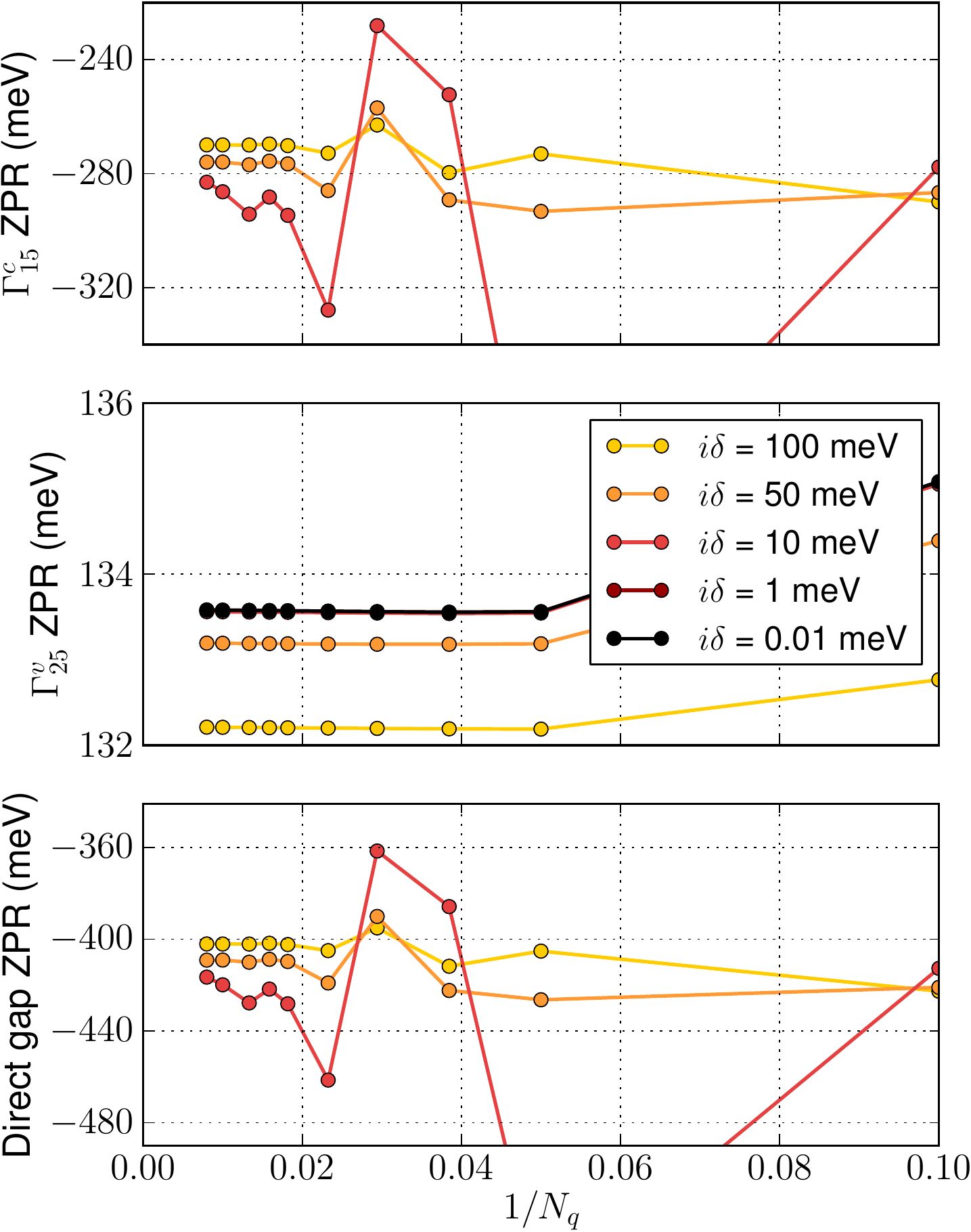}} 
\subfloat[$i\delta$ extrapolation of the ZPR.]
{\includegraphics[width=.41\linewidth]{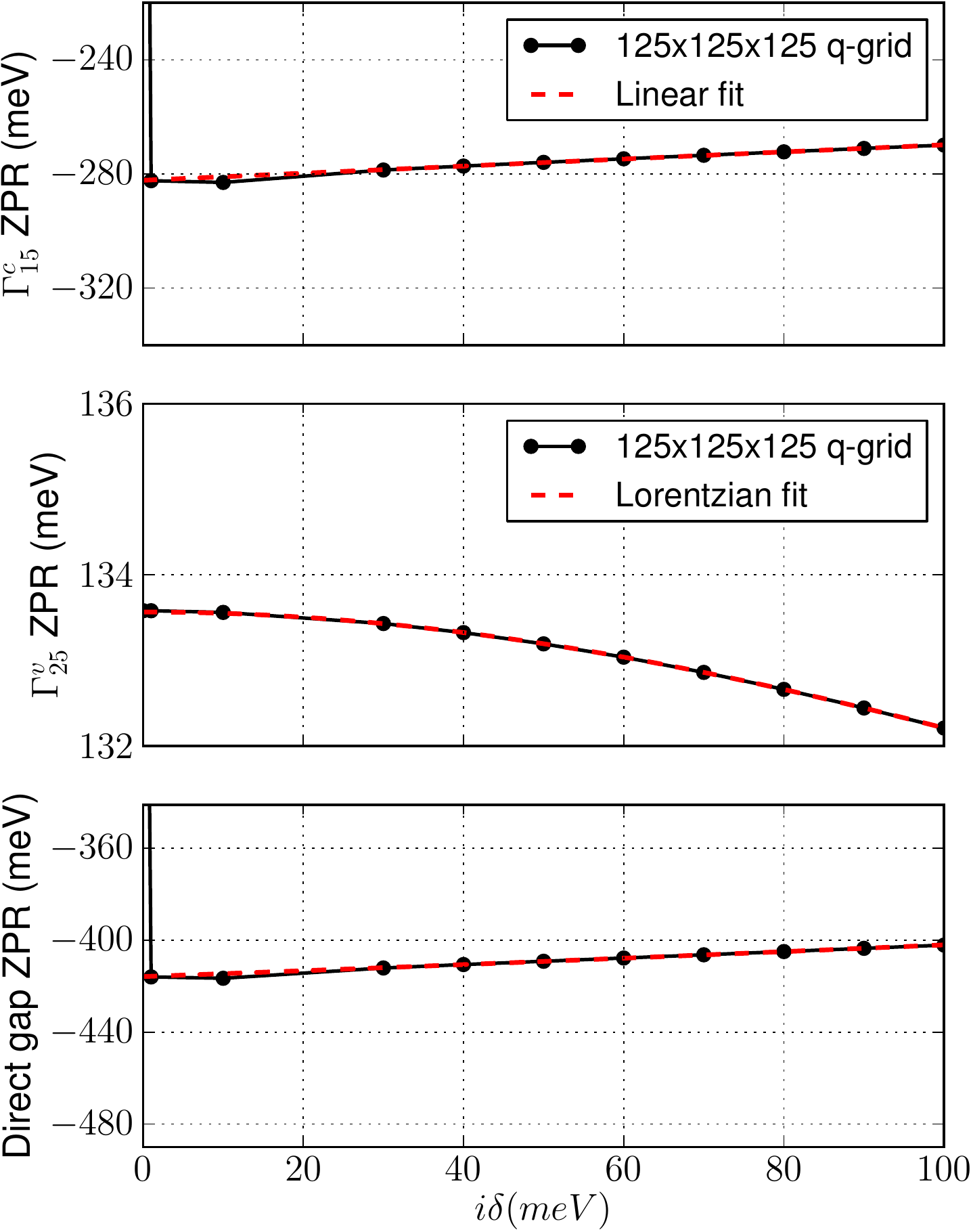}} 
\caption[Convergence study with respect to the $\mathbf{q}$-point grid density.]{\label{C1bis-conv} Convergence study for the non-adiabatic (a) $\mathbf{q}$-point grid density and (b) $i\delta$ parameter for the direct bandgap ZPR of diamond. The bottom Figures are the difference of the two Figures above them. The non-adiabatic ZPR of the direct bandgap of diamond is -415.8~meV.)}
\end{figure*}
For the non-adiabatic direct bandgap of diamond, the convergence can be found on Figure~\ref{C1bis-conv} and shows that the  $\Gamma_{15}^c$ state converges similarly but the VBM has a rapid convergence in $\mathbf{q}$-point integration and a Lorentzian behavior for the $\delta$ extrapolation. The fitted Lorentzian have three fitting parameters: a multiplicative constant $A$, the full width at half maximum (FWHM) and an additive constant $B$
\begin{equation}
A\frac{\frac{\Gamma}{2\pi}}{(\frac{\Gamma}{2})^2+x^2}+B,
\end{equation}  
where here $A = 10.11$, $\Gamma = 0.55$ and $B = 121.90$. The extrapolated ZPR is of -283.23~meV and 133.57~meV for the $\Gamma_{15}^c$ and $\Gamma_{25}^v$  states, respectively. This leads to a reduction of -415.8~meV of the direct bandgap due to electron-phonon interaction at 0~K. 

The convergences of the CBM for diamond are given in Figures 1 and 2 of the supplemental materials~\cite{supplemental} using the adiabatic and non-adiabatic equations, respectively. 
For the adiabatic case, the fact that the $\mathbf{q}$-convergence is not smooth for relatively small $i\delta$ is due to the finite $\mathbf{k+q}$ sampling. Indeed, when we compute the renormalization at one of the 6 symmetry equivalent CBM $\mathbf{k}$-points, the $\mathbf{k+q}$ sampling is such that the other five equivalent $\mathbf{k}$-points are not sampled exactly (not the numerically accurate minimum).  
The extrapolated ZPR of the CBM state is -219.24~meV using a square root fit for the adiabatic equation and -196.22~meV using a Lorentzian fit for the non-adiabatic equation (see Table~\ref{table:zpr_list} for more information).

\begin{figure}[bth]
\includegraphics[width=.99\linewidth]{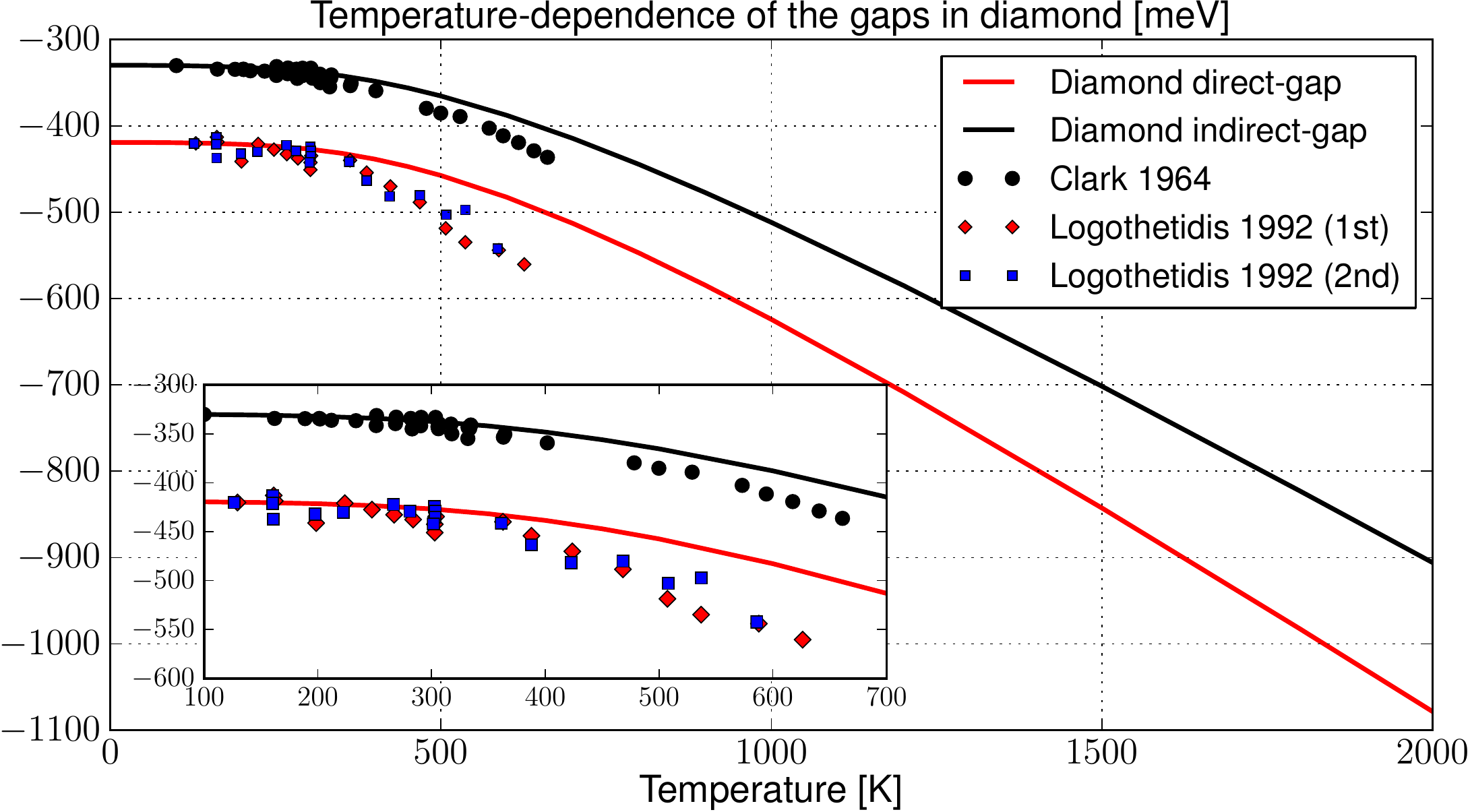}
\caption[Temperature dependence of bandgaps in diamond.]{\label{diamond-temp} Temperature dependence of the diamond gaps using the non-adiabatic temperature dependence on a 75x75x75 $\mathbf{q}$-grid with Lorentzian extrapolation to vanishing imaginary parameter $\delta$. The slopes at high temperature are -0.504~meV/K for the direct gap of diamond and -0.435~meV/K for the indirect one. The experimental points from Clark~\textit{et al.}~\cite{Clark1964} and Logothetidis~\textit{et al.}~\cite{Logothetidis1992} (using first or second-derivative line-shape analysis) are shifted so that the lowest temperature point matches the theoretical line.}
\end{figure}

The temperature dependence of the direct and indirect bandgaps is reported on Figure~\ref{diamond-temp} for a 75x75x75 $\mathbf{q}$-grid and shows that the slope at high temperature for the non-adiabatic renormalization with a Lorentzian extrapolation to vanishing imaginary parameter $\delta$ is -0.504~meV/K for the direct bandgap and -0.435~meV/K for the indirect one. 
The phonon-induced broadening $\frac{1}{2\tau_{n\mathbf{k}}^{(adiabatic,RIA)}}$ of Eq.~\eqref{phonon_induced_lifetime_eq} is calculated for the 75x75x75 $\mathbf{q}$-grid to be 180~meV and 63~meV for the direct and indirect bandgap of diamond at 0~K, respectively. 

Our result underestimates the experimental ZPR of the diamond indirect bandgap of -364~meV~\cite{Cardona2005} by 9.4\%. Since our calculation neglects several effects like anharmonicity, non-rigid-ion terms or many-body $GW$ corrections, we are rather close to the experimental value.  
The measured linear slope at high temperature for the indirect bandgap of diamond is -0.54~meV/K~\cite{Cardona2005}. The measured linear slope at high temperature for the direct bandgap is $-0.60$ or $-0.69$~meV/K~\cite{Logothetidis1992}, depending on the analysis\footnote{Logothetidis~\textit{et al.} deduced the temperature dependence of the direct bandgap of diamond from first and second-derivative line-shape analysis, see Ref.~\onlinecite{Logothetidis1992} for more details.}. 
Our theoretical values for the indirect and direct bandgaps underestimate the experimental ones by 19\% and 16\%, respectively. We hypothesize that this underestimation of the linear slope at high temperature for the direct bandgap of diamond is due to the underestimation of the ZPR within DFT. Indeed, as discussed in Ref.~\onlinecite{Antonius2014}, the correction brought by $GW$ to the ZPR is quite substantial for the direct bandgap (-209~meV). Since the ZPR is directly linked with the slope at high temperature, it is not surprising that we witness such an underestimation with respect to the experimental results. 
We did not compute such GW correction for ZPR of the indirect bandgap of diamond but expect from the results of Figure~\ref{diamond-temp} to have a smaller correction.  

It is worthwhile to note that the complete lack of experimental data for low temperature (T $<$ 100~K) and the relatively large error bars (up to $\pm$10~meV) between 200~K and 350~K generate an uncertainty of several meV on the experimental ZPR~\cite{Passler1999}. This calls for new, reliable and wide range temperature measurement of the evolution of the bandgap with temperature in diamond. We hope that our theoretical study will stimulate such experimental interest.  

\begin{figure}[bth]
\subfloat[Diamond bandstructure at 0K.]
{\includegraphics[width=.48\linewidth]{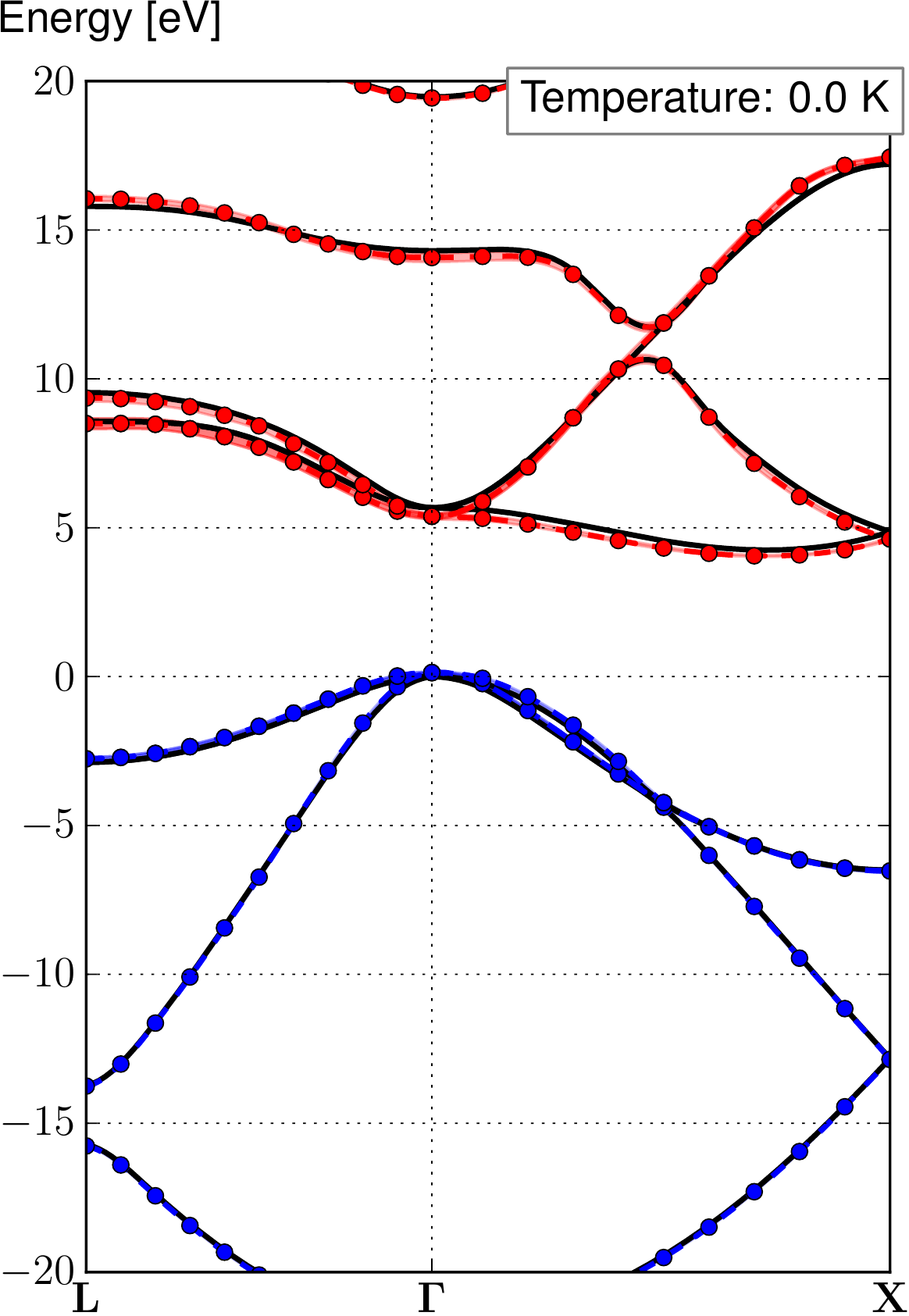}} \quad
\subfloat[Zoom of the left-hand side figure.]
{\includegraphics[width=.48\linewidth]{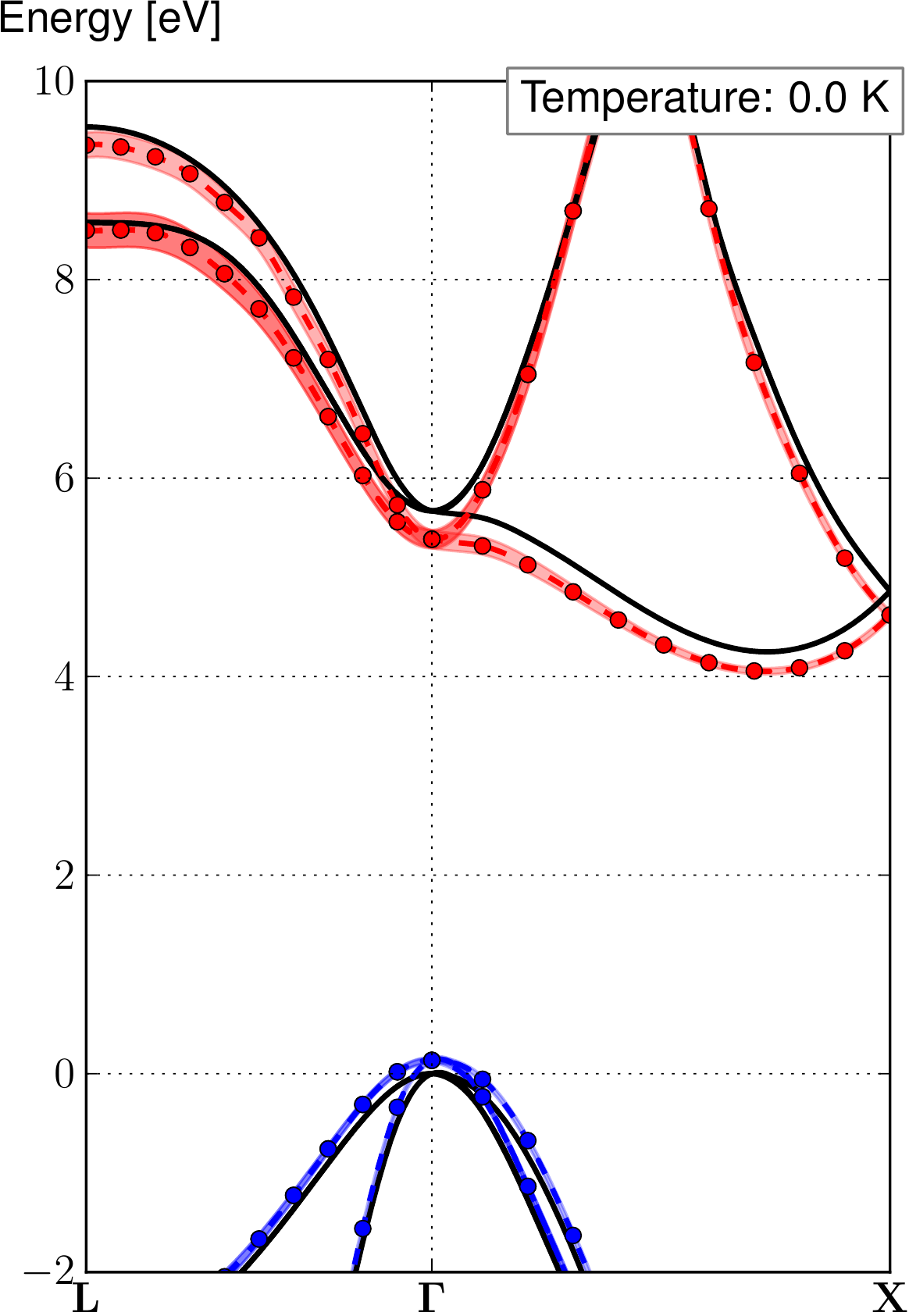}}
\caption[Electronic bandstructure of diamond.]{\label{C-bs} Electronic bandstructure (plain black line), renormalisation (dashed line) and phonon-induced broadening (envelope around the dashed line) at 0K using the non-adiabatic ZPR integrated on a 75x75x75 $\mathbf{q}$-point grid for $\delta$ extrapolated to zero for diamond, where the dots are the actual renormalization calculation. A spline function is used to connect the renormalization dots.}
\end{figure}

Finally, the non-adiabatically renormalized electronic bandstructure of diamond at 0K along the $\mathbf{L}-\boldsymbol{\Gamma}-\mathbf{X}$ high symmetry line is shown on Figure~\ref{C-bs} for a 75x75x75 $\mathbf{q}$-point grid for $\delta$ extrapolated to zero linearly and with a Lorentzian for the VBM and CBM.

\subsubsection{Silicon}

Silicon is a tetravalent metalloid widely used in integrated circuits and semiconductor electronics. It has a diamond cubic crystal structure with a Fd$\bar{3}$m (cubic, 227) space group. Pure silicon is usually not found in nature but manufactured to get monocrystalline silicon for use in computer microchips. Although the manufacturing process is rather expensive, this material is so important that about 50.000 metric tons are produced per year worldwide~\cite{Corathers2009}.  

The pseudopotential used for silicon was generated using the \texttt{fhi98PP} code~\cite{Fuchs1999} with a 1.0247 atomic unit cut-off radius for pseudization. The pseudopotential is a Troullier-Martins with the Perdew/Wang~\cite{Perdew1992} parametrization of LDA. The valence electrons of silicon, treated explicitly in the \textit{ab-initio} calculations, are the 3s$^{2}$3p$^{2}$ orbitals.

Careful convergence checks (error below 0.5~mHa per atom on the total energy) lead to the use of a 6x6x6 $\boldsymbol{\Gamma}$-centered Monkhorst-Pack $\mathbf{k}$-point sampling~\cite{Monkhorst1976} of the BZ and an energy cut-off of 20~Hartree for the truncation of the plane wave basis set. The relaxed lattice parameter is calculated to be 10.170~Bohr, 0.9\% below the experimental value, measured at room temperature~\cite{Wyckoff1963} (see Table~\ref{table:converegence_compounds} for more information on the structural properties).

The electronic bandstructure was computed at the DFT level and gave a direct gap at $\boldsymbol{\Gamma}$ of 2.567~eV and an indirect $\boldsymbol{\Gamma}-0.848\mathbf{X}$ bandgap of 0.463~eV intrinsically below the experimental bandgaps of 3.378 and 1.17~eV at $\sim$10K~\cite{Jellison1983} for the direct and indirect bandgap, respectively  (see Table~\ref{table:band_structure_compounds}).  
Many \textit{ab-initio} simulations have been performed on silicon and give similar values to ours. For example, Ref.~\onlinecite{Gillet2013} got a direct bandgap of 2.52~eV and an indirect one of 0.45~eV, also using the Abinit software. 

For the calculations of the ZPR, we used 10 bands to describe the active space in Eqs.~\eqref{equationsumoverstate_id} and \eqref{equation_dynamical}. 
The convergence with respect to $\mathbf{q}$-point integration for the direct bandgaps of silicon is shown on Figure~3 of the supplemental materials~\cite{supplemental} for the adiabatic equation and gives a linearly extrapolated ZPR of -6.23~meV for the $\Gamma_{15}^c$ state and a square-root extrapolation of 40.87~meV for the ZPR of the VBM, thus leading to a ZPR of -47.1~meV for the direct bandgap. 
The non-adiabatic direct bandgap ZPR shown on Figure~4 of the supplemental materials~\cite{supplemental} of the $\Gamma_{15}^c$ state is calculated to be -7.36~meV and 34.87~meV for the VBM. The non-adiabatic bandgap ZPR is therefore slightly smaller than the adiabatic one with a value of -42.1~meV. 
The densest grid computed is a 100x100x100 $\mathbf{q}$-grid (22776 $\mathbf{q}$-points in the IBZ).
The convergences with respect to the indirect bandgap of silicon are shown in Figures~5 and 6 of the supplemental materials~\cite{supplemental} for the adiabatic and non-adiabatic equations respectively. The ZPR of the first one can be extrapolated to -23.28~meV for the CBM and 40.87~meV for the VBM, leading to -64.3~meV renormalization of the bandgap. The second one can be extrapolated to -21.43~meV for the CBM and 34.75~meV for the VBM, leading to a smaller -56.2~meV renormalization of the bandgap.

Such values can be compared with those mentioned in the recently published paper by Patrick and Giustino~\cite{Patrick2014} who obtain values of -57 and -22~meV for the indirect and direct bandgap renormalization of silicon using a 4x4x4 supercell within the AHC framework (adiabatic equation).  Their results matches ours for the adiabatic 4x4x4 $\mathbf{q}$-grid ($1/N_q = 0.25$) with -52 and -29~meV for the indirect and direct bandgaps, respectively (as can also be seen on Figures~3 and 5 of the supplemental materials~\cite{supplemental}).

\begin{figure}[bth]
\includegraphics[width=.95\linewidth]{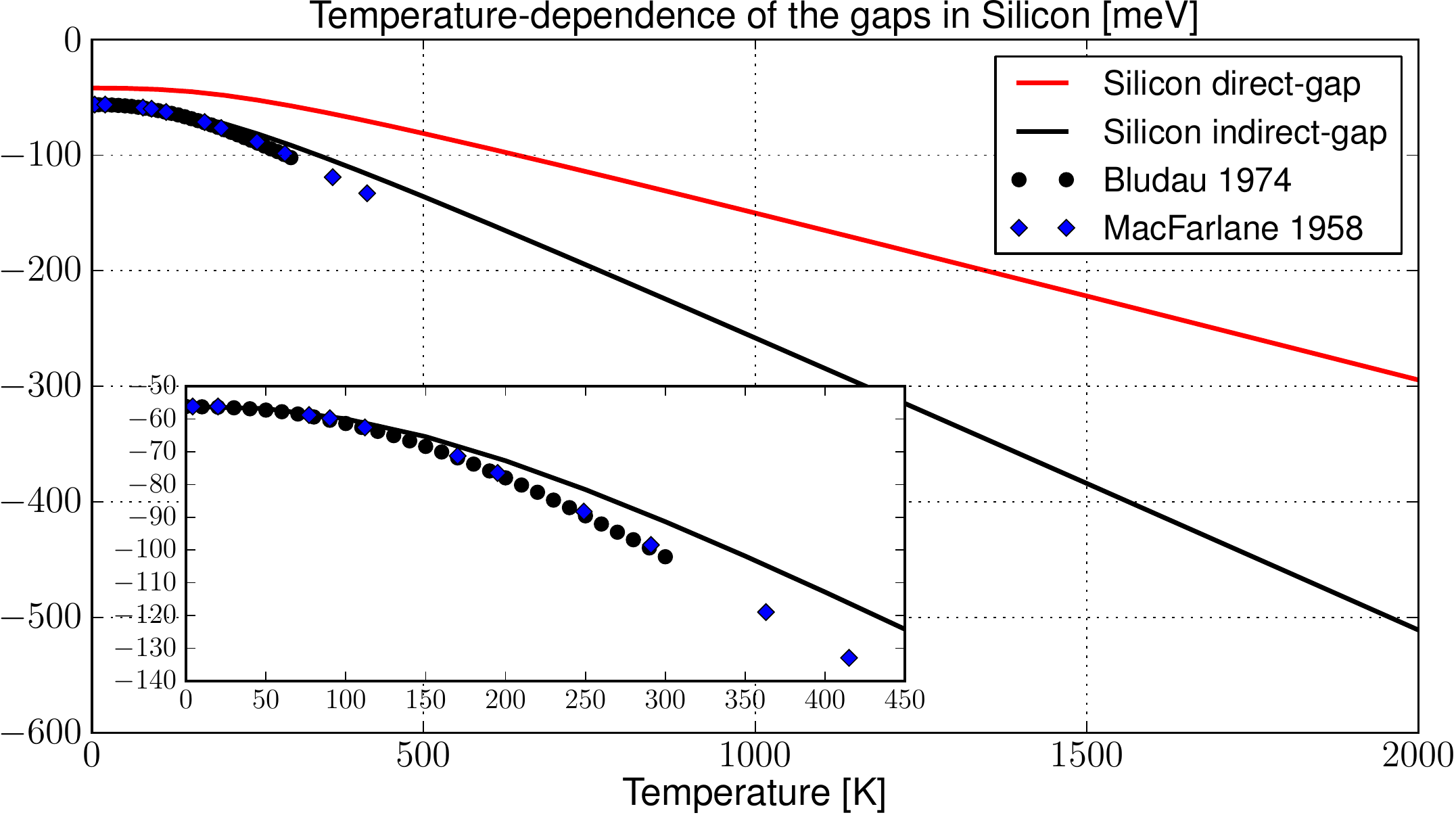}
\caption[Temperature dependence of bandgaps in Silicon.]{\label{Si-temp} Temperature dependence of the silicon gaps using the adiabatic temperature dependence on a 75x75x75 $\mathbf{q}$-grid with a Lorentzian extrapolation to vanishing imaginary parameter $\delta$. The slopes at high temperature are -0.147~meV/K for the direct gap of silicon and -0.255~meV/K for the indirect one. The experimental data from Ref.~\cite{Bludau1974} (black circles) and \cite{Macfarlane1958} (blue diamond).}
\end{figure}

The non-adiabatic temperature dependence of direct and indirect $\boldsymbol{\Gamma}-0.848\mathbf{X}$ bandgaps integrated over a 75x75x75 $\mathbf{q}$-grid with a Lorentzian extrapolation to vanishing imaginary parameter $\delta$ is reported on Figure~\ref{Si-temp} and gives slopes at high temperature of $-0.147$ and $-0.255$~meV/K, respectively.

The experimental zero-point motion renormalization of the silicon indirect bandgap is $-62$~meV obtained from mass derivative of the gap or $-64$~meV obtained from linear extrapolation to 0~K~\cite{Cardona2005}. The measured linear slope at high temperature is $-0.32$~meV/K~\cite{Bludau1974,Macfarlane1958}. 
Those experimental results are larger than our theoretical ones using the non-adiabatic extension to the AHC equations, as expected from DFT calculations. This is linked with the fact that we underestimate the ZPR with respect to GW calculations. 

\begin{figure}[bth]
\includegraphics[width=.95\linewidth]{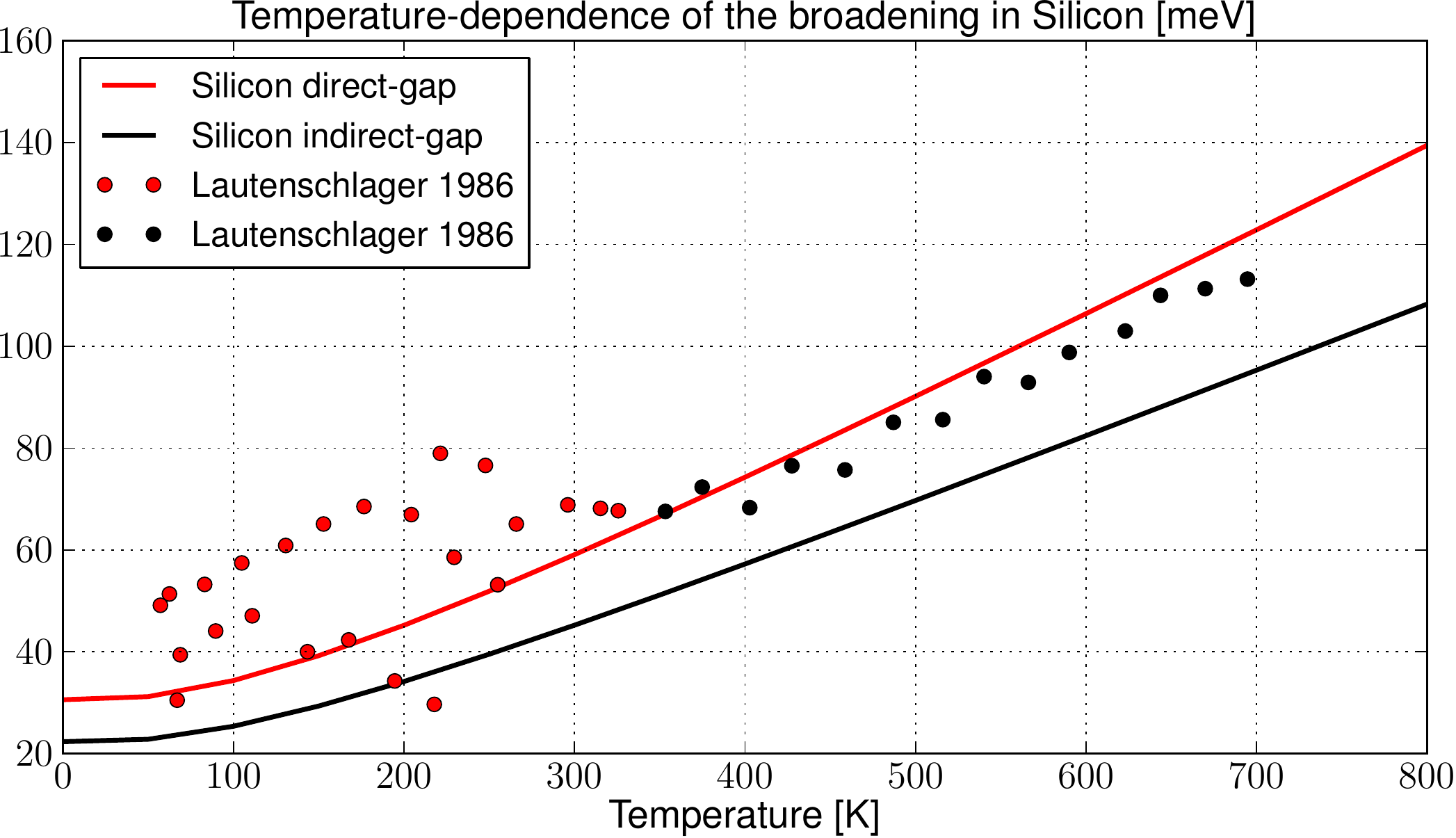}
\caption[Temperature dependence of bandgaps in Silicon.]{\label{Si-broad} Temperature dependence of the silicon broadening of the direct and indirect bandgap using the adiabatic temperature dependence on a 75x75x75 $\mathbf{q}$-grid. The experimental data (red and black dots) are from Ref.~\onlinecite{Lautenschlager1986}. See additional discussion in the text.}
\end{figure}

Additionally, we present on Figure~\ref{Si-broad} the phonon-induced broadening of the direct and indirect bandgap of silicon with temperature. The direct and indirect bandgap broadening at 0~K are computed to be 31~meV and 23~meV, respectively. The experimental broadening, measured with spectroscopic ellipsometry in Ref.~\onlinecite{Lautenschlager1986} (red dots), is attributed to the broadening of the E$_1$=$\boldsymbol{\Lambda}_{3}^v-\boldsymbol{\Lambda}_{1}^c$ direct transition with temperature. 
In Ref.~\onlinecite{Lautenschlager1986}, it is also mentioned that the measured values at higher temperature (black dots) are difficult to attribute to the broadening of one particular transition because the E$_1$ gap is nearly degenerate with the E$_0'=\boldsymbol{\Gamma}_{25'}^v-\boldsymbol{\Gamma}_{15}^c$.
Since the ellipsometry measurement is a spectroscopic measurement, it can only probe direct transitions. In consequence, we should compare the broadening results (both black and red dots) with the silicon red line (direct-gap E$_0'$).   

\begin{figure}[bth]
\subfloat[Silicon bandstructure at 0K.]
{\includegraphics[width=.48\linewidth]{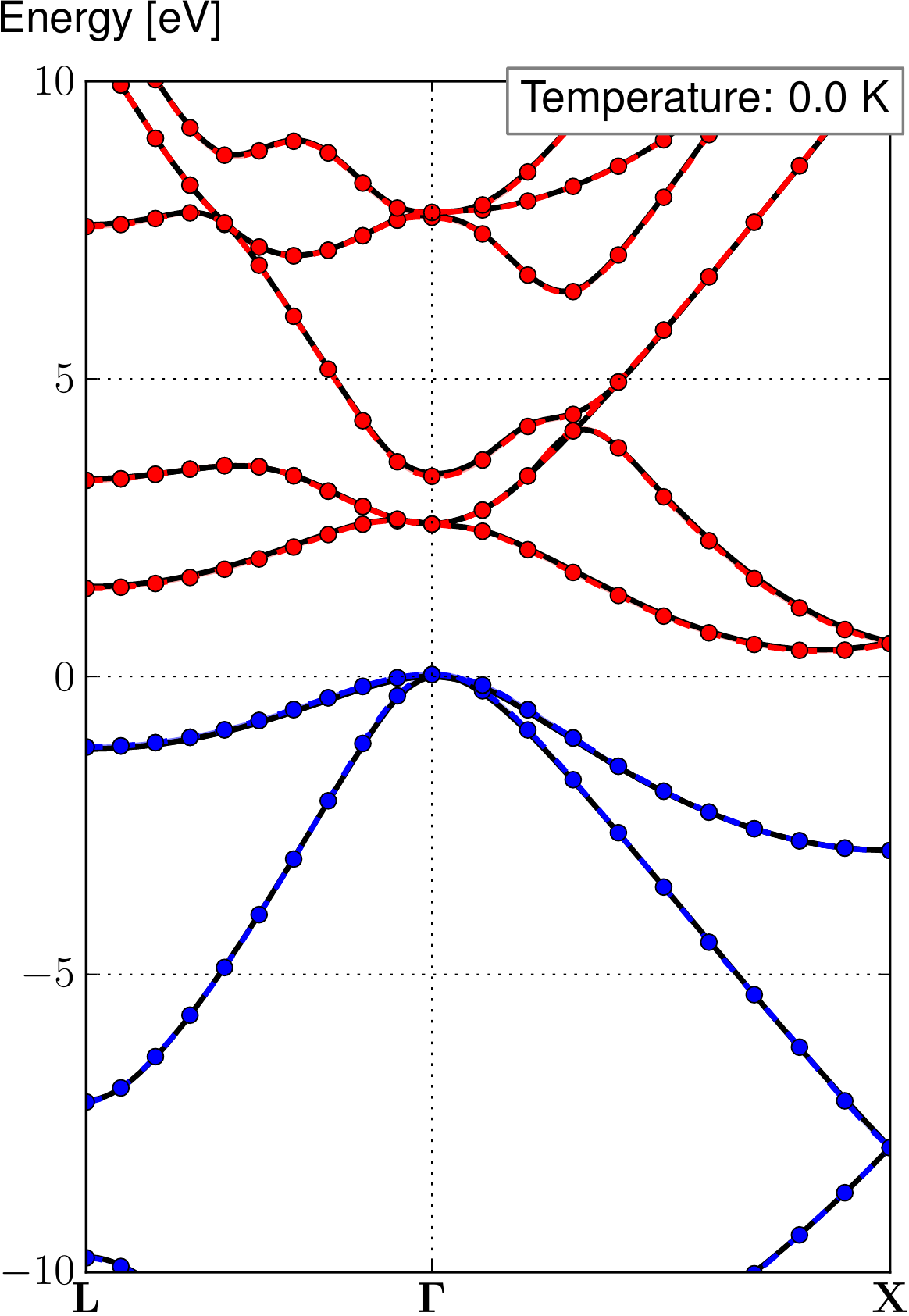}} \quad
\subfloat[Zoom of the left-hand side figure.]
{\includegraphics[width=.48\linewidth]{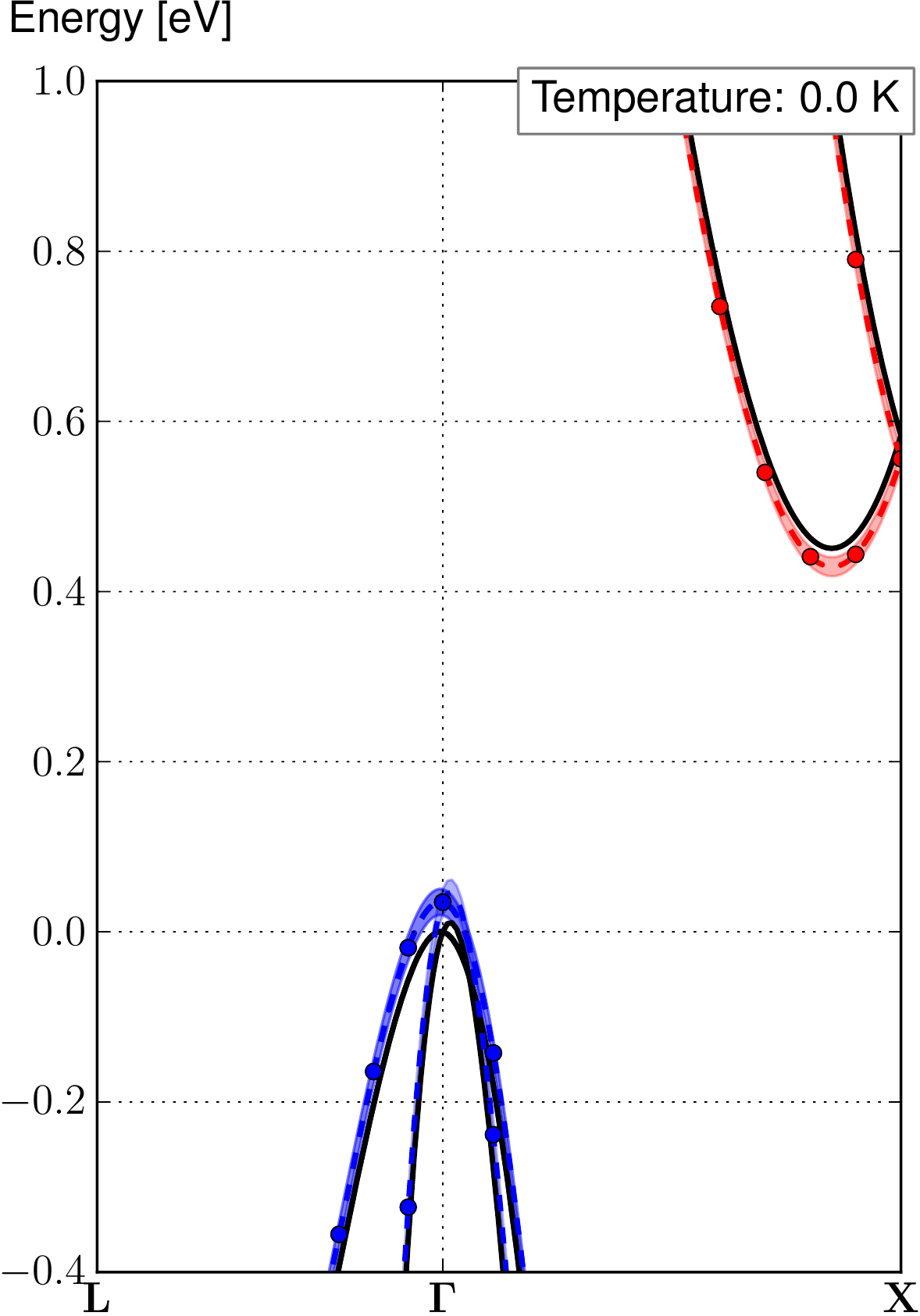}}
\caption[Electronic bandstructure of diamond.]{\label{Si-bs} Electronic bandstructure (plain black line), renormalization (dashed line) and phonon-induced broadening (envelope around the dashed line) at 0K using the non-adiabatic ZPR integrated on a 75x75x75 $\mathbf{q}$-point grid for $\delta$ extrapolated to zero for silicon, where the dots are the actual renormalization calculation. A spline function is used to connect the renormalization dots.}
\end{figure}

Finally, the non-adiabatically renormalized electronic bandstructure of silicon at 0~K along the $\mathbf{L}-\boldsymbol{\Gamma}-\mathbf{X}$ high symmetry line is shown on Figure~\ref{Si-bs} for a 75x75x75 $\mathbf{q}$-point grid with $\delta$ extrapolated to zero.

\subsection{Polar materials}
\subsubsection{Aluminum Nitride}

Aluminum nitride in the wurtzite structure ($\alpha$-AlN) is one of the widest bandgap nitride semiconductor. It has a P6$_3$mc (hexagonal, 186) space group. $\alpha$-AlN is used for high-temperature electronics and opto-elec\-tro\-nic applications~\cite{Litimein2002} with a melting temperature of 3273~K~\cite{MacChesney1970}.
The zinc\-blende form of aluminum nitride ($\beta$-AlN) has a F$\bar{4}3$m (cubic, 216) space group and has been reported to be experimentally metastable~\cite{Petrov1992}. We will study the temperature-dependence properties of these two phases using the equations mentioned above to compute the ZPR.

Concerning the numerical details of the calculations, the aluminium and nitrogen pseudopotentials were generated using the \texttt{fhi98PP} code \cite{Fuchs1999} with a 1.0247 atomic unit cut-off radius for pseudization and a maximum angular channel of $l=2$. Both of them are Troullier-Martins pseudopotentials with the Perdew/Wang~\cite{Perdew1992} parametrization of LDA.
The valence electrons of aluminum and nitrogen, treated explicitly in the \textit{ab-initio} calculations, are generated for the 3s$^{2}$3p$^{1}$ and 2s$^{2}$2p$^{3}$ configuration, respectively.

Convergence checks (error below 0.5~mHa per atom on the total energy) lead to the use of a 6x6x6 $\boldsymbol{\Gamma}$-centered Monkhorst-Pack $\mathbf{k}$-point sampling~\cite{Monkhorst1976} of the BZ  and an energy cut-off of 35~Hartree for the truncation of the plane wave basis set. 

The relaxed lattice parameters are calculated to be a=5.783 and c=9.255 Bohr for $\alpha$-AlN and a=8.130~Bohr for $\beta$-AlN. These values are at maximum 2.3\% below the experimental one (see Table~\ref{table:converegence_compounds}).

The electronic bandstructures were computed at the LDA level. For $\alpha$-AlN, there is a direct gap at $\boldsymbol{\Gamma}$ of 4.691~eV. This is well above the 4.056~eV bandgap computed within the Material's Project using GGA~\cite{Jain2013} but well in the range of other LDA bandgaps (see Table~\ref{table:band_structure_compounds}). 
 These DFT values naturally underestimate the experimental bandgap of 6.28~eV at 5K~\cite{Perry1978,Christensen1994}.
The zincblende $\beta$-AlN turns out to have an indirect $\boldsymbol{\Gamma}-\mathbf{X}$ bandgap of 3.308~eV, almost identical to the Materials Project value (see Table~\ref{table:band_structure_compounds}). The direct gap at $\boldsymbol{\Gamma}$ of 4.677~eV is a bit above most the the LDA values reported in the Table. We have nonetheless to bear in mind that the 4.2~eV LDA direct gap calculation performed in Ref.~\onlinecite{Litimein2002} is done at the experimental lattice parameter.

For the calculations of the ZPR we used 18 bands to describe the active space in Eqs.~\eqref{equationsumoverstate_id} and \eqref{equation_dynamical}. 

The $\mathbf{q}$-point integration for the direct bandgaps of $\alpha$-AlN is shown on Figure~7 of the supplemental materials~\cite{supplemental}  for the adiabatic equation and diverges for dense $\mathbf{q}$-grid as AlN is a polar material. 
The non-adiabatic direct bandgap ZPR shown on Figure~8 of the supplemental materials~\cite{supplemental} converges linearly with the $\mathbf{q}$-point grid and the $\delta$ behavior can be fitted by a Lorentzian function to 0. It gives a ZPR of -183.5~meV for the CBM and 194.2~meV for the VBM, leading to a ZPR of the direct bandgap of $\alpha$-AlN of -377.7~meV. 
The densest grid computed is a 34x34x34 $\mathbf{q}$-grid (2052 $\mathbf{q}$-points in the IBZ).

For the same reason as $\alpha$-AlN, the $\beta$-AlN diverges for the adiabatic equation and the divergence is shown on Figures~9 and 11 of the supplemental materials~\cite{supplemental}. 
The non-adiabatic direct bandgap ZPR shown on Figure~10 of the supplemental materials~\cite{supplemental} converges linearly with the $\mathbf{q}$-point grid and the $\delta$ behavior can be fitted by a linear or Lorentzian function to 0. It gives a ZPR of -187.54~meV for the CBM and 226.08~meV for the VBM, thus leading to a ZPR of the direct bandgap of $\beta$-AlN of -413.62~meV. 
The non-adiabatic indirect bandgap ZPR of $\beta$-AlN is shown on Figure~12 of the supplemental materials~\cite{supplemental} and converges linearly with the $\mathbf{q}$-point grid and the $\delta$ behavior can be fitted by a Lorentzian function to 0. This results into a ZPR of -108.36~meV for the CBM, resulting in a -334.4~meV ZPR of the indirect bandgap of $\beta$-AlN (see Table~\ref{table:zpr_list} for more information).
The densest computed grid  is a 100x100x1000 $\mathbf{q}$-grid ( 22776 $\mathbf{q}$-points in the IBZ).


\begin{figure}[bth]
\includegraphics[width=.99\linewidth]{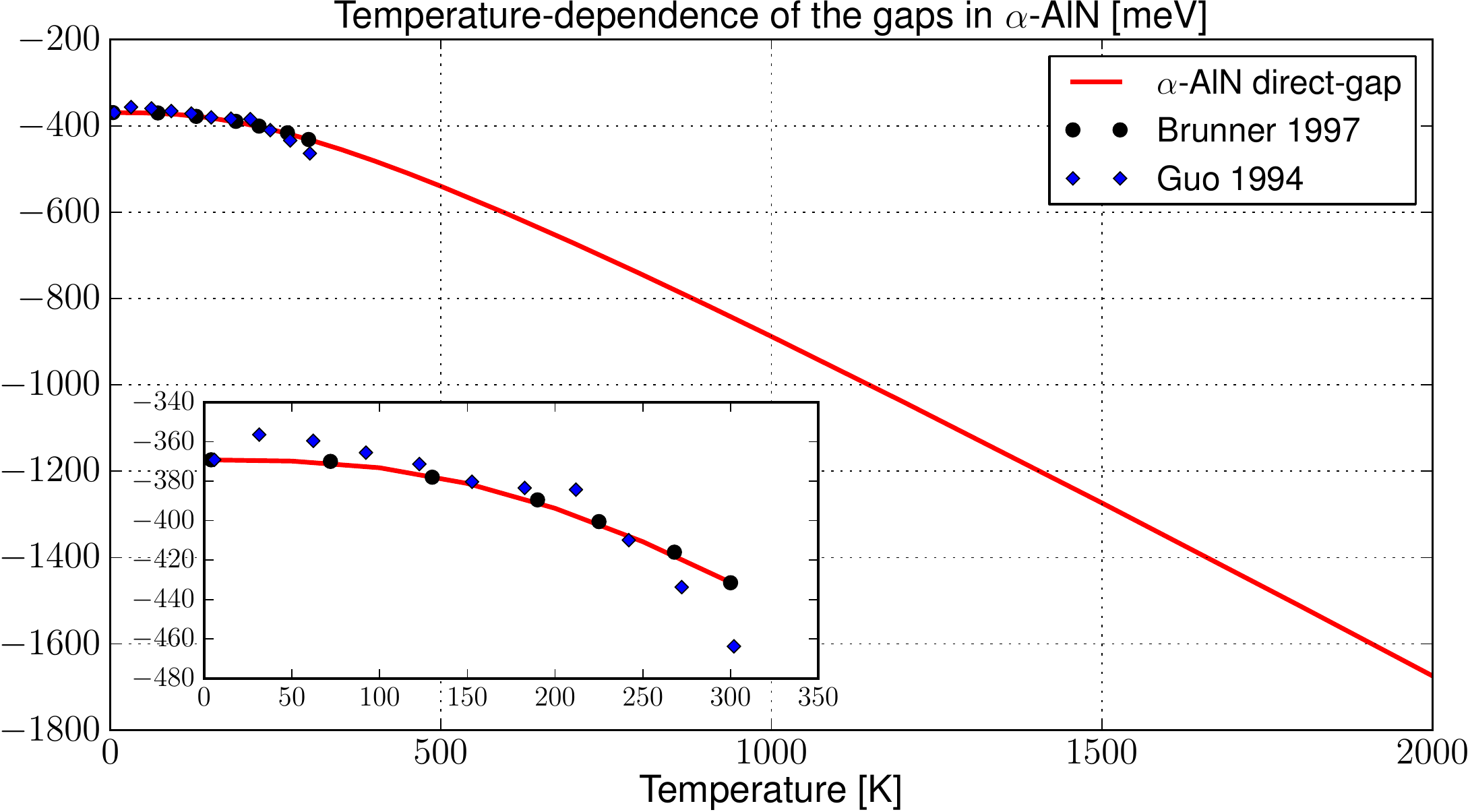}
\caption[Temperature dependence of bandgaps in diamond.]{ \label{aAlN-temp} Temperature dependence of the $\alpha$-AlN gaps using the non-adiabatic temperature dependence on a 34x34x34 $\mathbf{q}$-grid with Lorentzian extrapolation to vanishing imaginary parameter $\delta$. The slopes at high temperature is -0.772~meV/K  for the direct gap of $\alpha$-AlN.  The experimental data from Ref.~\onlinecite{Brunner1997} (black circles) and \onlinecite{Guo1994} (blue diamond).}
\end{figure}

\begin{figure}[bth]
\includegraphics[width=.99\linewidth]{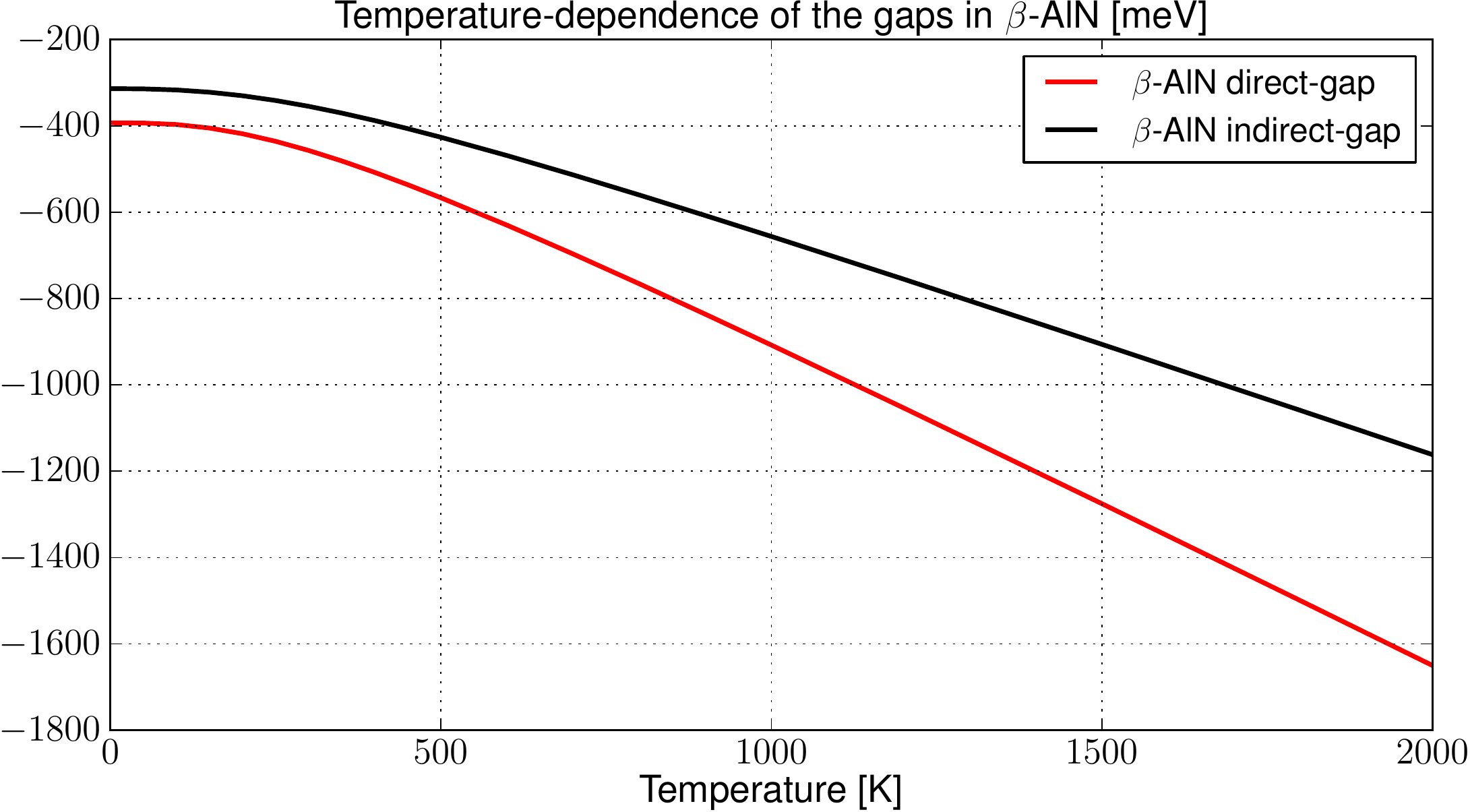}
\caption[Temperature dependence of bandgaps in diamond.]{\label{bAlN-temp} Temperature dependence of the $\beta$-AlN gaps using the non-adiabatic temperature dependence on a 75x75x75 $\mathbf{q}$-grid with Lorentzian extrapolation to vanishing imaginary parameter $\delta$. The slopes at high temperature are -0.763~meV/K for the direct gap of $\beta$-AlN and -0.521~meV/K for the indirect one. No experimental data were found in literature.}
\end{figure}

The temperature dependence of the three gaps is reported on Figures~\ref{aAlN-temp} and \ref{bAlN-temp}. The linear slopes at high temperature can be extracted to be -0.772, -0.521 and -0.763 meV/K for the direct gap of $\alpha$-AlN, the indirect bandgap of $\beta$-AlN and the direct bandgap of $\beta$-AlN, respectively.

The phonon-induced broadening is calculated for the 34x34x34  $\mathbf{q}$-grid to be 117~meV for the direct bandgap of $\alpha$-AlN. The broadening of the direct and indirect bandgaps of $\beta$-AlN at 0~K integrated on a 75x75x75 $\mathbf{q}$-grid  are 118~meV and 108~meV, respectively.

The experimental ZPR for $\alpha$-AlN has been obtained from linear extrapolation to 0~K of the change of the direct bandgap with temperature and yield a value of -239~meV with a linear slope at high temperature of -0.83~meV/K~\cite{Brunner1997,Passler1999,Cardona2005}, in relatively good agreement with our $\text{-}0.772$~meV/K value. 
The obvious disagreement with our theoretical value for the direct bandgap ZPR (-369~meV versus -239~meV) can be attributed to the fact that the experimental data set measured by Brunner \textit{et al.}~\cite{Brunner1997} are very scarce and on a narrow temperature range (4-298~K). As pointed out by P\"assler~\cite{Passler1999} for this compounds: \textit{``this illustrates the great importance of extending experimental measurements in wide bandgap materials far beyond room temperature"}. 

\begin{figure}[bth]
\subfloat[$\alpha$-AlN bandstructure at 0K.]
{\includegraphics[width=.48\linewidth]{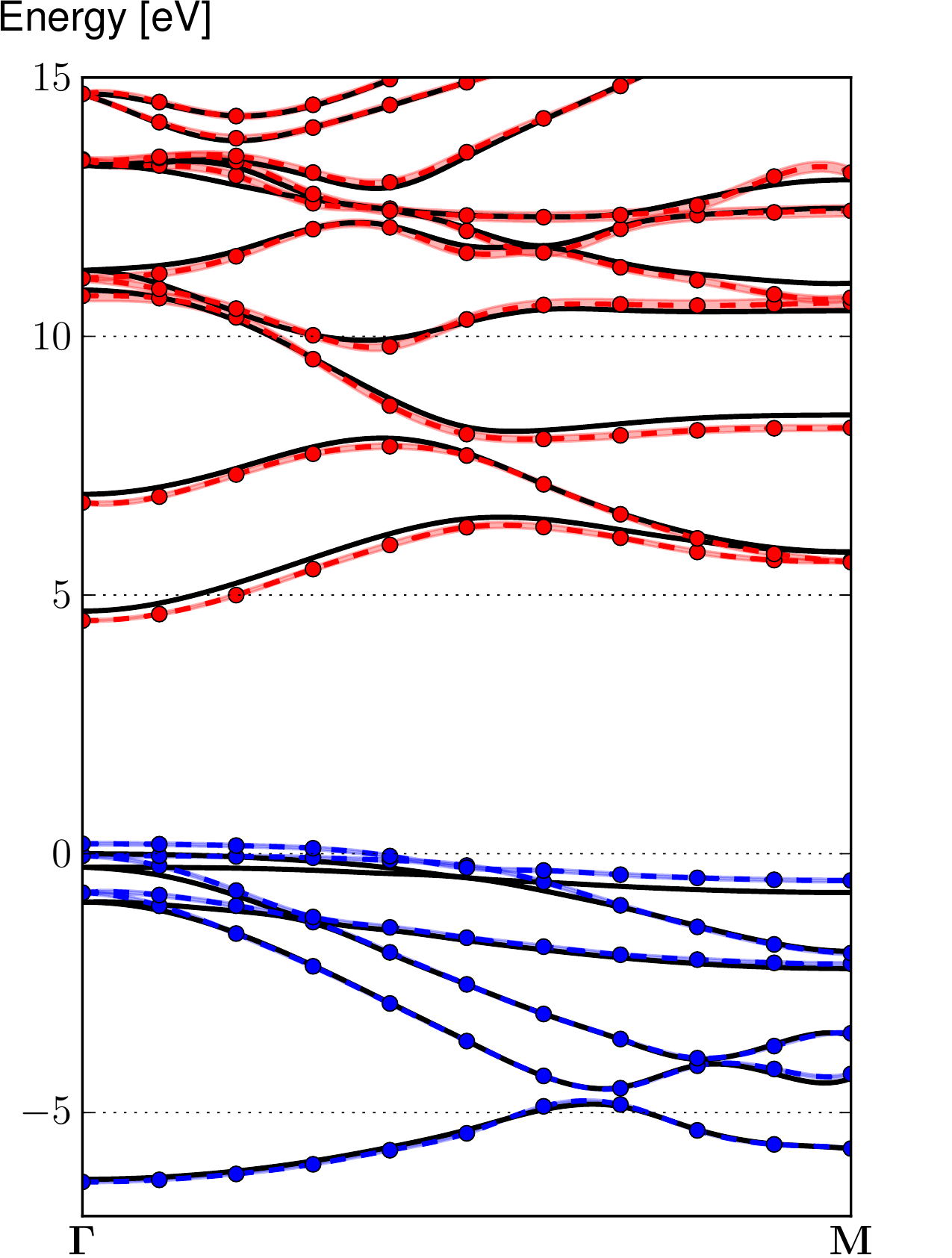}} \quad
\subfloat[Zoom of the left-hand side figure.]
{\includegraphics[width=.48\linewidth]{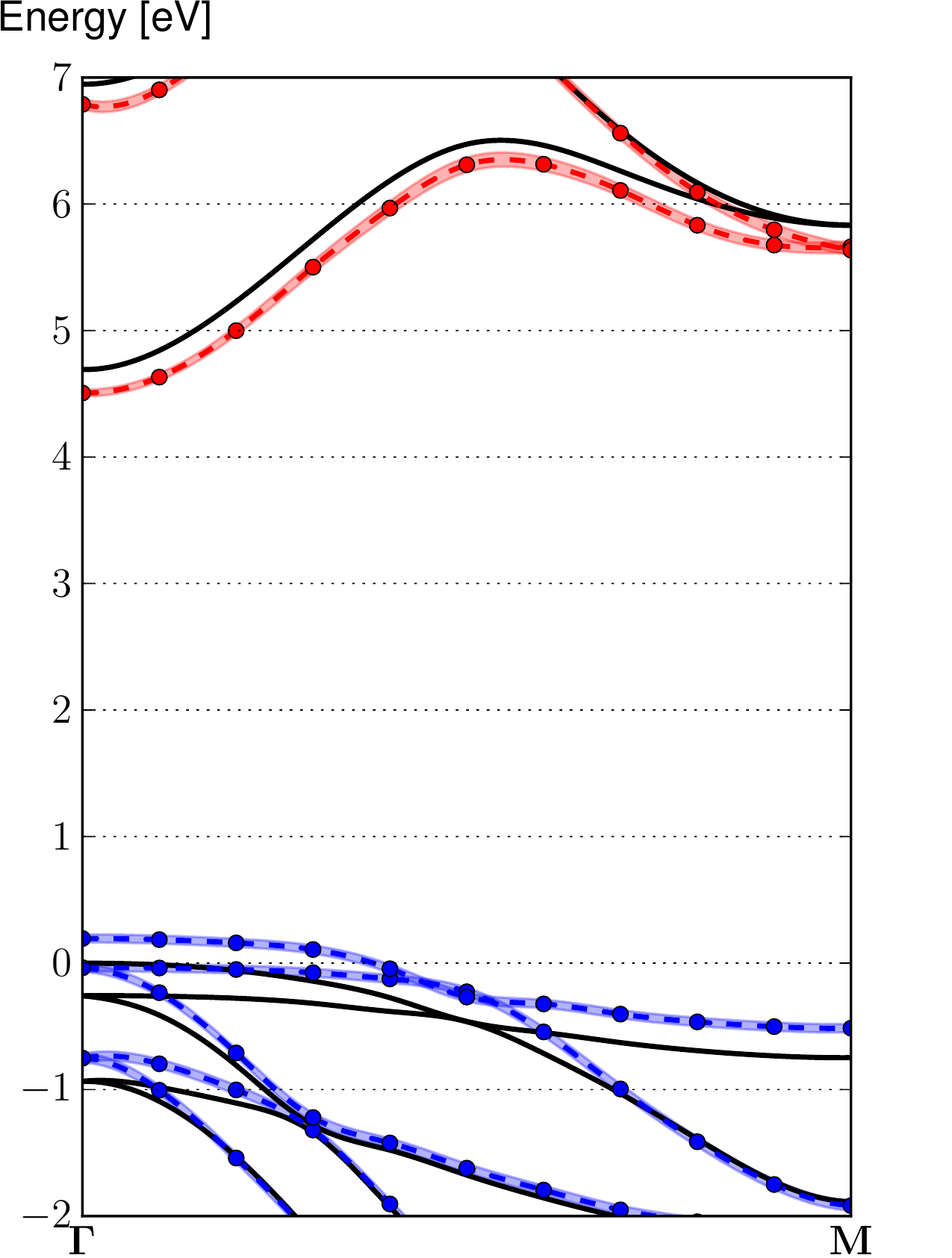}}
\caption[Electronic bandstructure of diamond.]{\label{aAlN-bs} Electronic bandstructure (plain black line), renormalization (dashed line) and phonon-induced broadening (envelope around the dashed line) at 0K using the non-adiabatic ZPR integrated on a 34x34x34 $\mathbf{q}$-point grid for $\delta$ extrapolated to zero for $\alpha$-AlN, where the dots are the actual renormalization calculation. A spline function is used to connect the renormalization dots.}
\end{figure}

\begin{figure}[bth]
\subfloat[$\beta$-AlN bandstructure at 0K.]
{\includegraphics[width=.48\linewidth]{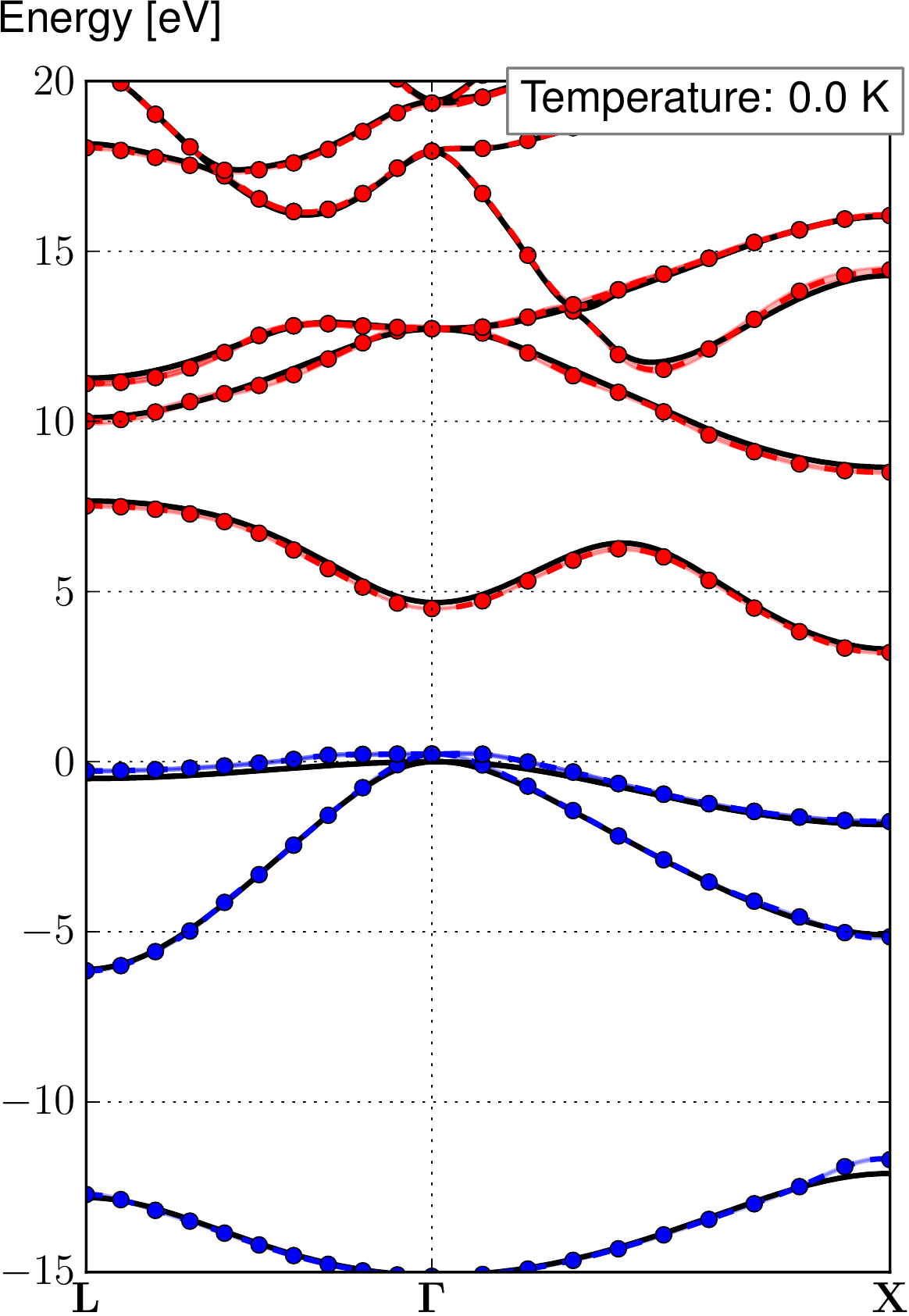}} \quad
\subfloat[Zoom of the left-hand side figure.]
{\includegraphics[width=.48\linewidth]{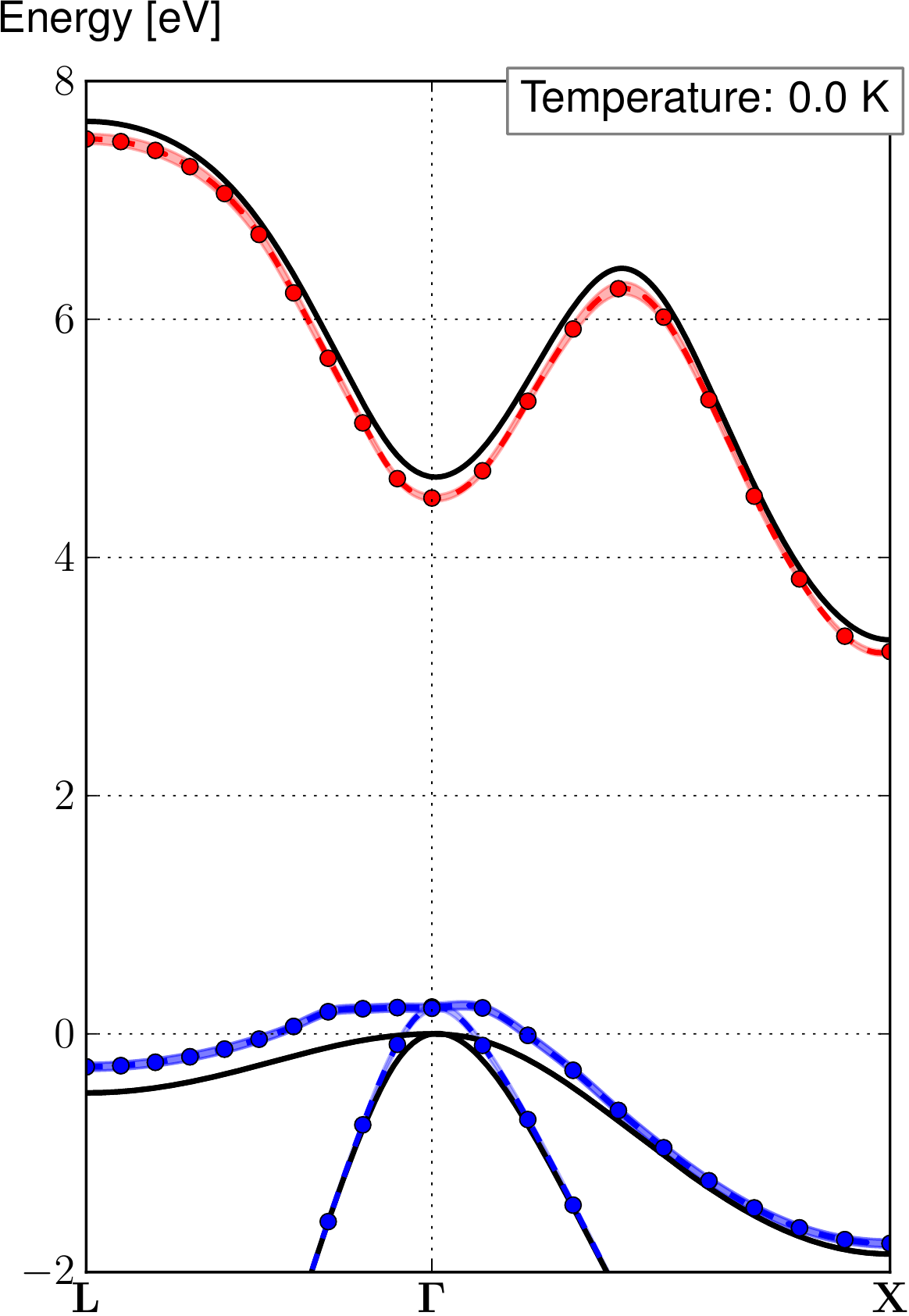}}
\caption[Electronic bandstructure of diamond.]{\label{bAlN-bs} Electronic bandstructure (plain black line), renormalization (dashed line) and phonon-induced broadening (envelope around the dashed line) at 0K using the non-adiabatic ZPR integrated on a 75x75x75 $\mathbf{q}$-point grid for $\delta$ extrapolated to zero for $\beta$-AlN, where the dots are the actual renormalization calculation. A spline function is used to connect the renormalization dots.}
\end{figure}

Finally, we show in Figures~\ref{aAlN-bs} and \ref{bAlN-bs} the non-adiabaticly renormalized electronic bandstructure at 0~K along the highest symmetry $\boldsymbol{\Gamma}-\mathbf{M}$ path for $\alpha$-AlN and along the $\mathbf{L}-\boldsymbol{\Gamma}-\mathbf{X}$ path of the $\beta$ phase of AlN. The thickness of the lines is associated with the lifetime of the electronic state computed with Eq.~\eqref{phonon_induced_lifetime_eq}.

\subsubsection{Boron Nitride}

\begin{figure*}[ht]
\subfloat[Adiabatic ZPR of c-BN.]
{\includegraphics[width=.41\linewidth]{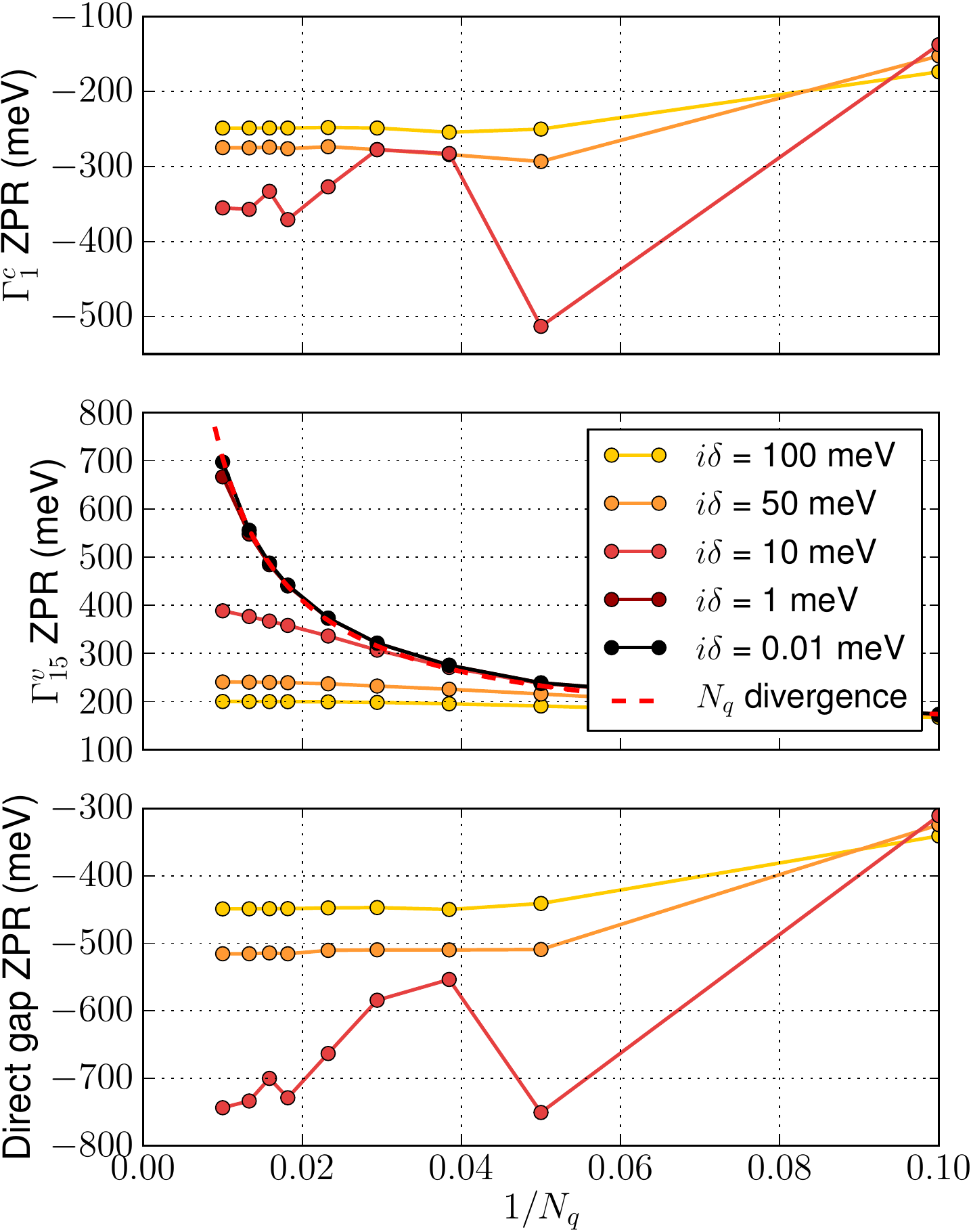}} 
\subfloat[$i\delta$ extrapolation of the ZPR. \label{BN1-conv1}]
{\includegraphics[width=.41\linewidth]{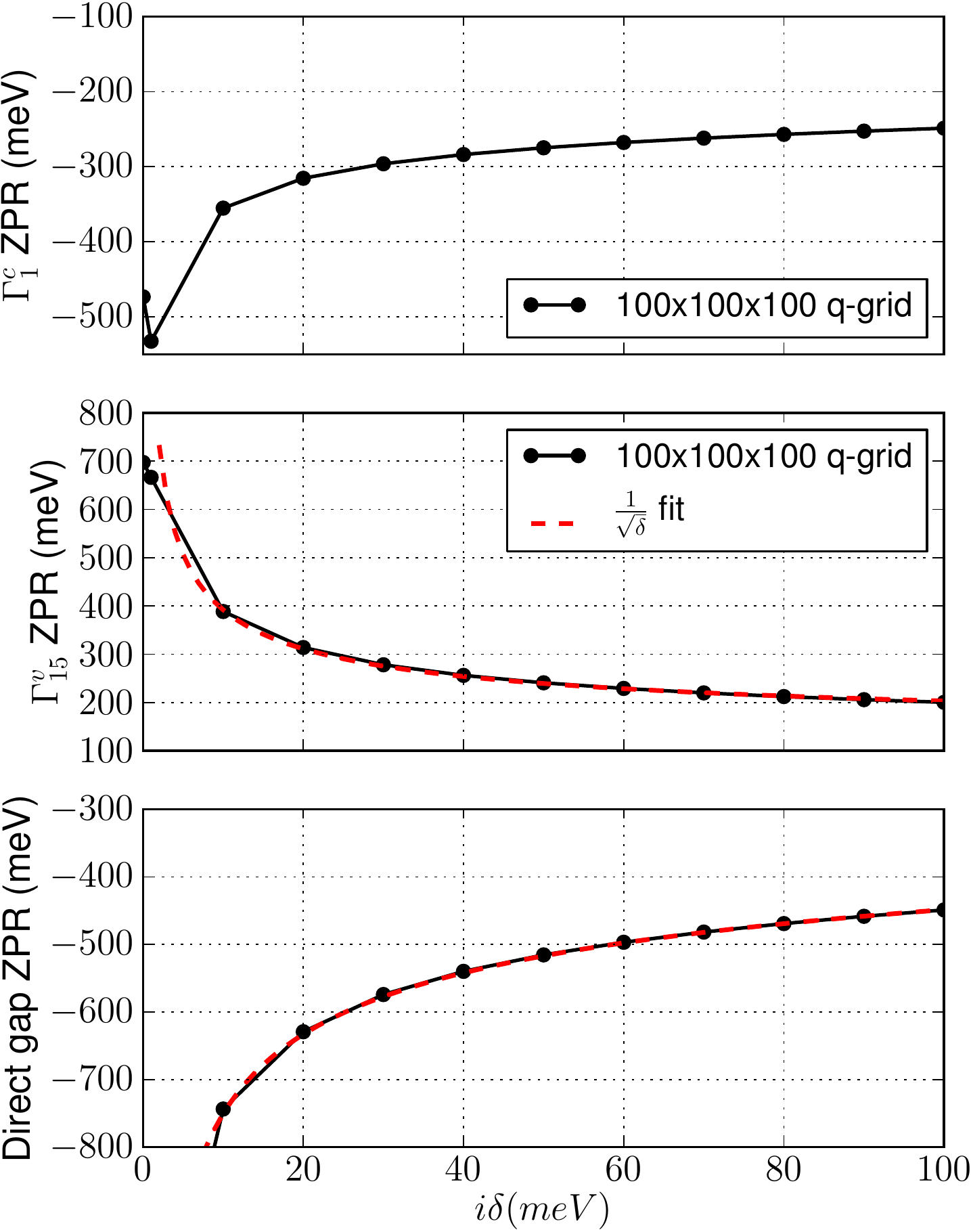}} 
\caption[Convergence study with respect to the $\mathbf{q}$-point grid density.]{\label{BN1-conv} Convergence study for the adiabatic (a) $\mathbf{q}$-point grid density and (b) $i\delta$ parameter for the direct bandgap ZPR of c-BN. The bottom Figures are the difference of the two Figures above them.  }
\end{figure*}

Boron nitride (BN) exists in various crystalline forms. Its most stable phase under normal condition is an hexagonal layered arrangement. It used as a lubricant and an additive to cosmetic products. The cubic boron nitride (c-BN) has a zincblende structure and is isoelectronic to diamond. It is the second hardest material below diamond, but its chemical stability is far superior with high thermal conductivity and low dielectric constant~\cite{Yu2003}.
Boron nitride is not found in nature and is therefore produced synthetically from boric acid or boron trioxide. We will only study the c-BN polymorph here. 

The boron and nitrogen pseudopotentials were also generated using the \texttt{fhi98PP} code~\cite{Fuchs1999} with a 1.0247 atomic unit cut-off radius for pseu\-di\-za\-tion and a maximum angular channel of $l=2$. Both of them are Troullier-Martins pseudopotential with the Perdew/Wang~\cite{Perdew1992} pa\-ra\-me\-tri\-za\-tion of LDA.
The valence electrons of boron and nitrogen, treated explicitly in the \textit{ab-initio} calculations, are the 2s$^{2}$2p$^{1}$ and 2s$^{2}$2p$^{3}$ orbitals, respectively.

Careful convergence checks (error below 0.5~mHa per atom on the total energy) lead to the use of a 8x8x8 $\boldsymbol{\Gamma}$-centered Monkhorst-Pack $\mathbf{k}$-point sampling~\cite{Monkhorst1976} of the BZ and an energy cut-off of 35~Hartree for the truncation of the plane wave basis set. 

The relaxed lattice parameter is calculated to be 6.746~Bohr, 1.3\% below the experimental value of 6.833~Bohr~\cite{Soma1974} (see Table~\ref{table:converegence_compounds} for more information on the structural properties).

The electronic bandstructure was computed at the DFT level and gave a direct gap at $\boldsymbol{\Gamma}$ of 8.890~eV and an indirect $\boldsymbol{\Gamma}-\mathbf{X}$ bandgap of 4.446~eV, intrinsically below the experimental bandgap of 6.4~eV at 300K~\cite{Chrenko1974} (see Table~\ref{table:band_structure_compounds}).   
For the calculations of the ZPR, we used 18 bands to describe the active space in Eqs.~\eqref{equationsumoverstate_id} and \eqref{equation_dynamical}. 


The c-BN is also a polar material and therefore diverges for the adiabatic equation as shown on Figures~\ref{BN1-conv} and 14 of the supplemental materials~\cite{supplemental}. 
The non-adiabatic direct bandgap ZPR shown on Figure~13 of the supplemental materials~\cite{supplemental} converges linearly with the $\mathbf{q}$-point grid and the $\delta$ behavior can be fitted by a linear or Lorentzian function to 0 and gives a ZPR of -301.48~meV for the $\Gamma_{1}^c$ state and 200.5~meV for the VBM, thus leading to a ZPR of the direct bandgap of c-BN of -502.0~meV. 
The non-adiabatic indirect bandgap ZPR of c-BN is shown on Figure~15 of the supplemental materials~\cite{supplemental} and converges linearly with the $\mathbf{q}$-point grid and the $\delta$ behavior can be fitted by a Lorentzian function to 0 and gives a ZPR of -205.08~meV for the CBM, thus leading to a ZPR of the indirect bandgap of c-BN of -405.6~meV (see Table~\ref{table:zpr_list} for more information).
The densest grid computed is a 100x100x1000 $\mathbf{q}$-grid ( 22776 $\mathbf{q}$-points in the IBZ).

\begin{figure}[bth]
\includegraphics[width=.99\linewidth]{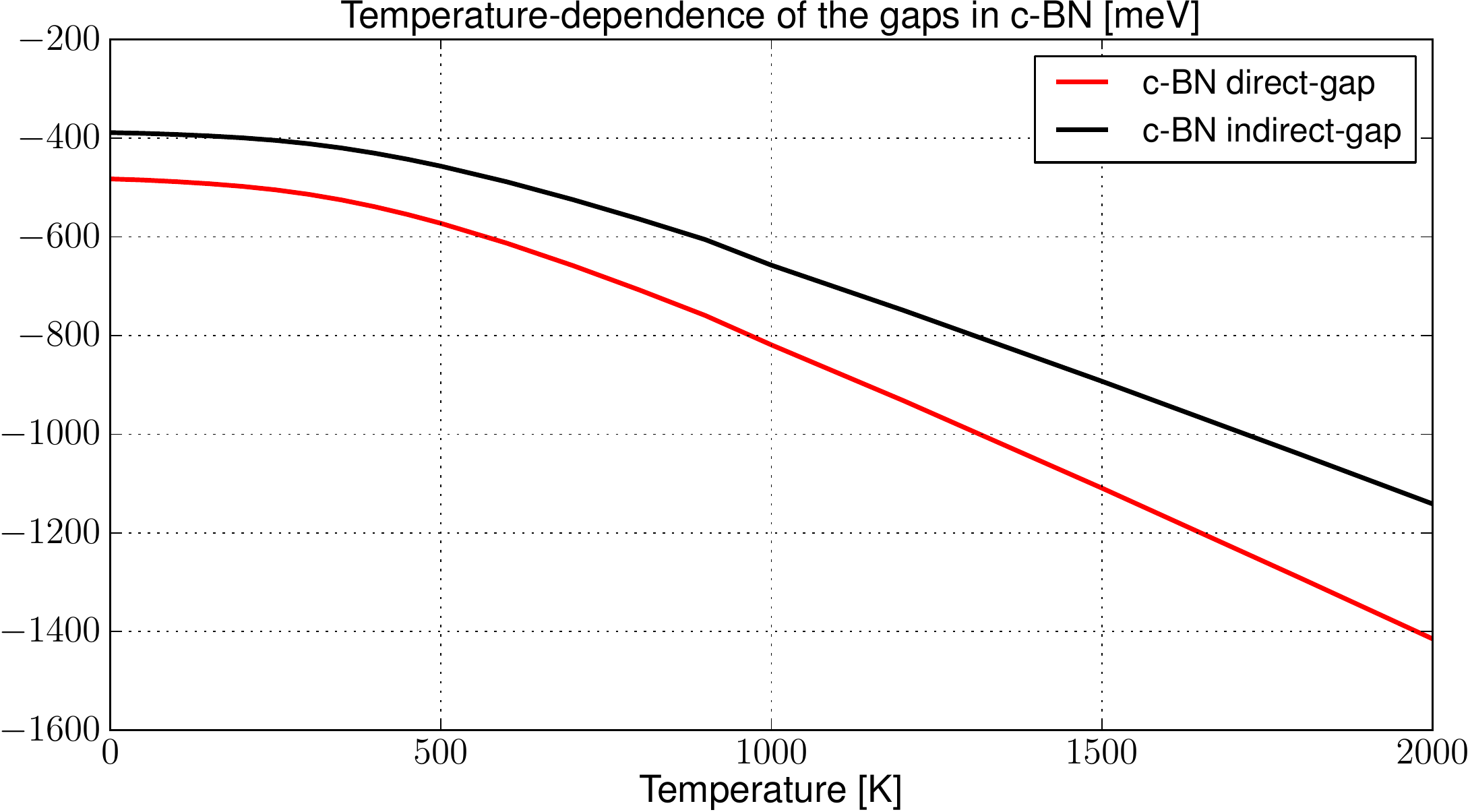}
\caption[Temperature dependence of bandgaps in diamond.]{\label{BN-temp} Temperature dependence of the c-BN gaps using the non-adiabatic temperature dependence on a 75x75x75 $\mathbf{q}$-grid with Lorentzian extrapolation to vanishing imaginary parameter $\delta$. The slopes at high temperature is -0.639~meV/K for the direct gap of c-BN and -0.521~meV/K for the indirect one. No experimental data were found in literature.}
\end{figure}

The non-adiabatic temperature dependence of direct and indirect bandgaps are reported on Figure~\ref{BN-temp} for a 75x75x75 $\mathbf{q}$-grid with extrapolation to zero $\delta$ and give slopes at high temperature of -0.639 and -0.521~meV/K, respectively.
The phonon-induced broadening is calculated for the 75x75x75 $\mathbf{q}$-grid to be 315~meV and 136~meV for the direct and indirect bandgap of c-BN at 0~K, respectively.

\begin{figure}[bth]
\subfloat[c-BN bandstructure at 0K.]
{\includegraphics[width=.48\linewidth]{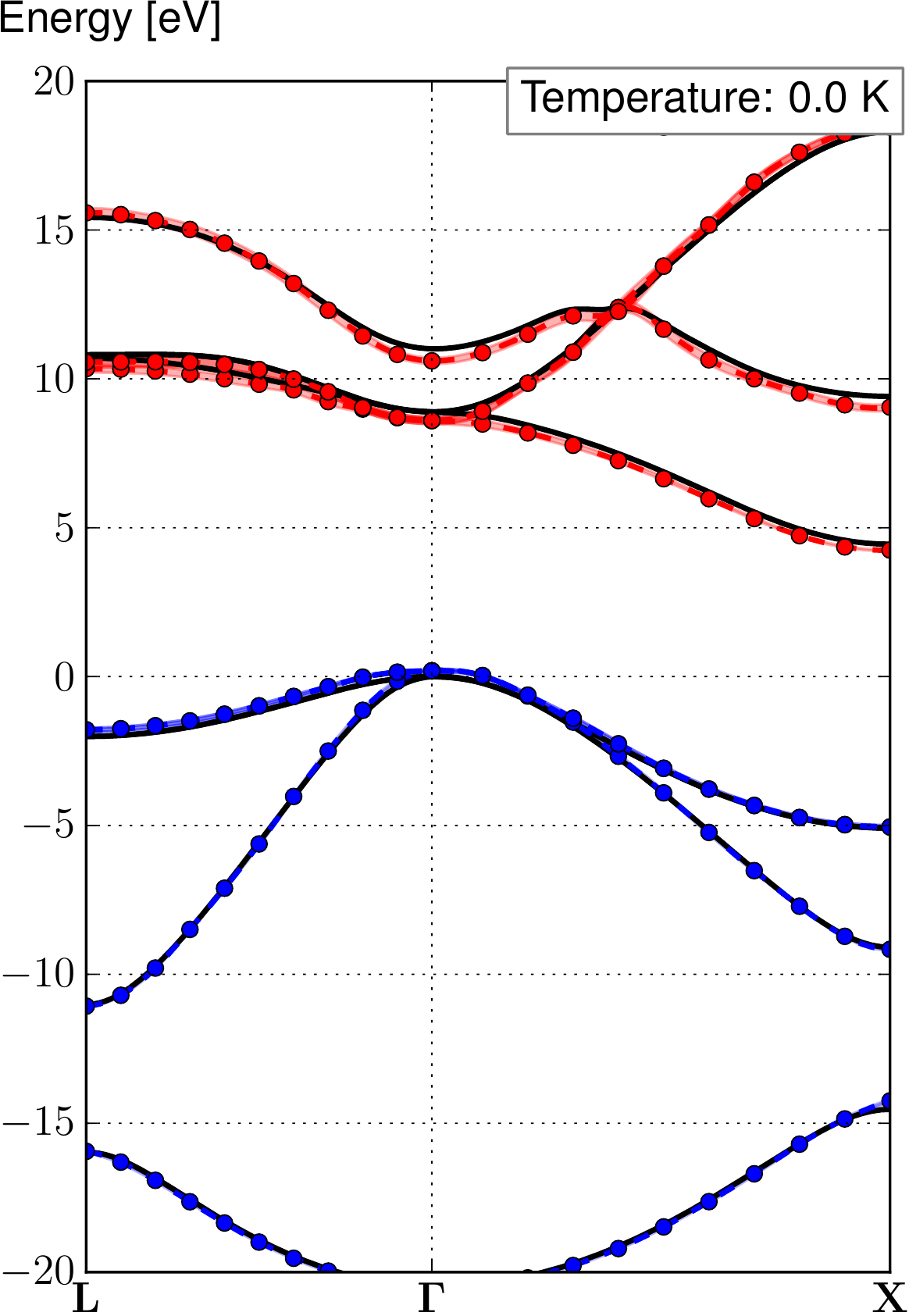}} \quad
\subfloat[Zoom of the left-hand side figure.]
{\includegraphics[width=.48\linewidth]{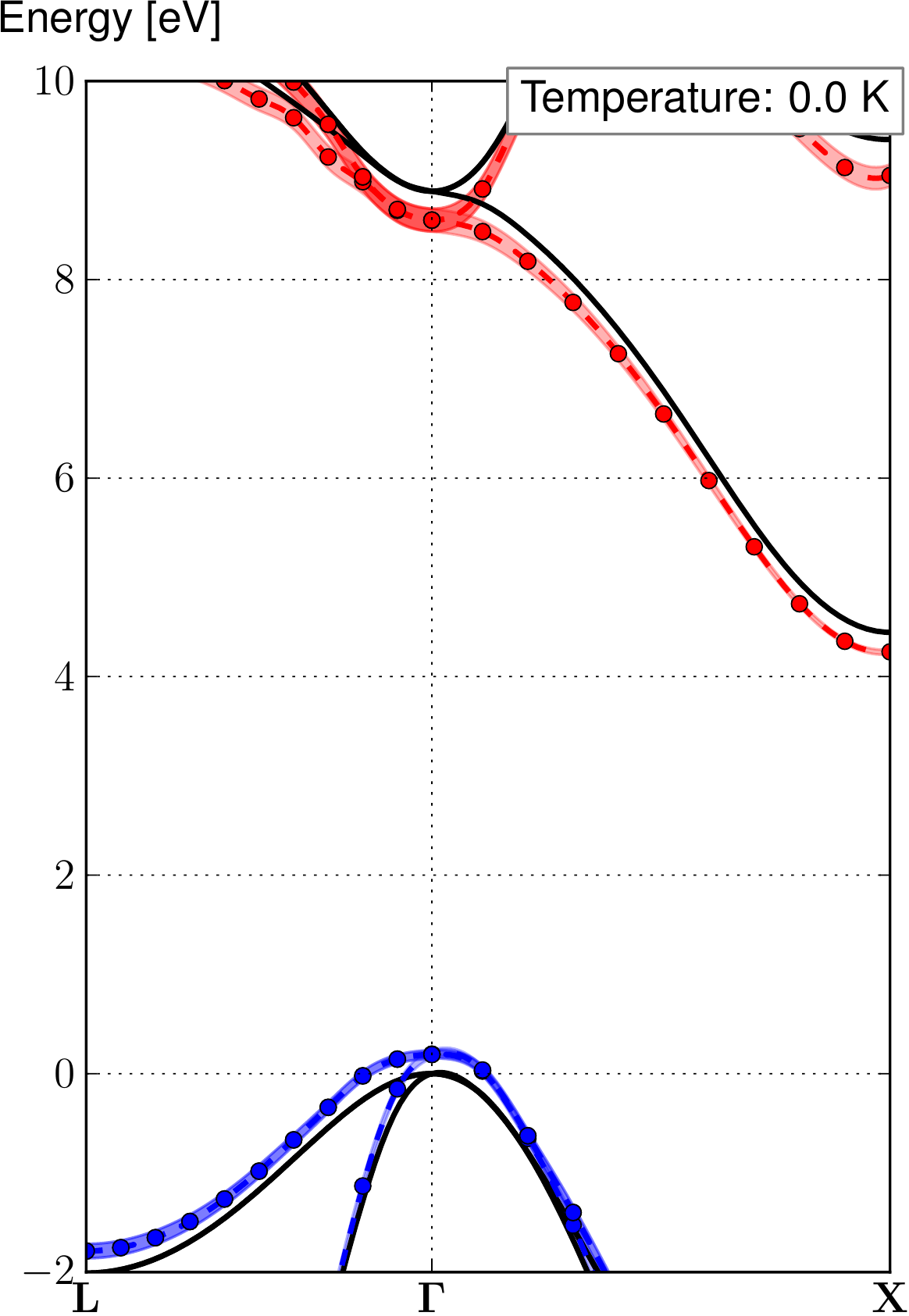}}
\caption[Electronic bandstructure of diamond.]{\label{BN-bs} Electronic bandstructure (plain black line), renormalization (dashed line) and phonon-induced broadening (envelope around the dashed line) at 0K using the non-adiabatic ZPR integrated on a 75x75x75 $\mathbf{q}$-point grid for $\delta$ extrapolated to zero for $\beta$-AlN, where the dots are the actual renormalization calculation. A spline function is used to connect the renormalization dots.}
\end{figure}

Finally, the non-adiabatic renormalized electronic bandstructure of c-BN at 0K along the $\mathbf{L}-\boldsymbol{\Gamma}-\mathbf{X}$ high symmetry line is shown on Figure~\ref{BN-bs} for a 75x75x75 $\mathbf{q}$-point grid with $\delta$ extrapolated to zero.

\begin{table*}[ht]
\begin{ruledtabular}
\begin{tabular}{l c c c c c c c c c}
             &               & \multicolumn{3}{c}{ZPR [meV]}  & \multicolumn{2}{c}{$(d Gap/dT)_{T\rightarrow \infty}$ [meV/K]} &  \multicolumn{2}{c}{Broadening [meV]} \\
\cline{2-5} \cline{6-7}\cline{8-9}
 Compounds   &  Gap & Adiabatic & Non-adiabatic & Experimental & Non-adiabatic & Experimental & Adiabatic limit & Experimental \\
\hline
$\alpha$-AlN & $\boldsymbol{\Gamma}-\boldsymbol{\Gamma}$   & -   & -377.7 & -239~\cite{Brunner1997}  & -0.772 & -0.83~\cite{Brunner1997,Passler1999,Cardona2005} & 117    & \\
$\beta$-AlN  & $\boldsymbol{\Gamma}-\boldsymbol{\Gamma}$   &  -  & -413.6 &                          & -0.763 & & 118 &  \\
             & $\boldsymbol{\Gamma}-\mathbf{X}$            & -   & -334.4 &                          & -0.521 & & 108 &  \\
c-BN         & $\boldsymbol{\Gamma}-\boldsymbol{\Gamma}$   & -   & -502.0 &                          & -0.639 & & 315 &   \\
             & $\boldsymbol{\Gamma}-\mathbf{X}$            & -   & -405.6 &                          & -0.521 & & 136 &  \\  
C            & $\boldsymbol{\Gamma}-\boldsymbol{\Gamma}$   & -438.6 & -415.8 & -320~\cite{Logothetidis1992},-450~\cite{Logothetidis1992}  & -0.504 & -0.60~\cite{Logothetidis1992},  -0.69~\cite{Logothetidis1992} &  180 & \\
             & $\boldsymbol{\Gamma}-0.727\mathbf{X}$       & -379.3 & -329.8   & -364~\cite{Cardona2005}  & -0.435 & -0.54~\cite{Cardona2005} &  63 & \\
Si           & $\boldsymbol{\Gamma}-\boldsymbol{\Gamma}$   &  -47.1 &  -42.1   &                          & -0.147 & & 31 &  \\ 
             & $\boldsymbol{\Gamma}-0.848\mathbf{X}$       &  -64.3 &  -56.2   & -62~\cite{Cardona2005},-64~\cite{Cardona2005}   & -0.255 & -0.32~\cite{Bludau1974,Macfarlane1958} & 22 & $\sim$35~\cite{Lautenschlager1986} \\       
\end{tabular}
\caption[ZPR for the different studied compounds.]{ZPR for the different studied compounds as well as the broadening in the static limit at 0~K. The two experimental results for the direct bandgap renormalization of diamond from Ref.~\onlinecite{Logothetidis1992} are extracted using first or second-derivative line-shape analysis. The two experimental ZPR for the indirect bandgap of silicon from Ref.~\onlinecite{Cardona2005} are obtained  from mass derivative of the gap and from linear extrapolation to 0~K.}
\label{table:zpr_list}
\end{ruledtabular}
\end{table*}

\section{Conclusions}
\label{Conclusions}
 
 After a brief reminder of the theory, we present a solution to the divergence problem due to a residual Born effective charge stemming from the finite $\mathbf{k}$-point grid in numerical $\textit{ab-initio}$ calculations. We analyze theoretically the $\mathbf{q}$-point convergence for a polar or non-polar material; in the adiabatic or non-adiabatic approximation; for the renormalization of the band extrema or other states. 
We propose an equivalent analysis for the convergence of the imaginary parameter $\delta$ tending to zero. We demonstrate that the adiabatic AHC formalism breaks down for polar materials and the non-adiabatic AHC formalism should therefore be used for these materials. We then propose a systematic procedure to converge the zero-point motion renormalization (ZPR) and apply it for five semiconductors and insulators (diamond, silicon, the $\alpha$ and $\beta$ phase of aluminum nitride and boron nitride).
For these materials, we present the non-adiabatic renormalized electronic bandstructure (at the density functional theory level) due to electron-phonon coupling as well as the phonon induced lifetime in the adiabatic limit. 
We also show the temperature dependence of their direct and indirect bandgaps and compare  them with experiment whenever available. 
The non-adiabatic ZPR at the DFT level systematically underestimates the experimental results (by less than 10\%) except in $\alpha$-AlN where the theoretical value is larger than the experimental one. 
We strongly question the validity of the experimental result in this case as the experimental ZPR was obtained by linear extrapolation to 0~K on a very limited temperature range (0-300~K) where the linear regime was not yet achieved. We therefore believe that the experimental result for $\alpha$-AlN from Ref.~\onlinecite{Brunner1997} is underestimated. 
On the one hand, we hope that our work will emulate experimental work on wider temperature range.
On the other hand, this approach might also be used in the future to compute more evolved temperature-dependent properties depending on the electronic structure, as e.g. optical properties.

\section{Acknowledgements}
\label{Acknowledgements}
This work was supported by the FRS-FNRS through a FRIA fellowship (S.P.) and a FNRS fellowship (Y.G.) as well as the FRQNT through a postdoctoral research fellowship (J.L.J.). Moreover, A. M. would like to acknowledge financial support from the Futuro in Ricerca grant No.~RBFR12SW0J of the Italian Ministry of Education, University and Research.
The authors would like to thank Yann Pouillon and Jean-Michel Beuken for their valuable technical support and help with the test and build systems of ABINIT.
Computational resources have been provided by the supercomputing facilities of the Universit\'e catholique de Louvain (CISM/UCL) and the Consortium des \'Equipements de Calcul Intensif en F\'ed\'eration Wallonie Bruxelles (CECI) funded by the Fonds de la Recherche Scientifique de Belgique (FRS-FNRS) under Grant No.~2.5020.11.

\appendix
\section{Derivation of the renormalization factor}
\label{Technicalities}

\subsection{Minimization of the variational second-order electronic energy}
The variational second-order electronic energy without non-linear core correction can be written as (see Eq.~(60) of Ref.~\onlinecite{Gonze1997})
\begin{widetext}
\begin{multline}\label{etotEQ}
E_{\mathbf{-q},\mathbf{q}}^{(2)} = \frac{\Omega_0}{(2\pi)^3}\int_{BZ} d\mathbf{k} \sum_{n}^{occ} s_n \Big( \langle u_{n\mathbf{k},\mathbf{q}}^{(1)}|H_{\mathbf{k+q,k+q}}^{(0)}-\varepsilon_{n\mathbf{k}}^{(0)} | u_{n\mathbf{k,q}}^{(1)}\rangle + \langle u_{n\mathbf{k,q}}^{(1)}| v_{ext,\mathbf{k+q,k}}^{(1)}| u_{n\mathbf{k}}^{(0)}\rangle +\langle u_{n\mathbf{k}}^{(0)}| v_{ext,\mathbf{k,k+q}}^{(1)}| u_{n\mathbf{k,q}}^{(1)}\rangle \\
+ \langle u_{n\mathbf{k}}^{(0)}| v_{ext,\mathbf{k,k}}^{(2)}| u_{n\mathbf{k}}^{(0)}\rangle \Big)  + \frac{1}{2}\int_{\Omega_0} \frac{d^2(n\varepsilon_{xc})}{dn^2}\bigg|_{n^{(0)}(\mathbf{r})}|n_{\mathbf{q}}^{(1)}(\mathbf{r})|^2d\mathbf{r}+2\pi \Omega_0\sum_{\mathbf{G}\neq0}  \frac{|n_{\mathbf{q}}^{(1)}(\mathbf{G})|^2}{|\mathbf{q+G}|^2}+n_{\mathbf{q}}^{(1)*}(\mathbf{G}=0) \frac{2\pi}{q}w_{\mathbf{q}}^{(1)} \\
+n_{\mathbf{q}}^{(1)}(\mathbf{G}=0)\frac{2\pi}{q}w_{\mathbf{q}}^{(1)*} + 2\pi \Omega_0 \frac{|n_{\mathbf{q}}^{(1)}(\mathbf{G}=0)|^2}{q^2},
\end{multline}
\end{widetext}
where we follow the notation of Ref.~\onlinecite{Gonze1997}, \emph{i.e.} the superscripts $(0)$ and $(1)$ refer to the unperturbed and first-order perturbation (here in nucleus motion) of the periodic part of the wavefunction, $s_n$ is the spin degeneracy factor, $v_{ext,\mathbf{k+q},\mathbf{k}}^{(1)}$ is the first-order perturbed potential external to the electronic system that includes the nucleus one  
\begin{equation}
v_{ext,\mathbf{k+q,k}}^{(1)}=  v_{sep,\mathbf{k+q,k}}^{(1)} +  \bar v_{loc,\mathbf{q}}^{(1)},
\end{equation}
and $\varepsilon_{xc}$ is the exchange-correlation energy per electron
\begin{equation}
\frac{d^2 (n\varepsilon_{xc})}{dn^2}\bigg|_{n^{(0)}}=\frac{dv_{xc}}{dn}\bigg|_{n^{(0)}}.
\end{equation} 
The second-order change of the nonlocal potential $v_{sep,\mathbf{k},\mathbf{k}}^{(2)}$ is given in Eq.~(54) of Ref.~\onlinecite{Gonze1997}.

The first-order change of the local potential of Eq.~\eqref{localpotential1} for $\mathbf{G}=0$ can be written as  
\begin{equation}
\lim_{q \to 0} \bar v_{loc,\mathbf{q}}^{(1)}(0) = \frac{4\pi}{\Omega_0}\frac{1}{q} w_{\mathbf{q}}^{(1)},
\end{equation}
with
\begin{equation}\label{w1}
w_{\mathbf{q}}^{(1)} = -i \frac{q_{\alpha}}{q}e^{-i\mathbf{q}\cdot \boldsymbol{\tau}_\kappa} \Big( -Z_\kappa +\frac{q^2}{4\pi}C_\kappa + O(q^4) \Big).
\end{equation}

To write Eq.~\eqref{etotEQ} is a more compact form, we define the following vectors
\begin{align}
\label{vec1} \mathbf{x}_i &\triangleq \braket{\mathbf{G} | u_{n\mathbf{k},\mathbf{q}}^{(1)}} \\
\label{vec2} \mathbf{u}_i &\triangleq s_n w_{n\mathbf{k}} \braket{\mathbf{G} | u_{n\mathbf{k}}^{(0)}} \\
\label{vec3} \mathbf{w}_i &\triangleq s_n w_{n\mathbf{k}} \braket{\mathbf{G} | v_{ext,\mathbf{k+q},\mathbf{k}}^{(1)} | u_{n\mathbf{k}}^{(0)}}
\end{align}
where $w_{n\mathbf{k}}$ is the weight of the $\mathbf{k}$-point and includes a band $n$ dependence to indicate that it is zero for unoccupied states and where the index $i$ stands for the combined planewave component $\mathbf{G}$, band index $n$ and wave-vector $\mathbf{k}$ indices.  
It will later be useful to note that 
\begin{equation}
\label{RESOLUTION_1}
\begin{split}
\mathbf{u}^\dagger\!\mathbf{x} &= \frac{\Omega_0}{(2\pi)^3}\int_{BZ} d\mathbf{k} \sum_{n}^{occ} s  \langle u_{n\mathbf{k}}^{(0)}| u_{n\mathbf{k},\mathbf{q}}^{(1)}\rangle \\
&= \frac{\Omega_0}{2}n_{\mathbf{q}}^{(1)}(\mathbf{G}=0).
\end{split}
\end{equation}
In addition, we define the following scalars
\begin{align}
\label{sca1} a &\triangleq \frac{8\pi}{\Omega_0}\frac{1}{q^2}, \\
\label{sca2} b &\triangleq \frac{4\pi}{\Omega_0 q} w_{\mathbf{q}}^{(1)}, \\
\label{sca3} c &\triangleq \frac{\Omega_0}{(2\pi)^3}\int_{BZ} d\mathbf{k} \sum_n^{occ} s_n \langle u_{n\mathbf{k}}^{(0)}|v_{ext,\mathbf{k},\mathbf{k}}^{(2)} | u_{n\mathbf{k}}^{(0)}\rangle. 
\end{align}
Finally, we introduce the matrix $\mathbf{A}$ so that 
\begin{multline}
\label{mat1}
\mathbf{x}^{\dagger}\!\mathbf{A}\mathbf{x} \triangleq \frac{\Omega_0}{(2\pi)^3}\int_{BZ} d\mathbf{k} \sum_{n}^{occ} s_n \langle u_{n\mathbf{k},\mathbf{q}}^{(1)}|H_{\mathbf{k+q,k+q}}^{(0)}\\
-\varepsilon_{n\mathbf{k}}^{(0)} | u_{n\mathbf{k,q}}^{(1)}\rangle 
 +\frac{1}{2}\int_{\Omega_0} \frac{d^2(n\varepsilon_{xc})}{dn^2}\bigg|_{n^{(0)}(\mathbf{r})}|n_{\mathbf{q}}^{(1)}(\mathbf{r})|^2d\mathbf{r}\\
 +2\pi \Omega_0\sum_{\mathbf{G}\neq0}  \frac{|n_{\mathbf{q}}^{(1)}(\mathbf{G})|^2}{|\mathbf{q+G}|^2}.
\end{multline}
Using the above definitions (Eqs.~\eqref{vec1}-\eqref{vec3} and \eqref{sca1}-\eqref{mat1}), Eq.~\eqref{etotEQ} can be re-written in short-hand notation
\begin{multline}\label{equationofE}
E_{\mathbf{-q},\mathbf{q}}^{(2)} = \mathbf{x}^{\dagger}\!\mathbf{A}\mathbf{x} + a(\mathbf{u}^\dagger\!\mathbf{x})(\mathbf{x}^\dagger\!\mathbf{u}) \\
+ (b\mathbf{x}^\dagger\mathbf{u}+ \mathbf{x}^\dagger\mathbf{w} + (c.c.)) + c.
\end{multline}

The physical value of the first-order perturbed periodic part of the wavefunction $\mathbf{x}$ is the one that minimizes $E_{\mathbf{-q},\mathbf{q}}^{(2)}$ (we will refer to it as $\mathbf{x}_1$)
\begin{multline}\label{Eqdelta}
\delta E_{\mathbf{-q},\mathbf{q}}^{(2)} = ( \{\delta\mathbf{x}_1^\dagger (\mathbf{A}\mathbf{x}_1 + (b+a(\mathbf{u}^\dagger\!\mathbf{x}_1))\mathbf{u} + \mathbf{w})\} \\
+ (c.c.))=0.
\end{multline}
The real part of the quantity between curly bracket \{\} is therefore zero
\begin{equation}
\Re\{ \delta\mathbf{x}_1^\dagger\! ( \mathbf{A}\mathbf{x}_1+(b+a(\mathbf{u}^\dagger\!\mathbf{x}_1))\mathbf{u} +\mathbf{w} )\} = 0.
\end{equation}
For the preceding relation to hold for any $\delta\mathbf{x}$, the quantity in parenthesis () must be zero, 
\begin{equation}\label{equationequal0A}
 \mathbf{A}\mathbf{x}_1+(b+a(\mathbf{u}^\dagger\!\mathbf{x}_1))\mathbf{u} +\mathbf{w} = 0,
\end{equation}
which leads to
\begin{equation}
 \mathbf{x}_1=-(b+a(\mathbf{u}^\dagger\!\mathbf{x}_1))\mathbf{A}^{-1}\mathbf{u} -\mathbf{A}^{-1}\mathbf{w}.
\end{equation}
We define 
\begin{equation}\label{defx}
\mathbf{x}_1 \triangleq \mathbf{x}_{1a} + \mathbf{x}_{1b},
\end{equation} 
where
\begin{equation}\label{Ax1-w}
\mathbf{x}_{1a} \triangleq -\mathbf{A}^{-1}\mathbf{w},
\end{equation}
and we are left with
\begin{equation}
\label{x1b_eq}
\mathbf{x}_{1b} = -(b+a(\mathbf{u}^\dagger\!(\mathbf{x}_{1a}+\mathbf{x}_{1b})))\mathbf{A}^{-1}\mathbf{u}. 
\end{equation}
By defining 
\begin{equation}
\label{b_prime}
b'=b+a(\mathbf{u}^\dagger\!\mathbf{x}_{1a}),
\end{equation}
Eq.~\eqref{x1b_eq} becomes
\begin{equation}\label{Eqlineaire}
\mathbf{x}_{1b} = -( b'+a(\mathbf{u}^\dagger\!\mathbf{x}_{1b}))\mathbf{A}^{-1}\mathbf{u}.
\end{equation}
By multiplying the preceding equation by $\mathbf{u}^\dagger$ and isolating $\mathbf{u}^\dagger\!\mathbf{x}_{1b}$, we obtain
\begin{equation}
\mathbf{u}^\dagger\!\mathbf{x}_{1b} = 
\frac{-b' \mathbf{u}^\dagger\!\mathbf{A}^{-1}\mathbf{u}}{1+a(\mathbf{u}^\dagger\!\mathbf{A}^{-1}\mathbf{u})}.
\end{equation}
Substituting this result back in Eq.~\eqref{Eqlineaire}, we obtain $\mathbf{x}_{1b}$
\begin{equation}
\mathbf{x}_{1b} = \frac{-b'}{1+a(\mathbf{u}^\dagger\!\mathbf{A}^{-1}\mathbf{u})}\mathbf{A}^{-1}\mathbf{u},
\end{equation}
and, using Eqs.~\eqref{defx}, \eqref{Ax1-w}, and \eqref{b_prime}, we finally have
\begin{equation}\label{RESOLUTION_2}
\mathbf{x}_1 = 
-\mathbf{A}^{-1}\mathbf{w}
+\frac{-b+a(\mathbf{u}^\dagger\!\mathbf{A}^{-1}\mathbf{w})}
     {1+a(\mathbf{u}^\dagger\!\mathbf{A}^{-1}\mathbf{u})}
     \mathbf{A}^{-1}\mathbf{u}.
\end{equation}
Substituting Eq.~\eqref{equationequal0A} into Eq.~\eqref{equationofE}, we obtain the value of $E_{\mathbf{-q},\mathbf{q}}^{(2)}$ at the minimum $\mathbf{x}_1$
\begin{equation} \label{Eqdildeequal0}
\tilde E_{\mathbf{-q},\mathbf{q}}^{(2)} = b^* \mathbf{u}^\dagger\! \mathbf{x}_1 + \mathbf{w}^\dagger\! \mathbf{x}_1 +c.
\end{equation}
Then, substituting Eq.~\eqref{RESOLUTION_2} into Eq.~\eqref{Eqdildeequal0}, we finally obtain
\begin{multline} \label{E0}
\tilde E_{\mathbf{-q},\mathbf{q}}^{(2)} = -\mathbf{w}^\dagger\mathbf{A}^{-1}\mathbf{w} + c + \frac{(-b\mathbf{w}^\dagger\mathbf{A}^{-1}\mathbf{u} + (c.c.))}
       {1+a\mathbf{u}^\dagger\mathbf{A}^{-1}\mathbf{u}}\\
+\frac{-|b|^2 \mathbf{u}^\dagger\mathbf{A}^{-1}\mathbf{u} + a|\mathbf{w}^\dagger\mathbf{A}^{-1}\mathbf{u}|^2}
       {1+a\mathbf{u}^\dagger\mathbf{A}^{-1}\mathbf{u}}.
\end{multline}

\subsection{Macroscopic dielectric constant}
From Eq.~(B3) of Ref.~\onlinecite{Gonze1997} and Eq.~\eqref{RESOLUTION_1}, we can deduce that the second-derivative of the total energy with respect to a monochromatic electric field of wavevector located inside the first Brillouin zone is
\begin{widetext}
\begin{multline}\label{totalenergy2}
E^{ef(2)}_{\mathbf{-q},\mathbf{q}} = \frac{\Omega_0}{(2\pi)^3}\int_{BZ} d\mathbf{k} \sum_{n}^{occ} s_n 
\Big( \langle u_{n\mathbf{k},\mathbf{q}}^{(1)}|H_{\mathbf{k+q,k+q}}^{(0)}-\varepsilon_{n\mathbf{k}}^{(0)} | u_{n\mathbf{k,q}}^{(1)}\rangle 
   + \langle u_{n\mathbf{k,q}}^{(1)} | u_{n\mathbf{k}}^{(0)}\rangle 
   + \langle u_{n\mathbf{k}}^{(0)} | u_{n\mathbf{k,q}}^{(1)}\rangle \Big) 
+ 2\pi \Omega_0 \frac{|n_{\mathbf{q}}^{(1)}(\mathbf{G}=0)|^2}{\mathbf{q}^2} \\
+ \frac{1}{2}\int_{\Omega_0} \frac{d^2(n\varepsilon_{xc})}{dn^2}\bigg|_{n^{(0)}(\mathbf{r})}|n_{\mathbf{q}}^{(1)}(\mathbf{r})|^2d\mathbf{r}
+ 2\pi \Omega_0 \sum_{\mathbf{G}\neq 0} \frac{|n_{\mathbf{q}}^{(1)}(\mathbf{G})|^2}{|\mathbf{q+G}|^2} = \mathbf{x}^{\dagger}\!\mathbf{A}\mathbf{x} + \mathbf{x}^\dagger\!\mathbf{u}+\mathbf{u}^\dagger\!\mathbf{x} +a(\mathbf{u}^\dagger\!\mathbf{x})(\mathbf{x}^\dagger\!\mathbf{u}), 
\end{multline}
\end{widetext}
where we have used the short-hand notation defined before. 
In the same spirit as Eq.~\eqref{Eqdelta}, we can find the value of $\mathbf{x}$ that minimizes Eq.~\eqref{totalenergy2} (that we will call $\mathbf{x}_2$) and deduce 
\begin{equation}\label{Axeq}
\mathbf{A}\mathbf{x}_2 + a(\mathbf{u}^\dagger\!\mathbf{x}_2)\mathbf{u} + \mathbf{u} = 0,
\end{equation}
which gives
\begin{equation}
\mathbf{x}_2 =(-a(\mathbf{u}^\dagger\!\mathbf{x}_2) - 1 )\mathbf{A}^{-1}\mathbf{u}.
\end{equation}
Multiplying Eq.~\eqref{Axeq} by $\mathbf{u}^\dagger$ and isolating $\mathbf{x}_2$  allows us to obtain 
\begin{equation}\label{x2}
\mathbf{x}_2 = - \frac{\mathbf{A}^{-1}\mathbf{u}}{1+a(\mathbf{u}^\dagger\!\mathbf{A}^{-1}\mathbf{u})}.
\end{equation}
Substituting Eq.~\eqref{Axeq} and then Eq.~\eqref{x2} into Eq.~\eqref{totalenergy2}, we obtain the value of $E^{ef(2)}_{\mathbf{-q},\mathbf{q}}$ at the minimum $\mathbf{x}_2$
\begin{equation}\label{non-stationary-expression}
\tilde E^{ef(2)}_{\mathbf{-q},\mathbf{q}} = \mathbf{u}^\dagger \mathbf{x}_2 
= - \frac{\mathbf{u}^\dagger\!\mathbf{A}^{-1}\mathbf{u}}{1+a(\mathbf{u}^\dagger\!\mathbf{A}^{-1}\mathbf{u})}.
\end{equation}

We can also define a total energy where the divergent $\mathbf{G}=0$ Hartree contribution has been removed. The resulting term is analytic 
\begin{align}\label{etilde1}
E_{\mathbf{-q},\mathbf{q}}^{ef,an(2)} &=  \mathbf{x}^{\dagger}\!\mathbf{A}\mathbf{x} + \mathbf{x}^\dagger\!\mathbf{u}+\mathbf{u}^\dagger\!\mathbf{x}, \\
&= \mathbf{u}^\dagger\!\mathbf{x} + \mathbf{x}^{\dagger}\!(\mathbf{A}\mathbf{x} + \mathbf{u}).
\end{align}
The location $\mathbf{x}_3$ of the minimum of $E_{\mathbf{-q},\mathbf{q}}^{ef,an(2)}$ can be obtained in a similar way to Eq.~\eqref{x2}
\begin{equation}
 \mathbf{A}\mathbf{x}_3 + \mathbf{u} = 0 \Rightarrow \mathbf{x}_3 = \mathbf{A}^{-1} \mathbf{u}. \label{x3}
\end{equation}
Substituting Eq.~\eqref{x3} into Eq.~\eqref{etilde1}, we obtain the value of $E_{\mathbf{-q},\mathbf{q}}^{ef,an(2)}$ at the minimum $\mathbf{x}_3$
\begin{equation}\label{etilde2}
\tilde E_{\mathbf{-q},\mathbf{q}}^{ef,an(2)} = - \mathbf{u}^\dagger \mathbf{A}^{-1} \mathbf{u}.
\end{equation}
Comparing Eqs.~\eqref{non-stationary-expression} and \eqref{etilde2}, we deduce
\begin{equation}\label{EetalphaEtilde}
\tilde E_{\mathbf{-q},\mathbf{q}}^{ef(2)} = \frac{\tilde E_{\mathbf{-q},\mathbf{q}}^{ef,an(2)}}{1-a\tilde E_{\mathbf{-q},\mathbf{q}}^{ef,an(2)}}.
\end{equation}

The polarizability $\chi(\mathbf{r,r'})$ is defined as the microscopic response to a change of external potential that gives the total change of electronic density $\delta n(\mathbf{r})$
\begin{align}\label{density1}
\delta n(\mathbf{r}) &= \int_{\Omega_0} \chi(\mathbf{r,r'})\delta v_{ext}(\mathbf{r'}), \\
\Rightarrow \frac{\delta n(\mathbf{r})}{\delta v_{ext}(\mathbf{r'})} &= \chi(\mathbf{r,r'}).
\end{align}
Transforming to reciprocal space and taking the $(\mathbf{G},\mathbf{G}')=(\mathbf{0},\mathbf{0})$ matrix element yields
\begin{equation} \label{chi-n1}
n^{(1)}_\mathbf{q}(\mathbf{G}=\mathbf{0}) = \chi_\mathbf{q}(\mathbf{G}=\mathbf{0},\mathbf{G}'=\mathbf{0}),
\end{equation}
where the $(1)$ superscript refers to the first-order perturbation in the external potential due to the electric field. 
Taking a long wavelength monochromatic electric field as the perturbation 
\begin{equation}
v^{(1)}_{ext,\mathbf{q}}(\mathbf{r,r'}) = e^{i\mathbf{q \cdot r}} \delta(\mathbf{r,r'}), 
\end{equation}
which gives in reciprocal space
\begin{equation}
v^{(1)}_{ext,\mathbf{k+q,k}}(\mathbf{G,G'}) = \delta_{\mathbf{G,G'}},
\end{equation}
 Eq.~\eqref{non-stationary-expression} tells us that 
\begin{align} 
\tilde E^{ef(2)}_{\mathbf{-q},\mathbf{q}} &= \mathbf{u}^\dagger \mathbf{x}_2, \\
&=\frac{\Omega_0}{2}n_{\mathbf{q}}^{(1)}(\mathbf{0}), \\
&=\frac{\Omega_0}{2}\chi_\mathbf{q}(\mathbf{0},\mathbf{0}),\label{E-chi}
\end{align}
where the second and third equalities stem from Eqs.~\eqref{RESOLUTION_1} and \eqref{chi-n1}, respectively.

The dielectric function is defined as (see Eq.~(23) of Ref.~\onlinecite{Hybertsen1987} for example)
\begin{equation}\label{chi-1}
\varepsilon_{\mathbf{q}}^{-1}(\mathbf{G,G'}) = \delta_{\mathbf{G,G'}} + \frac{4\pi}{|\mathbf{q+G}|^2}\chi_{\mathbf{q}}(\mathbf{G,G'}),
\end{equation}
and the macroscopic dielectric function, which is an average response to an applied field is (see Eq.~(15) of Ref.~\onlinecite{Hybertsen1987} for example)
\begin{equation}
\varepsilon_M (\mathbf{q}) = \frac{1}{\varepsilon_{\mathbf{q}}^{-1}(\mathbf{0,0})}.
\end{equation}
Therefore, using Eqs.~\eqref{sca1}, \eqref{E-chi}, and \eqref{chi-1}, we obtain 
\begin{equation}
\varepsilon_M (\mathbf{q}) = \frac{1}{1+a \tilde E^{ef(2)}_{\mathbf{-q},\mathbf{q}}}.
\end{equation} 
Using Eqs.~\eqref{etilde2} and \eqref{EetalphaEtilde} finally yields
\begin{align}
\varepsilon_M (\mathbf{q}) &= 1 - a \tilde E^{ef,an(2)}_{\mathbf{-q},\mathbf{q}} \\
&= 1 + a(\mathbf{u}^\dagger\!\mathbf{A}^{-1} \mathbf{u}).\label{macroepsilon}
\end{align}

\subsection{Born effective charge}

Following the phenomenological discussion of Born and Huang (see p.265 of Ref.~\onlinecite{Born1954}), we can extend the total energy density $E_{tot}$ (including the vacuum energy) in the long-wavelength limit quadratically in ionic displacement $U_{\kappa \alpha}$ for the atom $\kappa$ in the direction $\alpha$  and macroscopic electric field $\mathcal{E}_\alpha$ in the direction $\alpha$
\begin{multline} \label{Equad}
E_{tot} = \frac{1}{2} \sum_{\kappa\kappa'}\sum_{\gamma\delta}U_{\kappa\gamma}^* C_{\substack{\kappa\gamma\\ \kappa'\delta}}^{\rm an} U_{\kappa'\delta} - \frac{\Omega_0}{8\pi}\sum_{\gamma\delta}\mathcal{E}_\gamma \epsilon_{\gamma\delta}\mathcal{E}_\delta \\
- \sum_{\kappa\gamma,\delta} U_{\kappa\gamma}^* Z^*_{\kappa\gamma,\delta}\mathcal{E}_\delta, 
\end{multline}
where $\Omega_0$ is the volume, $C_{\substack{\kappa\alpha\\ \kappa'\beta}}^{\rm an} = \frac{\partial^2 E}{\partial R_{\kappa\alpha}R_{\kappa'\beta}}$ is the analytic interatomic force constant (IFC), \text{$\epsilon_{\alpha\beta}=\frac{\partial^2 E}{\partial \mathcal{E}_\alpha \mathcal{E}_\beta}$} is the dielectric function and \text{$Z^*_{\kappa\alpha,\beta}=\frac{\partial^2 E}{\partial R_{\kappa\alpha}\mathcal{E}_\beta}$} the Born effective charge. 
$C_{\substack{\kappa\alpha\\ \kappa'\beta}}^{\rm an}$ is associated to the second-order energy (see Eq.~\eqref{etotEQ}) where the non-analytic terms in $\mathbf{q}$ have been removed
\begin{equation} \label{Ean}
E_{\mathbf{-q},\mathbf{q}}^{an(2)} \triangleq \mathbf{x}\!^\dagger\!\mathbf{Ax}+ \mathbf{x}\!^\dagger\!\mathbf{w} + \mathbf{w}^\dagger\!\mathbf{x}+c.
\end{equation}
The solution $\mathbf{x}_4$ that minimizes Eq.~\eqref{Ean} is
\begin{equation}
\mathbf{x}_4 = - \mathbf{A}^{-1}\mathbf{w},
\end{equation}
which leads to the following value for $E_{\mathbf{-q},\mathbf{q}}^{an(2)}$ at the variational minimum 
\begin{equation}\label{Ean0}
\tilde E_{\mathbf{-q},\mathbf{q}}^{an(2)} = - \mathbf{w}^\dagger\!\mathbf{A}^{-1}\mathbf{w} + c = C_{\substack{\kappa\alpha\\ \kappa\alpha}}^{\rm an}.
\end{equation}

The definitions of the electric displacement 
\text{$\boldsymbol{\mathcal{D}} \triangleq \boldsymbol{\mathcal{E}} + 4\pi \boldsymbol{\mathcal{P}}$} 
and the polarizability 
\begin{equation}
\mathcal{P}_\alpha \triangleq -\frac{1}{\Omega_0} \frac{\partial (E+\frac{\Omega_0}{8\pi}\mathcal{E}^2)}{\partial \mathcal{E}_\alpha},
\end{equation}
where we have excluded the energy of the electric field in vacuum. We can combine this with the absence of free charge 
\text{($\boldsymbol{\nabla} \cdot \boldsymbol{\mathcal{D}}(\mathbf{r}) = 0 \Rightarrow \mathbf{q} \cdot \boldsymbol{\mathcal{D}}(\mathbf{q}) = 0 $)}
and magnetic field 
\text{($\boldsymbol{\nabla} \times \boldsymbol{\mathcal{E}}(\mathbf{r}) = 0 \Rightarrow \mathbf{q} \times \boldsymbol{\mathcal{E}}(\mathbf{q}) = \mathbf{0} $)} to 
allow us to deduce the form of the electric field from Eq.~\eqref{Equad}
\begin{equation} \label{Efield}
\mathcal{E}_\alpha = -\frac{4\pi}{\Omega_0} \frac{\sum_{\kappa\gamma,\delta}U^*_{\kappa\gamma} Z^*_{\kappa\gamma,\delta} q_\delta}{\sum_{\gamma\delta} q_\gamma \epsilon_{\gamma\delta} q_\delta}q_\alpha.
\end{equation}
Substituting Eq.~\eqref{Efield} into Eq.~\eqref{Equad} then allows to obtain 
\begin{multline}\label{effectivecharge}
E_{\mathbf{-q},\mathbf{q}}^{(2)} = \tilde E_{\mathbf{-q},\mathbf{q}}^{an(2)} + \frac{2\pi}{\Omega_0}\frac{(\sum_\gamma Z^*_{\kappa\alpha,\gamma} q_\gamma)^2}{\sum_{\gamma\delta} q_\gamma \epsilon_{\gamma\delta} q_\delta}.
\end{multline}

We also introduce a mixed term
\begin{equation} \label{Emix0}
\tilde E_{\mathbf{-q},\mathbf{q}}^{mix(2)} \triangleq - \mathbf{w}^\dagger\!\mathbf{A}^{-1}\mathbf{u}.
\end{equation}
Injecting Eqs.~\eqref{Emix0}, \eqref{Ean0}, and \eqref{etilde2} into Eq.~\eqref{E0}, we can express the the total $E_{\mathbf{-q},\mathbf{q}}^{(2)}$ as 
\begin{multline}\label{eqderivated}
\tilde E_{\mathbf{-q},\mathbf{q}}^{(2)} =
  \tilde E_{\mathbf{-q},\mathbf{q}}^{an(2)} + \frac{(b \tilde E_{\mathbf{-q},\mathbf{q}}^{mix(2)} + (c.c.))}{1-a \tilde E_{\mathbf{-q},\mathbf{q}}^{ef,an(2)} }\\
 + \frac{a | \tilde E_{\mathbf{-q},\mathbf{q}}^{mix(2)}|^2 + |b|^2 \tilde E_{\mathbf{-q},\mathbf{q}}^{ef,an(2)}}
       {1-a \tilde E_{\mathbf{-q},\mathbf{q}}^{ef,an(2)} },
\end{multline}
where the numerator of this equation should be a square to establish the connection with the effective charges $Z^*_{\kappa\alpha,\gamma}$ of Eq.~\eqref{effectivecharge}.
To make the link, we can compare Eqs.~\eqref{sca1} and \eqref{sca2} to find
\begin{equation}
|b|^2 = \frac{2\pi}{\Omega_0}a |w_{\mathbf{q}}^{(1)}|^2.
\end{equation}
Therefore Eq.~\eqref{eqderivated} becomes
\begin{multline}\label{eqderivated2}
 \tilde E_{\mathbf{-q},\mathbf{q}}^{(2)} =  
 \tilde E_{\mathbf{-q},\mathbf{q}}^{an(2)} 
 - \frac{2\pi}{\Omega_0}| w_{\mathbf{q}}^{(1)}|^2 
 + \frac{\Big( b \tilde E_{\mathbf{-q},\mathbf{q}}^{mix(2)} + (c.c.)\Big)}{\varepsilon_M (\mathbf{q})}\\
+ \frac{a |\tilde E_{\mathbf{-q},\mathbf{q}}^{mix(2)}|^2 + \frac{2\pi}{\Omega_0}| w_{\mathbf{q}}^{(1)}|^2 }{\varepsilon_M (\mathbf{q})},
\end{multline}
where we have introduced the macroscopic dielectric function using Eqs.~\eqref{macroepsilon} and \eqref{etilde2}. 
The second term will be canceled by a contribution from the Ewald ion-ion energy to the lowest order in $\mathbf{q}$ and can therefore be included in the analytic part~\cite{Gonze1997}
\begin{equation}
\tilde{E}_{\mathbf{-q},\mathbf{q}}^{Ew,an(2)} = \tilde E_{\mathbf{-q},\mathbf{q}}^{an(2)} - \frac{2\pi}{\Omega_0}| w_{\mathbf{q}}^{(1)}|^2.
\end{equation}
The non-analytic term in $\mathbf{q}=0$, i.e. the remainder of $\tilde E_{\mathbf{-q},\mathbf{q}}^{(2)}$, can be written as
\begin{equation}
\frac{1}{q^2 \varepsilon_M (\mathbf{q})}\frac{2\pi}{\Omega_0} | q w_{\mathbf{q}}^{(1)} + 2 \tilde E_{\mathbf{-q},\mathbf{q}}^{mix(2)*}|^2,
\end{equation}
where we have replaced $a$ and $b$ by their definition Eqs.~\eqref{sca1} and \eqref{sca2}.

For vanishing $\mathbf{q}$ we have, at the lowest order (see Eq.~\eqref{w1})
\begin{equation}
w_{\mathbf{q}}^{(1)} = \frac{q_\alpha}{q}iZ_{\kappa} + \mathcal{O}(\mathbf{q}).
\end{equation} 
Moreover, the macroscopic dielectric constant can be written as
\begin{equation}\label{emacroscopic}
\varepsilon_M(\mathbf{q}) = \frac{1}{q^2}\sum_{\gamma\delta}q_\gamma \varepsilon_{\gamma\delta}q_\delta.
\end{equation}

Eq.~\eqref{eqderivated2} thus becomes
\begin{equation}\label{theqmix}
 \tilde E_{\mathbf{-q},\mathbf{q}}^{(2)} =  \tilde{E}_{\mathbf{-q},\mathbf{q}}^{Ew,an(2)} 
 + \frac{2\pi}{\Omega_0} \frac{|q_\alpha Z_{\kappa} - 2i \tilde E_{\mathbf{-q},\mathbf{q}}^{mix(2)*}|^2}{\sum_{\gamma\delta}q_\gamma \varepsilon_{\gamma\delta}q_\delta}.
\end{equation}

By identification with Eq.~\eqref{effectivecharge}, we deduce
\begin{equation}
\sum_{\gamma} Z_{\kappa\alpha,\gamma}^* q_\gamma =q_\alpha Z_\kappa + 2i  \tilde E_{\mathbf{-q},\mathbf{q}}^{mix(2)},
\end{equation}
where we took the complex conjugate of the quantity between the norm of Eq.~\eqref{theqmix}.

The total Born effective charge is the sum of the ionic charge on the atom $\kappa$ and the electronic charge belonging to this atom 
\begin{equation}
\sum_{\gamma} Z_{\kappa\alpha,\gamma}^* q_\gamma =\sum_{\gamma} ( Z_\kappa \delta_{\alpha\gamma} + \Delta Z_{\kappa\alpha,\gamma}) q_\gamma, \label{ZZZ} 
\end{equation}
which naturally gives
\begin{equation}
2i \tilde E_{\mathbf{-q},\mathbf{q}}^{mix(2)} = \sum_\gamma \Delta Z_{\kappa\alpha,\gamma} q_\gamma.
\end{equation}
The last equation leads, in short hand notation, to the following relation (see Eq.~\eqref{Emix0})
\begin{equation}\label{RESOLUTION_3}
-\mathbf{u}^\dagger\!\mathbf{A}^{-1}\mathbf{w} = \frac{i}{2} \sum_\gamma \Delta Z_{\kappa\alpha,\gamma} q_\gamma.
\end{equation}
Finally, to the lowest order in $\mathbf{q}$, we deduce from the preceding relation, Eq.~\eqref{sca1} and Eq.~\eqref{sca2} 
\begin{align}
b-a\mathbf{u}^\dagger \mathbf{A}^{-1}\mathbf{w} &= \sum_\gamma \frac{4\pi i q_\gamma}{\Omega_0 q^2}(Z_\kappa+\Delta Z_{\kappa\alpha,\gamma})\\
&=\sum_\gamma \frac{4\pi i q_\gamma}{\Omega_0 q^2} Z_{\kappa\alpha,\gamma}^*. \label{RESOLUTION_4}
\end{align}

\subsection{Derivation of Eq.~\eqref{RenormEq}}
The first-order Hartree potential diverges as $1/q$ because of a residual electric charge in the first-order density. 
The first-order density at $\mathbf{G}=0$ can be written using Eqs.~\eqref{RESOLUTION_1} as 
\begin{equation}
n_{\mathbf{q}}^{(1)}(0) = \frac{2}{\Omega_0}\mathbf{u}^\dagger\!\mathbf{x}_1, 
\end{equation}
and \eqref{RESOLUTION_2} as
\begin{multline}
n_{\mathbf{q}}^{(1)}(0) =  \frac{2}{\Omega_0} \bigg(- \mathbf{u}^\dagger\!\mathbf{A}^{-1}\mathbf{w} \\
- \frac{(b-a(\mathbf{u}^\dagger\!\mathbf{A}^{-1}\mathbf{w}))}{1+a(\mathbf{u}^\dagger\!\mathbf{A}^{-1}\mathbf{u})}\mathbf{u}^\dagger\!\mathbf{A}^{-1}\mathbf{u} \bigg). 
\end{multline}

Using Eqs.~\eqref{macroepsilon}, \eqref{emacroscopic}, \eqref{RESOLUTION_3}, and \eqref{RESOLUTION_4}, we have
\begin{multline}
n_{\mathbf{q}}^{(1)}(0) =  \frac{2}{\Omega_0} \bigg(\frac{i}{2} \sum_\gamma \Delta Z_{\kappa\alpha,\gamma} q_\gamma \\
- \frac{ \sum_\gamma \frac{4\pi i}{\Omega_0 q^2}Z_{\kappa\alpha,\gamma}^* q_\gamma}{\frac{1}{q^2}\sum_{\gamma\delta}q_\gamma \epsilon_{\gamma\delta}q_\delta}\mathbf{u}^\dagger\!\mathbf{A}^{-1}\mathbf{u} \bigg). 
\end{multline}
Finally using Eq.~\eqref{macroepsilon} to replace $\mathbf{u}^\dagger\!\mathbf{A}^{-1}\mathbf{u}$ by $\frac{\varepsilon_M(\mathbf{q})-1}{a}$ and then Eqs.~\eqref{sca1}, \eqref{emacroscopic}, and \eqref{ZZZ}, we deduce
\begin{equation}\label{renormdensity}
n_{\mathbf{q}}^{(1)}(0) = -\sum_{\gamma}\frac{iq_\gamma}{\Omega_0}\Big( Z_\kappa \delta_{\alpha\gamma} -\frac{Z_{\kappa\alpha,\gamma}^*}{\frac{1}{q^2} \sum_{\gamma\delta}q_\gamma \epsilon_{\gamma\delta} q_\delta}\Big).
\end{equation}

The first-order Hartree term 
\begin{equation}
v_{H,\mathbf{q}}^{(1)}(\mathbf{0})= 4\pi \frac{n_{\mathbf{q}}^{(1)}}{q^2}, 
\end{equation}
can then be renormalized to account for the slow $\mathbf{k}$-point convergence of the Born effective charges by enforcing effective charge neutrality within the primitive cell.
To do so, we introduce the average Born effective charge per atom
\begin{equation}
\bar{Z}_{\alpha\gamma} = \frac{1}{N_{at}}\sum_\kappa Z_{\kappa\alpha,\gamma}^{*},
\end{equation}
where $N_{at}$ is the number of atoms in the primitive cell, and subtract it from $Z_{\kappa\alpha,\gamma}^{*}$
\begin{multline}
v_{H,\mathbf{q}}^{ren(1)}(\mathbf{0}) \\
= -\sum_{\gamma} \frac{4 \pi iq_\gamma}{\Omega_0 q^2}\Big( Z_\kappa \delta_{\alpha\gamma} -\frac{(Z_{\kappa\alpha,\gamma}^* - \bar{Z}_{\alpha\gamma})}{ \frac{1}{q^2} \sum_{\gamma\delta}q_\gamma \epsilon_{\gamma\delta} q_\delta}\Big), 
\end{multline}
which finally gives
\begin{multline}
v_{H,\mathbf{q}}^{ren(1)}(\mathbf{0}) = \\
v_{H,\mathbf{q}}^{(1)}(\mathbf{0})
\frac{\sum_\gamma q_\gamma \Big (Z_{\kappa}\delta_{\alpha\gamma} -\frac{(Z_{\kappa\alpha,\gamma}^{*} -\bar{Z}_{\alpha\gamma}) }{\frac{1}{q^2} \sum_{\delta,\xi}q_{\delta}\epsilon_{\delta\xi}q_{\xi}}\Big)}
{\sum_\gamma q_\gamma \Big(Z_{\kappa}\delta_{\alpha\gamma}-\frac{Z_{\kappa\alpha,\gamma}^{*}}{ \frac{1}{q^2} \sum_{\delta,\xi}q_{\delta}\epsilon_{\delta\xi}q_{\xi}}\Big)}. \label{finalrenorm}
\end{multline}

\section{Behavior of the ZPR with the shifted parabola model}

In this appendix, we study the behavior of the shifted parabola energy model presented in Eq.~\eqref{eq:parabola_shifted} to mimic the ZPR of polar and non-polar materials at points of the BZ that are not VBM nor CBM.

In spherical coordinates, Eq. \eqref{eq:parabola_shifted} can be re-expressed as
\begin{equation}
\varepsilon(\mathbf{q}) = q^2 - 2 q q_0 \cos \theta
\end{equation}
where we have chosen the shift $\mathbf{q}_0$ along the $z$-Cartesian axis.
This function vanishes when $\mathbf{q} = 0$ or \text{$q=2q_0 \cos \theta$}. The last root is a sphere centered around $\mathbf{q_0}$ with radius $q_0$.

\subsection{Integration on the spherical shell of poles}
\label{annex_spher_poles}

In this section, the ZPR of a polar material is analyzed in the case $\delta = 0$. The set of poles of the ZPR is located on a sphere centered around $\mathbf{q} = \mathbf{q_0}$. We can introduce the new variable $\mathbf{\tilde{q}} = \mathbf{q} - \mathbf{q_0}$ and express the ZPR as an integral on a sphere of radius $q_c$
\begin{multline} \label{eq:int_sphe_shell}
\int_0^{q_c} d\tilde{q} \int_{0}^{\pi}d\theta  \int_{-\pi}^{\pi}  d\phi  \frac{ \sin(\theta)}{(\mathbf{\tilde{q}} + \mathbf{q_0})^2} \frac{\tilde{q}^2}{\tilde{q}^2 - q_0^2} \\
=
2\pi \int_0^{q_c} d\tilde{q} \frac{\tilde{q}^2}{\tilde{q}^2 - q_0^2} \int_0^\pi d\theta \frac{\sin \theta}{\tilde{q}^2 + 2 q q_0 \cos \theta + q_0^2}.
\end{multline}
The radial part of Eq.~\eqref{eq:int_sphe_shell} can be integrated to
\begin{equation}
- \frac{2\pi}{q_0} \int_0^{q_c} d\tilde{q} \frac{\tilde{q}}{\tilde{q}^2 - q_0^2} \ln \Big( \frac{|\tilde{q} + q_0|}{|\tilde{q} - q_0 |} \Big).
\end{equation}

As we would like to understand the behaviour of the poles when integrated, we restrict the integral on a small spherical shell around $\tilde{q} = 0$ with $q_0 - \Delta < \tilde{q} < q_0 + \Delta$. Expressing everything in terms of a new variable \text{$u = \tilde{q} - q_0$}, we deduce
\begin{equation}
- 2 \pi \int_{-\Delta}^{\Delta} du \frac{1}{u}\frac{1 + \frac{u}{q_0}}{u + 2q_0}  \ln\Big( \frac{|2q_0+u|}{|u|} \Big),
\end{equation}
that may be rewritten as
\begin{equation}
\int_{-\Delta}^{\Delta} du F(u) \frac{1}{u} + G(u) \frac{\ln(|u|)}{u},  
\end{equation}
with
\begin{align}
F(u) &= - 2 \pi \ln( |2q_0+u|) \frac{1 + \frac{u}{q_0}}{u + 2q_0} \\
G(u) &= 2 \pi \frac{1 + \frac{u}{q_0}}{u + 2q_0}.
\end{align}

The $F(u)$ and $G(u)$ functions are analytic within the integration range and can be Taylor expanded. Restricting the expansion to first-order, one gets
\begin{multline} \label{eq:taylorFG}
\int_{-\Delta}^{\Delta} du F(0) \frac{1}{u} + G(0) \frac{\ln(|u|)}{u} \\
 + \int_{-\Delta}^{\Delta} du F'(0) + G(0) \ln(|u|).
\end{multline}

As $1/u$ and $\ln(|u|)/u$ are odd functions of $u$, the first two terms of Eq.~\eqref{eq:taylorFG} are zero. The first contributing terms arise from $F'(0)$ and $G'(0)$
\begin{align}
F'(0) \int_{-\Delta}^{\Delta} du &= 2 F'(0) \Delta \\
G'(0) \int_{-\Delta}^{\Delta} du \ln(|u|) &= 2 G'(0) (\Delta \ln \Delta - \Delta),
\end{align}
which shows that the integral on the spherical shell behaves linearly with the width of the shell, as would any regular function do.

In the case of non-polar materials, the $1/q^2$ prefactors is not present, and this makes the derivation easier as no angular dependence is present $G(u) = 0$. The conclusion is nonetheless the same
\begin{equation}
F(u) = \frac{(1 + \frac{u}{q_0})^2}{u + 2q_0} = F(0) + F'(0) u + \mathcal{O}(u^2).
\end{equation}

The first non-zero term in the integral of Eq.~\eqref{eq:taylorFG} is linear in $2 F'(0) \Delta$.

\subsection{Integration of the $q = 0$ pole}
\label{annex_int_0}
For non-polar material in the non-adiabatic approximation, at a point different from the CBM or VBM, the function that should radially be integrated is
\begin{equation}\label{eq:int_polar_shifted}
\Re \iiint_0^{q_c} dq^3 \frac{1}{(q^2 - q q_0 \cos \theta+\omega +i\delta)}.
\end{equation}
The last integral leads to
\begin{equation}\label{eq:int_dtheta}
2\pi \int_0^{q_c} dq q^2 \int_0^\pi d\theta \frac{\sin \theta (q^2 - q q_0 \cos \theta+\omega)}{((q^2 - q q_0 \cos \theta+\omega)^2 +\delta^2)},
\end{equation}
which gives the following radial integral
\begin{equation}
2\pi \int_0^{q_c} dq \frac{q}{4 q_0} \ln \bigg( \frac{(q^2+2qq_0+\omega)^2 +\delta^2}{(q^2-2qq_0+\omega)^2 +\delta^2}\bigg).
\end{equation}

When $\delta = 0$, this function behaves quadratically when $q$ tends to 0 because the lowest order of the Taylor expansion of the logarithm is linear in $q$. Therefore, the integral on a sphere of radius $\Delta$ of this function is
\begin{equation}
\int_0^\Delta dq \frac{q}{4 q_0} \ln \bigg( \frac{(q^2+2qq_0+\omega)^2}{(q^2-2qq_0+\omega)^2}\bigg) = C \Delta^3 + \mathcal{O}(\Delta^4).
\end{equation}

However, for polar materials, the function to integrate is similar to a Lindhard function
\begin{equation}\label{eq:int_polar}
\frac{1}{4 q q_0} \ln \bigg( \frac{(q^2+2qq_0+\omega)^2 +\delta^2}{(q^2-2qq_0+\omega)^2 +\delta^2}\bigg).
\end{equation}
This function actually tends to a finite value because of the linear behavior of the logarithm in $q$ that cancels the denominator. Eq.~\eqref{eq:int_polar} for $\delta = 0$ gives
\begin{equation}
\int_0^\Delta \frac{1}{4 q q_0} \ln \bigg( \frac{(q^2+2qq_0+\omega)^2}{(q^2-2qq_0+\omega)^2}\bigg) = A \Delta + \mathcal{O}(\Delta^2).
\end{equation}

\subsection{Integration on a spherical shell around $q = 0$}
\label{annexshifted}
In this section, we focus only on the $\delta$-behavior of polar materials in the non-adiabatic framework, where Eq.~\eqref{eq:int_polar} has to be integrated.

The integrand has 3 poles: when $q = 0$, and at the two real roots (if any) of $q^2-2qq_0+\omega$, which we call $q_1$ and $q_2$ with $q_1 < q_2$. 

Actually, when $q = 0$, the integrand does not diverge when $\delta \rightarrow 0$, as shown in the subsection \ref{annex_int_0}.
We will here focus on $q = q_2$ as the behavior with respect to $\delta$ in $q = q_1$ is similar.

The integration around $q_2$ is given by
\begin{equation}
2\pi \int_{q_2-\Delta}^{q_2+\Delta} dq \frac{1}{4 q q_0} \ln \bigg( \frac{(q^2+2qq_0+\omega)^2 +\delta^2}{(q^2-2qq_0+\omega)^2 +\delta^2}\bigg).
\end{equation}
We introduce the change of variable $u = q - q_2$, and we consider 
$\delta \ll \Delta \ll q_2$
\begin{equation}
2\pi \int_{-\Delta}^{\Delta} du \frac{1}{4 q_2 q_0} \ln \bigg( \frac{((q_1 + q_2)(2q_2))^2}{((q_2-q_1)(u))^2 +\delta^2}\bigg).
\end{equation}
This integral can be expressed as
\begin{multline}
\frac{2\pi}{4q_2q_0} \bigg( 2 \Delta \ln \Big(4(q_1+q_2)q_2^2\Big) - 2 \Delta \ln(\delta^2) \\
-  \int_{-\Delta}^{\Delta} du \ln \Big( \frac{ u^2 (q_2 - q_1)^2}{\delta^2} + 1 \Big) \bigg),
\end{multline}
and evaluated as
\begin{multline}\label{eq:bigint}
\frac{2\pi}{4q_2q_0} \Bigg( 2 \Delta \ln \Big(4(q_1+q_2)q_2^2\Big) - 2 \Delta \ln(\delta^2) \\
- \frac{\delta}{q_2 - q_1} \frac{2 \Delta (q_2 - q_1)}{\delta} \bigg( \ln \Big( \frac{ \Delta^2 (q_2 - q_1)^2}{\delta^2} + 1 \Big) - 2\bigg) \\
- \frac{\delta}{q_2 - q_1} 4 \tan^{-1}\Big(\frac{\Delta (q_2 - q_1)}{\delta}\Big) \Bigg).
\end{multline}

As $\Delta/\delta$ is large, Eq.~\eqref{eq:bigint} reduces to
\begin{multline}
\frac{2\pi}{4q_2q_0} \bigg( 2 \Delta \ln\big(4(q_1+q_2)q_2^2\big) - 2 \Delta \ln(\delta^2) \\
- \frac{\delta}{q_2 - q_1} \frac{2 \Delta (q_2 - q_1)}{\delta}  2\ln \Big(\frac{ \Delta (q_2 - q_1)}{\delta}\Big) \\
- \frac{\delta}{q_2 - q_1} 4 \frac{\pi}{2} \bigg),
\end{multline}
which behaves as
\begin{equation}
C_1 - C_2 \delta .
\end{equation}

The same reasoning applies when $\omega = 0$. In this case the poles are $q_1 = 0$ and $q_2 = 2 q_0$. The ZPR behavior of a non-extremum point in the Brillouin Zone is thus linear in the non-adiabatic and in the static case with respect to $\delta$.


\bibliography{article_v16}

\begin{thebibliography}{71}%
\makeatletter
\providecommand \@ifxundefined [1]{%
 \@ifx{#1\undefined}
}%
\providecommand \@ifnum [1]{%
 \ifnum #1\expandafter \@firstoftwo
 \else \expandafter \@secondoftwo
 \fi
}%
\providecommand \@ifx [1]{%
 \ifx #1\expandafter \@firstoftwo
 \else \expandafter \@secondoftwo
 \fi
}%
\providecommand \natexlab [1]{#1}%
\providecommand \enquote  [1]{``#1''}%
\providecommand \bibnamefont  [1]{#1}%
\providecommand \bibfnamefont [1]{#1}%
\providecommand \citenamefont [1]{#1}%
\providecommand \href@noop [0]{\@secondoftwo}%
\providecommand \href [0]{\begingroup \@sanitize@url \@href}%
\providecommand \@href[1]{\@@startlink{#1}\@@href}%
\providecommand \@@href[1]{\endgroup#1\@@endlink}%
\providecommand \@sanitize@url [0]{\catcode `\\12\catcode `\$12\catcode
  `\&12\catcode `\#12\catcode `\^12\catcode `\_12\catcode `\%12\relax}%
\providecommand \@@startlink[1]{}%
\providecommand \@@endlink[0]{}%
\providecommand \url  [0]{\begingroup\@sanitize@url \@url }%
\providecommand \@url [1]{\endgroup\@href {#1}{\urlprefix }}%
\providecommand \urlprefix  [0]{URL }%
\providecommand \Eprint [0]{\href }%
\providecommand \doibase [0]{http://dx.doi.org/}%
\providecommand \selectlanguage [0]{\@gobble}%
\providecommand \bibinfo  [0]{\@secondoftwo}%
\providecommand \bibfield  [0]{\@secondoftwo}%
\providecommand \translation [1]{[#1]}%
\providecommand \BibitemOpen [0]{}%
\providecommand \bibitemStop [0]{}%
\providecommand \bibitemNoStop [0]{.\EOS\space}%
\providecommand \EOS [0]{\spacefactor3000\relax}%
\providecommand \BibitemShut  [1]{\csname bibitem#1\endcsname}%
\let\auto@bib@innerbib\@empty
\bibitem [{\citenamefont {Ponc\'e}\ \emph
  {et~al.}(2014{\natexlab{a}})\citenamefont {Ponc\'e}, \citenamefont
  {Antonius}, \citenamefont {Gillet}, \citenamefont {Boulanger}, \citenamefont
  {Laflamme~Janssen}, \citenamefont {Marini}, \citenamefont {C\^ot\'e},\ and\
  \citenamefont {Gonze}}]{Ponce2014a}%
  \BibitemOpen
  \bibfield  {author} {\bibinfo {author} {\bibfnamefont {S.}~\bibnamefont
  {Ponc\'e}}, \bibinfo {author} {\bibfnamefont {G.}~\bibnamefont {Antonius}},
  \bibinfo {author} {\bibfnamefont {Y.}~\bibnamefont {Gillet}}, \bibinfo
  {author} {\bibfnamefont {P.}~\bibnamefont {Boulanger}}, \bibinfo {author}
  {\bibfnamefont {J.}~\bibnamefont {Laflamme~Janssen}}, \bibinfo {author}
  {\bibfnamefont {A.}~\bibnamefont {Marini}}, \bibinfo {author} {\bibfnamefont
  {M.}~\bibnamefont {C\^ot\'e}}, \ and\ \bibinfo {author} {\bibfnamefont
  {X.}~\bibnamefont {Gonze}},\ }\href@noop {} {\bibfield  {journal} {\bibinfo
  {journal} {Phys. Rev. B}\ }\textbf {\bibinfo {volume} {90}},\ \bibinfo
  {pages} {214304} (\bibinfo {year} {2014}{\natexlab{a}})}\BibitemShut
  {NoStop}%
\bibitem [{\citenamefont {Allen}\ and\ \citenamefont
  {Heine}(1976)}]{Allen1976}%
  \BibitemOpen
  \bibfield  {author} {\bibinfo {author} {\bibfnamefont {P.~B.}\ \bibnamefont
  {Allen}}\ and\ \bibinfo {author} {\bibfnamefont {V.}~\bibnamefont {Heine}},\
  }\href@noop {} {\bibfield  {journal} {\bibinfo  {journal} {Journal of Physics
  C: Solid State Physics}\ }\textbf {\bibinfo {volume} {9}},\ \bibinfo {pages}
  {2305} (\bibinfo {year} {1976})}\BibitemShut {NoStop}%
\bibitem [{\citenamefont {Allen}\ and\ \citenamefont
  {Cardona}(1981)}]{Allen1981}%
  \BibitemOpen
  \bibfield  {author} {\bibinfo {author} {\bibfnamefont {P.~B.}\ \bibnamefont
  {Allen}}\ and\ \bibinfo {author} {\bibfnamefont {M.}~\bibnamefont
  {Cardona}},\ }\href@noop {} {\bibfield  {journal} {\bibinfo  {journal} {Phys.
  Rev. B}\ }\textbf {\bibinfo {volume} {23}},\ \bibinfo {pages} {1495}
  (\bibinfo {year} {1981})}\BibitemShut {NoStop}%
\bibitem [{\citenamefont {Allen}\ and\ \citenamefont
  {Cardona}(1983)}]{Allen1983}%
  \BibitemOpen
  \bibfield  {author} {\bibinfo {author} {\bibfnamefont {P.~B.}\ \bibnamefont
  {Allen}}\ and\ \bibinfo {author} {\bibfnamefont {M.}~\bibnamefont
  {Cardona}},\ }\href@noop {} {\bibfield  {journal} {\bibinfo  {journal} {Phys.
  Rev. B}\ }\textbf {\bibinfo {volume} {27}},\ \bibinfo {pages} {4760}
  (\bibinfo {year} {1983})}\BibitemShut {NoStop}%
\bibitem [{\citenamefont {Marini}(2008)}]{Marini2008}%
  \BibitemOpen
  \bibfield  {author} {\bibinfo {author} {\bibfnamefont {A.}~\bibnamefont
  {Marini}},\ }\href@noop {} {\bibfield  {journal} {\bibinfo  {journal} {Phys.
  Rev. Lett.}\ }\textbf {\bibinfo {volume} {101}},\ \bibinfo {pages} {106405}
  (\bibinfo {year} {2008})}\BibitemShut {NoStop}%
\bibitem [{\citenamefont {Giustino}, \citenamefont {Louie},\ and\ \citenamefont
  {Cohen}(2010)}]{Giustino2010}%
  \BibitemOpen
  \bibfield  {author} {\bibinfo {author} {\bibfnamefont {F.}~\bibnamefont
  {Giustino}}, \bibinfo {author} {\bibfnamefont {S.~G.}\ \bibnamefont {Louie}},
  \ and\ \bibinfo {author} {\bibfnamefont {M.~L.}\ \bibnamefont {Cohen}},\
  }\href@noop {} {\bibfield  {journal} {\bibinfo  {journal} {Phys. Rev. Lett.}\
  }\textbf {\bibinfo {volume} {105}},\ \bibinfo {pages} {265501} (\bibinfo
  {year} {2010})}\BibitemShut {NoStop}%
\bibitem [{\citenamefont {Ponc\'e}\ \emph
  {et~al.}(2014{\natexlab{b}})\citenamefont {Ponc\'e}, \citenamefont
  {Antonius}, \citenamefont {Boulanger}, \citenamefont {Cannuccia},
  \citenamefont {Marini}, \citenamefont {C\^ot\'e},\ and\ \citenamefont
  {Gonze}}]{Ponce2014}%
  \BibitemOpen
  \bibfield  {author} {\bibinfo {author} {\bibfnamefont {S.}~\bibnamefont
  {Ponc\'e}}, \bibinfo {author} {\bibfnamefont {G.}~\bibnamefont {Antonius}},
  \bibinfo {author} {\bibfnamefont {P.}~\bibnamefont {Boulanger}}, \bibinfo
  {author} {\bibfnamefont {E.}~\bibnamefont {Cannuccia}}, \bibinfo {author}
  {\bibfnamefont {A.}~\bibnamefont {Marini}}, \bibinfo {author} {\bibfnamefont
  {M.}~\bibnamefont {C\^ot\'e}}, \ and\ \bibinfo {author} {\bibfnamefont
  {X.}~\bibnamefont {Gonze}},\ }\href@noop {} {\bibfield  {journal} {\bibinfo
  {journal} {Computational Materials Science}\ }\textbf {\bibinfo {volume}
  {83}},\ \bibinfo {pages} {341 } (\bibinfo {year}
  {2014}{\natexlab{b}})}\BibitemShut {NoStop}%
\bibitem [{\citenamefont {Gonze}, \citenamefont {Boulanger},\ and\
  \citenamefont {C\^ot\'e}(2011)}]{Gonze2011}%
  \BibitemOpen
  \bibfield  {author} {\bibinfo {author} {\bibfnamefont {X.}~\bibnamefont
  {Gonze}}, \bibinfo {author} {\bibfnamefont {P.}~\bibnamefont {Boulanger}}, \
  and\ \bibinfo {author} {\bibfnamefont {M.}~\bibnamefont {C\^ot\'e}},\
  }\href@noop {} {\bibfield  {journal} {\bibinfo  {journal} {Annalen der
  Physik}\ }\textbf {\bibinfo {volume} {523}},\ \bibinfo {pages} {168}
  (\bibinfo {year} {2011})}\BibitemShut {NoStop}%
\bibitem [{\citenamefont {Cannuccia}\ and\ \citenamefont
  {Marini}(2011)}]{Cannuccia2011}%
  \BibitemOpen
  \bibfield  {author} {\bibinfo {author} {\bibfnamefont {E.}~\bibnamefont
  {Cannuccia}}\ and\ \bibinfo {author} {\bibfnamefont {A.}~\bibnamefont
  {Marini}},\ }\href@noop {} {\bibfield  {journal} {\bibinfo  {journal} {Phys.
  Rev. Lett.}\ }\textbf {\bibinfo {volume} {107}},\ \bibinfo {pages} {255501}
  (\bibinfo {year} {2011})}\BibitemShut {NoStop}%
\bibitem [{\citenamefont {Cannuccia}\ and\ \citenamefont
  {Marini}(2012)}]{Cannuccia2012}%
  \BibitemOpen
  \bibfield  {author} {\bibinfo {author} {\bibfnamefont {E.}~\bibnamefont
  {Cannuccia}}\ and\ \bibinfo {author} {\bibfnamefont {A.}~\bibnamefont
  {Marini}},\ }\href {http://dx.doi.org/10.1140/epjb/e2012-30105-4} {\bibfield
  {journal} {\bibinfo  {journal} {The European Physical Journal B}\ }\textbf
  {\bibinfo {volume} {85}},\ \bibinfo {pages} {1} (\bibinfo {year}
  {2012})}\BibitemShut {NoStop}%
\bibitem [{\citenamefont {Marini}, \citenamefont {Ponc\'e},\ and\ \citenamefont
  {Gonze}(2015)}]{Marini2015}%
  \BibitemOpen
  \bibfield  {author} {\bibinfo {author} {\bibfnamefont {A.}~\bibnamefont
  {Marini}}, \bibinfo {author} {\bibfnamefont {S.}~\bibnamefont {Ponc\'e}}, \
  and\ \bibinfo {author} {\bibfnamefont {X.}~\bibnamefont {Gonze}},\ }\href
  {http://arxiv.org/abs/1503.00567} {\bibfield  {journal} {\bibinfo  {journal}
  {Physical Review B (submitted)}\ } (\bibinfo {year} {2015})}\BibitemShut
  {NoStop}%
\bibitem [{\citenamefont {Monserrat}, \citenamefont {Drummond},\ and\
  \citenamefont {Needs}(2013)}]{Monserrat2013}%
  \BibitemOpen
  \bibfield  {author} {\bibinfo {author} {\bibfnamefont {B.}~\bibnamefont
  {Monserrat}}, \bibinfo {author} {\bibfnamefont {N.~D.}\ \bibnamefont
  {Drummond}}, \ and\ \bibinfo {author} {\bibfnamefont {R.~J.}\ \bibnamefont
  {Needs}},\ }\href@noop {} {\bibfield  {journal} {\bibinfo  {journal} {Phys.
  Rev. B}\ }\textbf {\bibinfo {volume} {87}},\ \bibinfo {pages} {144302}
  (\bibinfo {year} {2013})}\BibitemShut {NoStop}%
\bibitem [{\citenamefont {Antonius}\ \emph {et~al.}(2014)\citenamefont
  {Antonius}, \citenamefont {Ponc\'e}, \citenamefont {Boulanger}, \citenamefont
  {C\^ot\'e},\ and\ \citenamefont {Gonze}}]{Antonius2014}%
  \BibitemOpen
  \bibfield  {author} {\bibinfo {author} {\bibfnamefont {G.}~\bibnamefont
  {Antonius}}, \bibinfo {author} {\bibfnamefont {S.}~\bibnamefont {Ponc\'e}},
  \bibinfo {author} {\bibfnamefont {P.}~\bibnamefont {Boulanger}}, \bibinfo
  {author} {\bibfnamefont {M.}~\bibnamefont {C\^ot\'e}}, \ and\ \bibinfo
  {author} {\bibfnamefont {X.}~\bibnamefont {Gonze}},\ }\href@noop {}
  {\bibfield  {journal} {\bibinfo  {journal} {Phys. Rev. Lett.}\ }\textbf
  {\bibinfo {volume} {112}},\ \bibinfo {pages} {215501} (\bibinfo {year}
  {2014})}\BibitemShut {NoStop}%
\bibitem [{\citenamefont {Ram\'irez}\ \emph {et~al.}(2008)\citenamefont
  {Ram\'irez}, \citenamefont {Herrero}, \citenamefont {Hern\'andez},\ and\
  \citenamefont {Cardona}}]{Ramirez2008}%
  \BibitemOpen
  \bibfield  {author} {\bibinfo {author} {\bibfnamefont {R.}~\bibnamefont
  {Ram\'irez}}, \bibinfo {author} {\bibfnamefont {C.~P.}\ \bibnamefont
  {Herrero}}, \bibinfo {author} {\bibfnamefont {E.~R.}\ \bibnamefont
  {Hern\'andez}}, \ and\ \bibinfo {author} {\bibfnamefont {M.}~\bibnamefont
  {Cardona}},\ }\href@noop {} {\bibfield  {journal} {\bibinfo  {journal} {Phys.
  Rev. B}\ }\textbf {\bibinfo {volume} {77}},\ \bibinfo {pages} {045210}
  (\bibinfo {year} {2008})}\BibitemShut {NoStop}%
\bibitem [{\citenamefont {Born}\ and\ \citenamefont {Huang}(1954)}]{Born1954}%
  \BibitemOpen
  \bibfield  {author} {\bibinfo {author} {\bibfnamefont {M.}~\bibnamefont
  {Born}}\ and\ \bibinfo {author} {\bibfnamefont {K.}~\bibnamefont {Huang}},\
  }\href@noop {} {\emph {\bibinfo {title} {Dynamical Theory of Crystal
  Lattices}}}\ (\bibinfo  {publisher} {Oxford University Press},\ \bibinfo
  {year} {1954})\BibitemShut {NoStop}%
\bibitem [{\citenamefont {Sternheimer}(1954)}]{Sternheimer1954}%
  \BibitemOpen
  \bibfield  {author} {\bibinfo {author} {\bibfnamefont {R.~M.}\ \bibnamefont
  {Sternheimer}},\ }\href@noop {} {\bibfield  {journal} {\bibinfo  {journal}
  {Phys. Rev.}\ }\textbf {\bibinfo {volume} {96}},\ \bibinfo {pages} {951}
  (\bibinfo {year} {1954})}\BibitemShut {NoStop}%
\bibitem [{\citenamefont {Zollner}, \citenamefont {Cardona},\ and\
  \citenamefont {Gopalan}(1992)}]{Zollner1992}%
  \BibitemOpen
  \bibfield  {author} {\bibinfo {author} {\bibfnamefont {S.}~\bibnamefont
  {Zollner}}, \bibinfo {author} {\bibfnamefont {M.}~\bibnamefont {Cardona}}, \
  and\ \bibinfo {author} {\bibfnamefont {S.}~\bibnamefont {Gopalan}},\
  }\href@noop {} {\bibfield  {journal} {\bibinfo  {journal} {Phys. Rev. B}\
  }\textbf {\bibinfo {volume} {45}},\ \bibinfo {pages} {3376} (\bibinfo {year}
  {1992})}\BibitemShut {NoStop}%
\bibitem [{\citenamefont {Allen}(1978)}]{Allen1978}%
  \BibitemOpen
  \bibfield  {author} {\bibinfo {author} {\bibfnamefont {P.~B.}\ \bibnamefont
  {Allen}},\ }\href {http://link.aps.org/doi/10.1103/PhysRevB.18.5217}
  {\bibfield  {journal} {\bibinfo  {journal} {Phys. Rev. B}\ }\textbf {\bibinfo
  {volume} {18}},\ \bibinfo {pages} {5217} (\bibinfo {year}
  {1978})}\BibitemShut {NoStop}%
\bibitem [{\citenamefont {Grimvall}(1981)}]{Grimvall1981}%
  \BibitemOpen
  \bibfield  {author} {\bibinfo {author} {\bibfnamefont {G.}~\bibnamefont
  {Grimvall}},\ }\href {http://iopscience.iop.org/1402-4896/14/1-2/013} {\emph
  {\bibinfo {title} {The electron-phonon interaction in metals}}}\ (\bibinfo
  {publisher} {North-Holland Publishing Company},\ \bibinfo {year}
  {1981})\BibitemShut {NoStop}%
\bibitem [{\citenamefont {Gonze}(1997)}]{Gonze1997}%
  \BibitemOpen
  \bibfield  {author} {\bibinfo {author} {\bibfnamefont {X.}~\bibnamefont
  {Gonze}},\ }\href@noop {} {\bibfield  {journal} {\bibinfo  {journal} {Phys.
  Rev. B}\ }\textbf {\bibinfo {volume} {55}},\ \bibinfo {pages} {10337}
  (\bibinfo {year} {1997})}\BibitemShut {NoStop}%
\bibitem [{\citenamefont {Gonze}\ and\ \citenamefont {Lee}(1997)}]{Gonze1997a}%
  \BibitemOpen
  \bibfield  {author} {\bibinfo {author} {\bibfnamefont {X.}~\bibnamefont
  {Gonze}}\ and\ \bibinfo {author} {\bibfnamefont {C.}~\bibnamefont {Lee}},\
  }\href@noop {} {\bibfield  {journal} {\bibinfo  {journal} {Phys. Rev. B}\
  }\textbf {\bibinfo {volume} {55}},\ \bibinfo {pages} {10355} (\bibinfo {year}
  {1997})}\BibitemShut {NoStop}%
\bibitem [{sup()}]{supplemental}%
  \BibitemOpen
  \href@noop {} {\enquote {\bibinfo {title} {See supplemental material at [url
  will be inserted by aip] for the convergence studies.}}\ }\BibitemShut
  {NoStop}%
\bibitem [{\citenamefont {Ramdas}(2000)}]{Ramdas2000}%
  \BibitemOpen
  \bibfield  {author} {\bibinfo {author} {\bibfnamefont {A.}~\bibnamefont
  {Ramdas}},\ }\href@noop {} {\emph {\bibinfo {title} {Properties, Growth and
  Applications of Diamond}}},\ edited by\ \bibinfo {editor} {\bibfnamefont
  {A.}~\bibnamefont {Neves}}\ and\ \bibinfo {editor} {\bibfnamefont {M.~H.}\
  \bibnamefont {Nazaré}}\ (\bibinfo  {publisher} {INSPEC},\ \bibinfo {year}
  {2000})\BibitemShut {NoStop}%
\bibitem [{\citenamefont {Wright}\ and\ \citenamefont
  {Nelson}(1995)}]{Wright1995}%
  \BibitemOpen
  \bibfield  {author} {\bibinfo {author} {\bibfnamefont {A.}~\bibnamefont
  {Wright}}\ and\ \bibinfo {author} {\bibfnamefont {J.}~\bibnamefont
  {Nelson}},\ }\href {http://link.aps.org/doi/10.1103/PhysRevB.51.7866}
  {\bibfield  {journal} {\bibinfo  {journal} {Phys. Rev. B}\ }\textbf {\bibinfo
  {volume} {51}},\ \bibinfo {pages} {7866} (\bibinfo {year}
  {1995})}\BibitemShut {NoStop}%
\bibitem [{\citenamefont {Jain}\ \emph {et~al.}(2013)\citenamefont {Jain},
  \citenamefont {Ong}, \citenamefont {Hautier}, \citenamefont {Chen},
  \citenamefont {Richards}, \citenamefont {Dacek}, \citenamefont {Cholia},
  \citenamefont {Gunter}, \citenamefont {Skinner}, \citenamefont {Ceder},\ and\
  \citenamefont {Persson}}]{Jain2013}%
  \BibitemOpen
  \bibfield  {author} {\bibinfo {author} {\bibfnamefont {A.}~\bibnamefont
  {Jain}}, \bibinfo {author} {\bibfnamefont {S.~P.}\ \bibnamefont {Ong}},
  \bibinfo {author} {\bibfnamefont {G.}~\bibnamefont {Hautier}}, \bibinfo
  {author} {\bibfnamefont {W.}~\bibnamefont {Chen}}, \bibinfo {author}
  {\bibfnamefont {W.~D.}\ \bibnamefont {Richards}}, \bibinfo {author}
  {\bibfnamefont {S.}~\bibnamefont {Dacek}}, \bibinfo {author} {\bibfnamefont
  {S.}~\bibnamefont {Cholia}}, \bibinfo {author} {\bibfnamefont
  {D.}~\bibnamefont {Gunter}}, \bibinfo {author} {\bibfnamefont
  {D.}~\bibnamefont {Skinner}}, \bibinfo {author} {\bibfnamefont
  {G.}~\bibnamefont {Ceder}}, \ and\ \bibinfo {author} {\bibfnamefont {K.~a.}\
  \bibnamefont {Persson}},\ }\href
  {http://link.aip.org/link/AMPADS/v1/i1/p011002/s1\&Agg=doi} {\bibfield
  {journal} {\bibinfo  {journal} {APL Materials}\ }\textbf {\bibinfo {volume}
  {1}},\ \bibinfo {pages} {011002} (\bibinfo {year} {2013})}\BibitemShut
  {NoStop}%
\bibitem [{\citenamefont {Yu}(2001)}]{Yu2001}%
  \BibitemOpen
  \bibfield  {author} {\bibinfo {author} {\bibfnamefont {G.}~\bibnamefont
  {Yu}},\ }\href
  {http://eu.wiley.com/WileyCDA/WileyTitle/productCd-0471358274.html} {\emph
  {\bibinfo {title} {Properties of Advanced SemiconductorMaterials GaN, AlN,
  InN, BN, SiC, SiGe}}},\ edited by\ \bibinfo {editor} {\bibfnamefont {S.~M.}\
  \bibnamefont {Levinshtein~M.E.}, \bibfnamefont {Rumyantsev~S.L.}}\ (\bibinfo
  {publisher} {John Wiley \& Sons},\ \bibinfo {year} {2001})\BibitemShut
  {NoStop}%
\bibitem [{\citenamefont {Petrov}\ \emph {et~al.}(1992)\citenamefont {Petrov},
  \citenamefont {Mojab}, \citenamefont {Powell}, \citenamefont {Greene},
  \citenamefont {Hultman},\ and\ \citenamefont {Sundgren}}]{Petrov1992}%
  \BibitemOpen
  \bibfield  {author} {\bibinfo {author} {\bibfnamefont {I.}~\bibnamefont
  {Petrov}}, \bibinfo {author} {\bibfnamefont {E.}~\bibnamefont {Mojab}},
  \bibinfo {author} {\bibfnamefont {R.~C.}\ \bibnamefont {Powell}}, \bibinfo
  {author} {\bibfnamefont {J.~E.}\ \bibnamefont {Greene}}, \bibinfo {author}
  {\bibfnamefont {L.}~\bibnamefont {Hultman}}, \ and\ \bibinfo {author}
  {\bibfnamefont {J.}~\bibnamefont {Sundgren}},\ }\href
  {http://ieeexplore.ieee.org/xpl/login.jsp?tp=&arnumber=4878318&url=http%3A%2F%2Fieeexplore.ieee.org%2Fiel5%2F4816218%2F4878302%2F04878318.pdf%3Farnumber%3D4878318}
  {\bibfield  {journal} {\bibinfo  {journal} {Applied Physics Letters}\
  }\textbf {\bibinfo {volume} {60}},\ \bibinfo {pages} {2491} (\bibinfo {year}
  {1992})}\BibitemShut {NoStop}%
\bibitem [{\citenamefont {Schulz}\ and\ \citenamefont
  {Thiemann}(1977)}]{Schulz1977}%
  \BibitemOpen
  \bibfield  {author} {\bibinfo {author} {\bibfnamefont {H.}~\bibnamefont
  {Schulz}}\ and\ \bibinfo {author} {\bibfnamefont {K.}~\bibnamefont
  {Thiemann}},\ }\href
  {http://www.sciencedirect.com/science/article/pii/0038109877909590}
  {\bibfield  {journal} {\bibinfo  {journal} {Solid State Communications}\
  }\textbf {\bibinfo {volume} {23}},\ \bibinfo {pages} {815 } (\bibinfo {year}
  {1977})}\BibitemShut {NoStop}%
\bibitem [{\citenamefont {Lambrecht}\ and\ \citenamefont
  {Segall}(1994)}]{Lambrecht1994}%
  \BibitemOpen
  \bibfield  {author} {\bibinfo {author} {\bibfnamefont {W.~R.~L.}\
  \bibnamefont {Lambrecht}}\ and\ \bibinfo {author} {\bibfnamefont
  {B.}~\bibnamefont {Segall}},\ }\href@noop {} {\emph {\bibinfo {title}
  {Properties of Group III Nitrides}}},\ edited by\ \bibinfo {editor}
  {\bibfnamefont {J.~H.}\ \bibnamefont {Edgar}}\ (\bibinfo  {publisher}
  {London: EMIS Datareviews Series},\ \bibinfo {year} {1994})\BibitemShut
  {NoStop}%
\bibitem [{\citenamefont {Kim}, \citenamefont {Lambrecht},\ and\ \citenamefont
  {Segall}(1996)}]{Kim1996}%
  \BibitemOpen
  \bibfield  {author} {\bibinfo {author} {\bibfnamefont {K.}~\bibnamefont
  {Kim}}, \bibinfo {author} {\bibfnamefont {W.}~\bibnamefont {Lambrecht}}, \
  and\ \bibinfo {author} {\bibfnamefont {B.}~\bibnamefont {Segall}},\ }\href
  {http://link.aps.org/doi/10.1103/PhysRevB.53.16310} {\bibfield  {journal}
  {\bibinfo  {journal} {Phys. Rev. B}\ }\textbf {\bibinfo {volume} {53}},\
  \bibinfo {pages} {16310} (\bibinfo {year} {1996})}\BibitemShut {NoStop}%
\bibitem [{\citenamefont {Stampfl}\ and\ \citenamefont {Van~de
  Walle}(1999)}]{Stampfl1999}%
  \BibitemOpen
  \bibfield  {author} {\bibinfo {author} {\bibfnamefont {C.}~\bibnamefont
  {Stampfl}}\ and\ \bibinfo {author} {\bibfnamefont {C.}~\bibnamefont {Van~de
  Walle}},\ }\href {http://link.aps.org/doi/10.1103/PhysRevB.59.5521}
  {\bibfield  {journal} {\bibinfo  {journal} {Phys. Rev. B}\ }\textbf {\bibinfo
  {volume} {59}},\ \bibinfo {pages} {5521} (\bibinfo {year}
  {1999})}\BibitemShut {NoStop}%
\bibitem [{\citenamefont {Cappellini}\ \emph {et~al.}(2001)\citenamefont
  {Cappellini}, \citenamefont {Satta}, \citenamefont {Palummo},\ and\
  \citenamefont {Onida}}]{Cappellini2001}%
  \BibitemOpen
  \bibfield  {author} {\bibinfo {author} {\bibfnamefont {G.}~\bibnamefont
  {Cappellini}}, \bibinfo {author} {\bibfnamefont {G.}~\bibnamefont {Satta}},
  \bibinfo {author} {\bibfnamefont {M.}~\bibnamefont {Palummo}}, \ and\
  \bibinfo {author} {\bibfnamefont {G.}~\bibnamefont {Onida}},\ }\href
  {http://link.aps.org/doi/10.1103/PhysRevB.64.035104} {\bibfield  {journal}
  {\bibinfo  {journal} {Phys. Rev. B}\ }\textbf {\bibinfo {volume} {64}},\
  \bibinfo {pages} {035104} (\bibinfo {year} {2001})}\BibitemShut {NoStop}%
\bibitem [{\citenamefont {S$\overline{o}$ma}, \citenamefont {Sawaoka},\ and\
  \citenamefont {Saito}(1974)}]{Soma1974}%
  \BibitemOpen
  \bibfield  {author} {\bibinfo {author} {\bibfnamefont {T.}~\bibnamefont
  {S$\overline{o}$ma}}, \bibinfo {author} {\bibfnamefont {A.}~\bibnamefont
  {Sawaoka}}, \ and\ \bibinfo {author} {\bibfnamefont {S.}~\bibnamefont
  {Saito}},\ }\href
  {http://www.sciencedirect.com/science/article/pii/002554087490110X}
  {\bibfield  {journal} {\bibinfo  {journal} {Materials Research Bulletin}\
  }\textbf {\bibinfo {volume} {9}},\ \bibinfo {pages} {755 } (\bibinfo {year}
  {1974})}\BibitemShut {NoStop}%
\bibitem [{\citenamefont {Furthm\"uller}, \citenamefont {Hafner},\ and\
  \citenamefont {Kresse}(1994)}]{Furthmuller1994}%
  \BibitemOpen
  \bibfield  {author} {\bibinfo {author} {\bibfnamefont {J.}~\bibnamefont
  {Furthm\"uller}}, \bibinfo {author} {\bibfnamefont {J.}~\bibnamefont
  {Hafner}}, \ and\ \bibinfo {author} {\bibfnamefont {G.}~\bibnamefont
  {Kresse}},\ }\href {http://link.aps.org/doi/10.1103/PhysRevB.50.15606}
  {\bibfield  {journal} {\bibinfo  {journal} {Phys. Rev. B}\ }\textbf {\bibinfo
  {volume} {50}},\ \bibinfo {pages} {15606} (\bibinfo {year}
  {1994})}\BibitemShut {NoStop}%
\bibitem [{\citenamefont {Madelung}(1972)}]{Madelung1972}%
  \BibitemOpen
  \bibinfo {editor} {\bibfnamefont {O.}~\bibnamefont {Madelung}},\ ed.,\
  \href@noop {} {\emph {\bibinfo {title} {Numerical Data and Functional
  Relationships in Science and Technology- Crystal and Solid State Physics,
  Vol. III of Landolt-B\"ornstein}}}\ (\bibinfo  {publisher} {Springer},\
  \bibinfo {year} {1972})\BibitemShut {NoStop}%
\bibitem [{\citenamefont {Wentzcovitch}, \citenamefont {Chang},\ and\
  \citenamefont {Cohen}(1986)}]{Wentzcovitch1986}%
  \BibitemOpen
  \bibfield  {author} {\bibinfo {author} {\bibfnamefont {R.}~\bibnamefont
  {Wentzcovitch}}, \bibinfo {author} {\bibfnamefont {K.}~\bibnamefont {Chang}},
  \ and\ \bibinfo {author} {\bibfnamefont {M.}~\bibnamefont {Cohen}},\ }\href
  {http://link.aps.org/doi/10.1103/PhysRevB.34.1071} {\bibfield  {journal}
  {\bibinfo  {journal} {Phys. Rev. B}\ }\textbf {\bibinfo {volume} {34}},\
  \bibinfo {pages} {1071} (\bibinfo {year} {1986})}\BibitemShut {NoStop}%
\bibitem [{\citenamefont {Xu}\ and\ \citenamefont {Ching}(1991)}]{Xu1991}%
  \BibitemOpen
  \bibfield  {author} {\bibinfo {author} {\bibfnamefont {Y.-N.}\ \bibnamefont
  {Xu}}\ and\ \bibinfo {author} {\bibfnamefont {W.}~\bibnamefont {Ching}},\
  }\href {http://link.aps.org/doi/10.1103/PhysRevB.44.7787} {\bibfield
  {journal} {\bibinfo  {journal} {Phys. Rev. B}\ }\textbf {\bibinfo {volume}
  {44}},\ \bibinfo {pages} {7787} (\bibinfo {year} {1991})}\BibitemShut
  {NoStop}%
\bibitem [{\citenamefont {Liu}\ \emph {et~al.}(2013)\citenamefont {Liu},
  \citenamefont {Li}, \citenamefont {Li}, \citenamefont {Li},\ and\
  \citenamefont {Lu}}]{Liu2013}%
  \BibitemOpen
  \bibfield  {author} {\bibinfo {author} {\bibfnamefont {X.}~\bibnamefont
  {Liu}}, \bibinfo {author} {\bibfnamefont {L.}~\bibnamefont {Li}}, \bibinfo
  {author} {\bibfnamefont {Q.}~\bibnamefont {Li}}, \bibinfo {author}
  {\bibfnamefont {Y.}~\bibnamefont {Li}}, \ and\ \bibinfo {author}
  {\bibfnamefont {F.}~\bibnamefont {Lu}},\ }\href
  {http://www.sciencedirect.com/science/article/pii/S1369800113001212}
  {\bibfield  {journal} {\bibinfo  {journal} {Materials Science in
  Semiconductor Processing}\ }\textbf {\bibinfo {volume} {16}},\ \bibinfo
  {pages} {1369 } (\bibinfo {year} {2013})}\BibitemShut {NoStop}%
\bibitem [{\citenamefont {Gildenblat}\ and\ \citenamefont
  {Schmidt}(1996)}]{Gildenblat1996}%
  \BibitemOpen
  \bibfield  {author} {\bibinfo {author} {\bibfnamefont {G.}~\bibnamefont
  {Gildenblat}}\ and\ \bibinfo {author} {\bibfnamefont {P.}~\bibnamefont
  {Schmidt}},\ }\href
  {http://www.worldscientific.com/worldscibooks/10.1142/2046} {\emph {\bibinfo
  {title} {Handbook Series on Semiconductor Parameters}}}\ (\bibinfo
  {publisher} {World Scientific},\ \bibinfo {year} {1996})\ pp.\ \bibinfo
  {pages} {58--76}\BibitemShut {NoStop}%
\bibitem [{\citenamefont {Patrick}\ and\ \citenamefont
  {Giustino}(2014)}]{Patrick2014}%
  \BibitemOpen
  \bibfield  {author} {\bibinfo {author} {\bibfnamefont {C.~E.}\ \bibnamefont
  {Patrick}}\ and\ \bibinfo {author} {\bibfnamefont {F.}~\bibnamefont
  {Giustino}},\ }\href {http://stacks.iop.org/0953-8984/26/i=36/a=365503}
  {\bibfield  {journal} {\bibinfo  {journal} {Journal of Physics: Condensed
  Matter}\ }\textbf {\bibinfo {volume} {26}},\ \bibinfo {pages} {365503}
  (\bibinfo {year} {2014})}\BibitemShut {NoStop}%
\bibitem [{\citenamefont {Wyckoff}(1963)}]{Wyckoff1963}%
  \BibitemOpen
  \bibfield  {author} {\bibinfo {author} {\bibfnamefont {R.}~\bibnamefont
  {Wyckoff}},\ }\href {https://archive.org/details/structureofcryst030914mbp}
  {\emph {\bibinfo {title} {Crystal Structures}}},\ Vol.~\bibinfo {volume} {1}\
  (\bibinfo  {publisher} {John Wiley \& Sons},\ \bibinfo {year}
  {1963})\BibitemShut {NoStop}%
\bibitem [{\citenamefont {Fuchs}\ and\ \citenamefont
  {Scheffler}(1999)}]{Fuchs1999}%
  \BibitemOpen
  \bibfield  {author} {\bibinfo {author} {\bibfnamefont {M.}~\bibnamefont
  {Fuchs}}\ and\ \bibinfo {author} {\bibfnamefont {M.}~\bibnamefont
  {Scheffler}},\ }\href
  {http://www.sciencedirect.com/science/article/pii/S001046559800201X}
  {\bibfield  {journal} {\bibinfo  {journal} {Computer Physics Communications}\
  }\textbf {\bibinfo {volume} {119}},\ \bibinfo {pages} {67 } (\bibinfo {year}
  {1999})}\BibitemShut {NoStop}%
\bibitem [{\citenamefont {Monkhorst}\ and\ \citenamefont
  {Pack}(1976)}]{Monkhorst1976}%
  \BibitemOpen
  \bibfield  {author} {\bibinfo {author} {\bibfnamefont {H.~J.}\ \bibnamefont
  {Monkhorst}}\ and\ \bibinfo {author} {\bibfnamefont {J.~D.}\ \bibnamefont
  {Pack}},\ }\href@noop {} {\bibfield  {journal} {\bibinfo  {journal} {Phys.
  Rev. B}\ }\textbf {\bibinfo {volume} {13}},\ \bibinfo {pages} {5188}
  (\bibinfo {year} {1976})}\BibitemShut {NoStop}%
\bibitem [{\citenamefont {Perdew}\ and\ \citenamefont
  {Zunger}(1981)}]{Perdew1981}%
  \BibitemOpen
  \bibfield  {author} {\bibinfo {author} {\bibfnamefont {J.~P.}\ \bibnamefont
  {Perdew}}\ and\ \bibinfo {author} {\bibfnamefont {A.}~\bibnamefont
  {Zunger}},\ }\href@noop {} {\bibfield  {journal} {\bibinfo  {journal} {Phys.
  Rev. B}\ }\textbf {\bibinfo {volume} {23}},\ \bibinfo {pages} {5048}
  (\bibinfo {year} {1981})}\BibitemShut {NoStop}%
\bibitem [{\citenamefont {Cardona}\ and\ \citenamefont
  {Thewalt}(2005)}]{Cardona2005a}%
  \BibitemOpen
  \bibfield  {author} {\bibinfo {author} {\bibfnamefont {M.}~\bibnamefont
  {Cardona}}\ and\ \bibinfo {author} {\bibfnamefont {M.~L.~W.}\ \bibnamefont
  {Thewalt}},\ }\href {http://link.aps.org/pdf/10.1103/RevModPhys.77.1173}
  {\bibfield  {journal} {\bibinfo  {journal} {Rev. Mod. Phys.}\ }\textbf
  {\bibinfo {volume} {77}},\ \bibinfo {pages} {1173} (\bibinfo {year}
  {2005})}\BibitemShut {NoStop}%
\bibitem [{\citenamefont {Vogel}, \citenamefont {Kr\"uger},\ and\ \citenamefont
  {Pollmann}(1997)}]{Vogel1997}%
  \BibitemOpen
  \bibfield  {author} {\bibinfo {author} {\bibfnamefont {D.}~\bibnamefont
  {Vogel}}, \bibinfo {author} {\bibfnamefont {P.}~\bibnamefont {Kr\"uger}}, \
  and\ \bibinfo {author} {\bibfnamefont {J.}~\bibnamefont {Pollmann}},\ }\href
  {http://link.aps.org/doi/10.1103/PhysRevB.55.12836} {\bibfield  {journal}
  {\bibinfo  {journal} {Phys. Rev. B}\ }\textbf {\bibinfo {volume} {55}},\
  \bibinfo {pages} {12836} (\bibinfo {year} {1997})}\BibitemShut {NoStop}%
\bibitem [{\citenamefont {Perry}\ and\ \citenamefont {Rutz}(1978)}]{Perry1978}%
  \BibitemOpen
  \bibfield  {author} {\bibinfo {author} {\bibfnamefont {P.~B.}\ \bibnamefont
  {Perry}}\ and\ \bibinfo {author} {\bibfnamefont {R.~F.}\ \bibnamefont
  {Rutz}},\ }\href@noop {} {\bibfield  {journal} {\bibinfo  {journal} {Applied
  Physics Letters}\ }\textbf {\bibinfo {volume} {33}},\ \bibinfo {pages} {319}
  (\bibinfo {year} {1978})}\BibitemShut {NoStop}%
\bibitem [{\citenamefont {Litimein}\ \emph {et~al.}(2002)\citenamefont
  {Litimein}, \citenamefont {Bouhafs}, \citenamefont {Dridi},\ and\
  \citenamefont {Ruterana}}]{Litimein2002}%
  \BibitemOpen
  \bibfield  {author} {\bibinfo {author} {\bibfnamefont {F.}~\bibnamefont
  {Litimein}}, \bibinfo {author} {\bibfnamefont {B.}~\bibnamefont {Bouhafs}},
  \bibinfo {author} {\bibfnamefont {Z.}~\bibnamefont {Dridi}}, \ and\ \bibinfo
  {author} {\bibfnamefont {P.}~\bibnamefont {Ruterana}},\ }\href
  {http://iopscience.iop.org/1367-2630/4/1/364} {\bibfield  {journal} {\bibinfo
   {journal} {New Journal of Physics}\ }\textbf {\bibinfo {volume} {4}},\
  \bibinfo {pages} {64} (\bibinfo {year} {2002})}\BibitemShut {NoStop}%
\bibitem [{\citenamefont {Christensen}\ and\ \citenamefont
  {Gorczyca}(1994)}]{Christensen1994}%
  \BibitemOpen
  \bibfield  {author} {\bibinfo {author} {\bibfnamefont {N.~E.}\ \bibnamefont
  {Christensen}}\ and\ \bibinfo {author} {\bibfnamefont {I.}~\bibnamefont
  {Gorczyca}},\ }\href
  {http://journals.aps.org/prb/abstract/10.1103/PhysRevB.50.4397} {\bibfield
  {journal} {\bibinfo  {journal} {Phys. Rev. B}\ }\textbf {\bibinfo {volume}
  {50}},\ \bibinfo {pages} {4397} (\bibinfo {year} {1994})}\BibitemShut
  {NoStop}%
\bibitem [{\citenamefont {Christensen}\ and\ \citenamefont
  {Gorczyca}(1993)}]{Christensen1993}%
  \BibitemOpen
  \bibfield  {author} {\bibinfo {author} {\bibfnamefont {N.~E.}\ \bibnamefont
  {Christensen}}\ and\ \bibinfo {author} {\bibfnamefont {I.}~\bibnamefont
  {Gorczyca}},\ }\href {http://link.aps.org/doi/10.1103/PhysRevB.47.4307}
  {\bibfield  {journal} {\bibinfo  {journal} {Phys. Rev. B}\ }\textbf {\bibinfo
  {volume} {47}},\ \bibinfo {pages} {4307} (\bibinfo {year}
  {1993})}\BibitemShut {NoStop}%
\bibitem [{\citenamefont {Rubio}\ \emph {et~al.}(1993)\citenamefont {Rubio},
  \citenamefont {Corkill}, \citenamefont {Cohen}, \citenamefont {Shirley},\
  and\ \citenamefont {Louie}}]{Rubio1993}%
  \BibitemOpen
  \bibfield  {author} {\bibinfo {author} {\bibfnamefont {A.}~\bibnamefont
  {Rubio}}, \bibinfo {author} {\bibfnamefont {J.}~\bibnamefont {Corkill}},
  \bibinfo {author} {\bibfnamefont {M.}~\bibnamefont {Cohen}}, \bibinfo
  {author} {\bibfnamefont {E.}~\bibnamefont {Shirley}}, \ and\ \bibinfo
  {author} {\bibfnamefont {S.}~\bibnamefont {Louie}},\ }\href
  {http://link.aps.org/doi/10.1103/PhysRevB.48.11810} {\bibfield  {journal}
  {\bibinfo  {journal} {Phys. Rev. B}\ }\textbf {\bibinfo {volume} {48}},\
  \bibinfo {pages} {11810} (\bibinfo {year} {1993})}\BibitemShut {NoStop}%
\bibitem [{\citenamefont {Prasad}\ and\ \citenamefont
  {Dubey}(1984)}]{Prasad1984}%
  \BibitemOpen
  \bibfield  {author} {\bibinfo {author} {\bibfnamefont {C.}~\bibnamefont
  {Prasad}}\ and\ \bibinfo {author} {\bibfnamefont {J.}~\bibnamefont {Dubey}},\
  }\href {http://onlinelibrary.wiley.com/doi/10.1002/pssb.2221250223/abstract}
  {\bibfield  {journal} {\bibinfo  {journal} {Phys. Status Solidi (b)}\
  }\textbf {\bibinfo {volume} {125}},\ \bibinfo {pages} {625} (\bibinfo {year}
  {1984})}\BibitemShut {NoStop}%
\bibitem [{\citenamefont {Coleburn}\ and\ \citenamefont
  {Forbes}(1968)}]{Coleburn1968}%
  \BibitemOpen
  \bibfield  {author} {\bibinfo {author} {\bibfnamefont {N.~L.}\ \bibnamefont
  {Coleburn}}\ and\ \bibinfo {author} {\bibfnamefont {J.~W.}\ \bibnamefont
  {Forbes}},\ }\href
  {http://scitation.aip.org/content/aip/journal/jcp/48/2/10.1063/1.1668682}
  {\bibfield  {journal} {\bibinfo  {journal} {The Journal of Chemical Physics}\
  }\textbf {\bibinfo {volume} {48}},\ \bibinfo {pages} {555} (\bibinfo {year}
  {1968})}\BibitemShut {NoStop}%
\bibitem [{\citenamefont {Chrenko}(1974)}]{Chrenko1974}%
  \BibitemOpen
  \bibfield  {author} {\bibinfo {author} {\bibfnamefont {R.}~\bibnamefont
  {Chrenko}},\ }\href
  {http://www.sciencedirect.com/science/article/pii/0038109874909788}
  {\bibfield  {journal} {\bibinfo  {journal} {Solid State Communications}\
  }\textbf {\bibinfo {volume} {14}},\ \bibinfo {pages} {511 } (\bibinfo {year}
  {1974})}\BibitemShut {NoStop}%
\bibitem [{\citenamefont {Gillet}, \citenamefont {Giantomassi},\ and\
  \citenamefont {Gonze}(2013)}]{Gillet2013}%
  \BibitemOpen
  \bibfield  {author} {\bibinfo {author} {\bibfnamefont {Y.}~\bibnamefont
  {Gillet}}, \bibinfo {author} {\bibfnamefont {M.}~\bibnamefont {Giantomassi}},
  \ and\ \bibinfo {author} {\bibfnamefont {X.}~\bibnamefont {Gonze}},\ }\href
  {http://link.aps.org/doi/10.1103/PhysRevB.88.094305} {\bibfield  {journal}
  {\bibinfo  {journal} {Phys. Rev. B}\ }\textbf {\bibinfo {volume} {88}},\
  \bibinfo {pages} {094305} (\bibinfo {year} {2013})}\BibitemShut {NoStop}%
\bibitem [{\citenamefont {Jellison}\ and\ \citenamefont
  {Modine}(1983)}]{Jellison1983}%
  \BibitemOpen
  \bibfield  {author} {\bibinfo {author} {\bibfnamefont {G.~E.}\ \bibnamefont
  {Jellison}}\ and\ \bibinfo {author} {\bibfnamefont {F.~A.}\ \bibnamefont
  {Modine}},\ }\href {http://link.aps.org/doi/10.1103/PhysRevB.27.7466}
  {\bibfield  {journal} {\bibinfo  {journal} {Phys. Rev. B}\ }\textbf {\bibinfo
  {volume} {27}},\ \bibinfo {pages} {7466} (\bibinfo {year}
  {1983})}\BibitemShut {NoStop}%
\bibitem [{\citenamefont {Clark}, \citenamefont {Dean},\ and\ \citenamefont
  {Harris}(1964)}]{Clark1964}%
  \BibitemOpen
  \bibfield  {author} {\bibinfo {author} {\bibfnamefont {C.~D.}\ \bibnamefont
  {Clark}}, \bibinfo {author} {\bibfnamefont {P.~J.}\ \bibnamefont {Dean}}, \
  and\ \bibinfo {author} {\bibfnamefont {P.~V.}\ \bibnamefont {Harris}},\
  }\href@noop {} {\bibfield  {journal} {\bibinfo  {journal} {Proceedings of the
  Royal Society of London Series A Mathematical and Physical Sciences
  (1934-1990)}\ }\textbf {\bibinfo {volume} {277}} (\bibinfo {year}
  {1964})}\BibitemShut {NoStop}%
\bibitem [{\citenamefont {Logothetidis}\ \emph {et~al.}(1992)\citenamefont
  {Logothetidis}, \citenamefont {Petalas}, \citenamefont {Polatoglou},\ and\
  \citenamefont {Fuchs}}]{Logothetidis1992}%
  \BibitemOpen
  \bibfield  {author} {\bibinfo {author} {\bibfnamefont {S.}~\bibnamefont
  {Logothetidis}}, \bibinfo {author} {\bibfnamefont {J.}~\bibnamefont
  {Petalas}}, \bibinfo {author} {\bibfnamefont {H.~M.}\ \bibnamefont
  {Polatoglou}}, \ and\ \bibinfo {author} {\bibfnamefont {D.}~\bibnamefont
  {Fuchs}},\ }\href {http://link.aps.org/doi/10.1103/PhysRevB.46.4483}
  {\bibfield  {journal} {\bibinfo  {journal} {Phys. Rev. B}\ }\textbf {\bibinfo
  {volume} {46}},\ \bibinfo {pages} {4483} (\bibinfo {year}
  {1992})}\BibitemShut {NoStop}%
\bibitem [{\citenamefont {Cardona}(2005)}]{Cardona2005}%
  \BibitemOpen
  \bibfield  {author} {\bibinfo {author} {\bibfnamefont {M.}~\bibnamefont
  {Cardona}},\ }\href@noop {} {\bibfield  {journal} {\bibinfo  {journal} {Solid
  State Communications}\ }\textbf {\bibinfo {volume} {133}},\ \bibinfo {pages}
  {3 } (\bibinfo {year} {2005})}\BibitemShut {NoStop}%
\bibitem [{Note1()}]{Note1}%
  \BibitemOpen
  \bibinfo {note} {Logothetidis~\protect \textit {et al.} deduced the
  temperature dependence of the direct bandgap of diamond from first and
  second-derivative line-shape analysis, see Ref.~\protect \rev@citealpnum
  {Logothetidis1992} for more details.}\BibitemShut {Stop}%
\bibitem [{\citenamefont {P\"assler}(1999)}]{Passler1999}%
  \BibitemOpen
  \bibfield  {author} {\bibinfo {author} {\bibfnamefont {R.}~\bibnamefont
  {P\"assler}},\ }\href
  {http://onlinelibrary.wiley.com/doi/10.1002/%28SICI%291521-3951%28199912%29216:2%3C975::AID-PSSB975%3E3.0.CO;2-N/abstract}
  {\bibfield  {journal} {\bibinfo  {journal} {physica status solidi (b)}\
  }\textbf {\bibinfo {volume} {216}} (\bibinfo {year} {1999})}\BibitemShut
  {NoStop}%
\bibitem [{\citenamefont {Corathers}(2009)}]{Corathers2009}%
  \BibitemOpen
  \bibfield  {author} {\bibinfo {author} {\bibfnamefont {L.~A.}\ \bibnamefont
  {Corathers}},\ }\href@noop {} {\emph {\bibinfo {title} {2009 Minerals
  Yearbook: Silicon}}}\ (\bibinfo  {publisher} {USGC},\ \bibinfo {year}
  {2009})\BibitemShut {NoStop}%
\bibitem [{\citenamefont {Perdew}\ and\ \citenamefont
  {Wang}(1992)}]{Perdew1992}%
  \BibitemOpen
  \bibfield  {author} {\bibinfo {author} {\bibfnamefont {J.~P.}\ \bibnamefont
  {Perdew}}\ and\ \bibinfo {author} {\bibfnamefont {Y.}~\bibnamefont {Wang}},\
  }\href {http://journals.aps.org/prb/abstract/10.1103/PhysRevB.45.13244}
  {\bibfield  {journal} {\bibinfo  {journal} {Phys. Rev. B}\ }\textbf {\bibinfo
  {volume} {45}},\ \bibinfo {pages} {13244} (\bibinfo {year}
  {1992})}\BibitemShut {NoStop}%
\bibitem [{\citenamefont {Bludau}, \citenamefont {Onton},\ and\ \citenamefont
  {Heinke}(1974)}]{Bludau1974}%
  \BibitemOpen
  \bibfield  {author} {\bibinfo {author} {\bibfnamefont {W.}~\bibnamefont
  {Bludau}}, \bibinfo {author} {\bibfnamefont {A.}~\bibnamefont {Onton}}, \
  and\ \bibinfo {author} {\bibfnamefont {W.}~\bibnamefont {Heinke}},\ }\href
  {http://scitation.aip.org/content/aip/journal/jap/45/4/10.1063/1.1663501}
  {\bibfield  {journal} {\bibinfo  {journal} {Journal of Applied Physics}\
  }\textbf {\bibinfo {volume} {45}},\ \bibinfo {pages} {1846} (\bibinfo {year}
  {1974})}\BibitemShut {NoStop}%
\bibitem [{\citenamefont {Macfarlane}\ \emph {et~al.}(1958)\citenamefont
  {Macfarlane}, \citenamefont {McLean}, \citenamefont {Quarrington},\ and\
  \citenamefont {Roberts}}]{Macfarlane1958}%
  \BibitemOpen
  \bibfield  {author} {\bibinfo {author} {\bibfnamefont {G.}~\bibnamefont
  {Macfarlane}}, \bibinfo {author} {\bibfnamefont {T.}~\bibnamefont {McLean}},
  \bibinfo {author} {\bibfnamefont {J.}~\bibnamefont {Quarrington}}, \ and\
  \bibinfo {author} {\bibfnamefont {V.}~\bibnamefont {Roberts}},\ }\href
  {http://link.aps.org/doi/10.1103/PhysRev.111.1245} {\bibfield  {journal}
  {\bibinfo  {journal} {Phys. Rev.}\ }\textbf {\bibinfo {volume} {111}},\
  \bibinfo {pages} {1245} (\bibinfo {year} {1958})}\BibitemShut {NoStop}%
\bibitem [{\citenamefont {Lautenschlager}, \citenamefont {Allen},\ and\
  \citenamefont {Cardona}(1986)}]{Lautenschlager1986}%
  \BibitemOpen
  \bibfield  {author} {\bibinfo {author} {\bibfnamefont {P.}~\bibnamefont
  {Lautenschlager}}, \bibinfo {author} {\bibfnamefont {P.~B.}\ \bibnamefont
  {Allen}}, \ and\ \bibinfo {author} {\bibfnamefont {M.}~\bibnamefont
  {Cardona}},\ }\href@noop {} {\bibfield  {journal} {\bibinfo  {journal} {Phys.
  Rev. B}\ }\textbf {\bibinfo {volume} {33}},\ \bibinfo {pages} {5501}
  (\bibinfo {year} {1986})}\BibitemShut {NoStop}%
\bibitem [{\citenamefont {MacChesney}, \citenamefont {Bridenbaugh},\ and\
  \citenamefont {O’Connor}(1970)}]{MacChesney1970}%
  \BibitemOpen
  \bibfield  {author} {\bibinfo {author} {\bibfnamefont {J.}~\bibnamefont
  {MacChesney}}, \bibinfo {author} {\bibfnamefont {P.}~\bibnamefont
  {Bridenbaugh}}, \ and\ \bibinfo {author} {\bibfnamefont {P.}~\bibnamefont
  {O’Connor}},\ }\href
  {http://www.ioffe.ru/SVA/NSM/Semicond/InN/reference.html} {\bibfield
  {journal} {\bibinfo  {journal} {Mater. Res. Bull.}\ }\textbf {\bibinfo
  {volume} {5}},\ \bibinfo {pages} {783} (\bibinfo {year} {1970})}\BibitemShut
  {NoStop}%
\bibitem [{\citenamefont {Brunner}\ \emph {et~al.}(1997)\citenamefont
  {Brunner}, \citenamefont {Angerer}, \citenamefont {Bustarret}, \citenamefont
  {Freudenberg}, \citenamefont {H\"opler}, \citenamefont {Dimitrov},\ and\
  \citenamefont {Stutzmann}}]{Brunner1997}%
  \BibitemOpen
  \bibfield  {author} {\bibinfo {author} {\bibfnamefont {D.}~\bibnamefont
  {Brunner}}, \bibinfo {author} {\bibfnamefont {H.}~\bibnamefont {Angerer}},
  \bibinfo {author} {\bibfnamefont {E.}~\bibnamefont {Bustarret}}, \bibinfo
  {author} {\bibfnamefont {F.}~\bibnamefont {Freudenberg}}, \bibinfo {author}
  {\bibfnamefont {R.}~\bibnamefont {H\"opler}}, \bibinfo {author}
  {\bibfnamefont {O.}~\bibnamefont {Dimitrov}, \bibfnamefont {R.and~Ambacher}},
  \ and\ \bibinfo {author} {\bibfnamefont {M.}~\bibnamefont {Stutzmann}},\
  }\href
  {http://scitation.aip.org/content/aip/journal/jap/82/10/10.1063/1.366309}
  {\bibfield  {journal} {\bibinfo  {journal} {Journal of Applied Physics}\
  }\textbf {\bibinfo {volume} {82}} (\bibinfo {year} {1997})}\BibitemShut
  {NoStop}%
\bibitem [{\citenamefont { Guo}\ and\ \citenamefont
  { Yoshida}(1994)}]{Guo1994}%
  \BibitemOpen
  \bibfield  {author} {\bibinfo {author} {\bibfnamefont {Q.}~\bibnamefont
  { Guo}}\ and\ \bibinfo {author} {\bibfnamefont {A.}~\bibnamefont
  { Yoshida}},\ }\href {http://stacks.iop.org/1347-4065/33/i=5R/a=2453}
  {\bibfield  {journal} {\bibinfo  {journal} {Japanese Journal of Applied
  Physics}\ }\textbf {\bibinfo {volume} {33}},\ \bibinfo {pages} {2453}
  (\bibinfo {year} {1994})}\BibitemShut {NoStop}%
\bibitem [{\citenamefont {Yu}\ \emph {et~al.}(2003)\citenamefont {Yu},
  \citenamefont {Lau}, \citenamefont {Chan}, \citenamefont {Liu},\ and\
  \citenamefont {Zheng}}]{Yu2003}%
  \BibitemOpen
  \bibfield  {author} {\bibinfo {author} {\bibfnamefont {W.~J.}\ \bibnamefont
  {Yu}}, \bibinfo {author} {\bibfnamefont {W.~M.}\ \bibnamefont {Lau}},
  \bibinfo {author} {\bibfnamefont {S.~P.}\ \bibnamefont {Chan}}, \bibinfo
  {author} {\bibfnamefont {Z.~F.}\ \bibnamefont {Liu}}, \ and\ \bibinfo
  {author} {\bibfnamefont {Q.~Q.}\ \bibnamefont {Zheng}},\ }\href
  {http://ieeexplore.ieee.org/xpl/login.jsp?tp=&arnumber=4847564&url=http%3A%2F%2Fieeexplore.ieee.org%2Fiel5%2F4816218%2F4847542%2F04847564.pdf%3Farnumber%3D4847564}
  {\bibfield  {journal} {\bibinfo  {journal} {Phys. Rev. B}\ }\textbf {\bibinfo
  {volume} {67}},\ \bibinfo {pages} {014108} (\bibinfo {year}
  {2003})}\BibitemShut {NoStop}%
\bibitem [{\citenamefont {Hybertsen}\ and\ \citenamefont
  {Louie}(1987)}]{Hybertsen1987}%
  \BibitemOpen
  \bibfield  {author} {\bibinfo {author} {\bibfnamefont {M.~S.}\ \bibnamefont
  {Hybertsen}}\ and\ \bibinfo {author} {\bibfnamefont {S.~G.}\ \bibnamefont
  {Louie}},\ }\href {\doibase 10.1103/PhysRevB.35.5585} {\bibfield  {journal}
  {\bibinfo  {journal} {Phys. Rev. B}\ }\textbf {\bibinfo {volume} {35}},\
  \bibinfo {pages} {5585} (\bibinfo {year} {1987})}\BibitemShut {NoStop}%
\end{thebibliography}%
\end{document}